\RequirePackage{fix-cm}
\documentclass{svjour3}                     
\smartqed  
\usepackage{graphicx}
\usepackage{natbib}
\usepackage{siunitx}
\usepackage{bm}
\usepackage{amsmath}
\usepackage{amssymb}
\usepackage{multirow} 
\usepackage{subfigure}
\usepackage[T1]{fontenc}

\begin{document}

\title{Transfer design between neighborhoods of planetary moons in the circular restricted three-body problem
}
\subtitle{The Moon-to-Moon Analytical Transfer Method}

\titlerunning{The Moon-to-Moon Analytical Transfer Method}        

\author{David Canales        \and
        Kathleen C. Howell     \and
        Elena Fantino
}


\institute{David Canales \at
              School of Aeronautics and Astronautics, Purdue University, West Lafayette, IN 47907\\
              Tel.: +34-628-929-331\\
              \email{dcanales@purdue.edu}            \\
	      ORCID: 0000-0003-0166-2391
           \and
           Kathleen C. Howell \at
              School of Aeronautics and Astronautics, Purdue University, West Lafayette, IN 47907\\
              Tel.: +1-765-494-5786 \\
              \email{howell@purdue.edu}           \\
              ORCID: 0000-0002-1298-5017
           \and
           Elena Fantino \at
              Aerospace Engineering Department, Khalifa University of Science and Technology, P.O. Box 127788, Abu Dhabi,
United Arab Emirates\\
              Tel.: +971-2-312-4014\\
              \email{elena.fantino@ku.ac.ae}            \\
              ORCID: 0000-0001-7633-8567
}

\date{Received: date / Accepted: date}

\maketitle

\begin{abstract}
Given the interest in future space missions devoted to the exploration of key moons in the solar system and that may involve libration point orbits, an efficient design strategy for transfers between moons is introduced that leverages the dynamics in these multi-body systems. The moon-to-moon analytical transfer (MMAT) method is introduced, comprised of a general methodology for transfer design between the vicinities of the moons in any given
system within the context of the circular restricted three-body problem, useful regardless of the orbital planes in which the moons reside. A simplified model enables analytical constraints to efficiently determine the feasibility of a transfer between two different moons moving in the vicinity of a common planet. In particular, connections between the periodic orbits of such two different moons are achieved. The strategy is applicable for any type of direct transfers that satisfy the analytical constraints. Case studies are presented for the Jovian and Uranian systems. The transition of the transfers into higher-fidelity ephemeris models confirms the validity of the MMAT method as a fast tool to provide possible transfer options between two consecutive moons.
\end{abstract}


\section{Introduction}
\label{intro}
The exploration of planetary moons has always played a crucial role for an improved understanding of the solar system and in the search for life beyond Earth. Past milestone missions to the gas giants incorporated tours
of the moons and, thus, satisfied multiple science objectives at different targets simultaneously. For example, for the Galileo spacecraft, launched in 1989, \citet{doi:10.2514/6.1983-101} designed a 23-month tour of the Jovian system that was successfully accomplished in the 1990s. For the Cassini-Huygens mission, launched in 1997, \citet{doi:10.2514/6.2002-4720} introduced a tour around the Saturnian system to meet scientific objectives that demonstrates the advantage of this category of trajectories for this type of missions. Scientific questions motivate new mission proposals from various worldwide space agencies. For example, ESA's JUICE mission, presented in \citet{JUICE}, aims at studying Ganymede, Callisto and Europa. NASA plans include Europa Clipper \citep{doi:10.1002/2014EO200002} for the exploration of Europa, as well as Dragonfly \citep{dragonfly}, whose
goal is landing a robot on Titan.

The construction of transfers between moons in a multi-body environment involves the design of increasingly complex mission scenarios that require effective trajectories with low propellant consumption. The potential to transfer from one moon to another is constrained by multiple parameters. For example, a critical constraint is the relative locations of the moons at a given epoch; transfers between moons are efficiently achieved only for certain relative geometries.  Also, even when the angle between the orbital planes is small, designing transfers between these moons is not trivial. As a result, for a feasible transfer, a combination of appropriate rendezvous maneuvers guarantees that the spacecraft arrives in the desired location at the correct epoch. The goal is a strategy to efficiently design these relative orientations such that both the propellant and the time required to accomplish the transfers are reduced. 

\citet{Koon2000} tackle this problem with a strategy that leverages the coupled circular restricted three-body problem (coupled CR3BP) and locates connections between moons using Poincar\'e
sections. \citet{Kakoi2014} and \citet{Short:2015aa} expand on this approach for different systems and types of orbits. Generally, in such analyses, the identification of a suitable relative phase between the moons
is not considered, and many studies assume that the moon orbits are coplanar. Therefore, for the selected relative orientations, the resulting transfers might not intersect in space since the moons are actually moving in distinct planes. An alternative approach, one that also assumes the moon orbits to be coplanar is presented by \citet{Fantino2016} with an expanded analysis in \citet{FantinoHaloTransfers} and \citet{Fantino2020}. In this analysis, departure and arrival trajectories are propagated to a certain distance from the moons, where the states that represent these trajectories are evaluated as conic arcs and any available tangential connections are determined, producing an optimal relative phase between the moons as well as an analytically minimum-$\Delta v_{tot}$ option. However, despite having evaluated the error in the conic approximation in the coupled planar CR3BP \citep{Fantino2016}, this result is not yet validated with a higher-fidelity
ephemeris model. To identify a connection between invariant manifolds or other orbits in the vicinity of the two moons at a given epoch, Lambert arcs are also proposed in \citet{articleLambert}. Yet, the resulting transfers are likely to be more expensive due to the relative orientation between the moons.

Another common approach to construct transfers between moons involves multi-gravity assists; strategies to exploit the moon gravity are examined via Tisserand
graphs in the two-body problem (2BP) allowing multiple flybys at low cost. Notable contributors include those of \citet{COLASURDO2014190} and \citet{Lynam:2011aa}. \citet{Campagnola:2012aa} inspired gravity assisted flyby analysis using Tisserand graphs in the CR3BP, expanding any 2BP analysis by including third-body perturbations. Also, utilizing a patched model that includes both the dynamics of the 2BP and the CR3BP, \citet{GroverRoss} introduce a semi-analytical method for multi-moon orbiters based on Keplerian maps to decrease the time involving the gravity assist sequences and \citet{LANTOINE20111361} combine the resonant gravity assist technique with manifolds that originate from Lyapunov periodic orbits. Most preliminary results using this technique include the assumption that the moons are in coplanar orbits. Implementation of such an approach in moon-to-moon transfer design reduces the $\Delta v_{tot}$ budget, but the transfer time-of-flight ($t_{tot}$) is sometimes considerably increased and the tour design process is computationally intensive. Although all the moons can be eventually encountered, the tour design is also challenging for other complex mission scenarios, e.g., leveraging libration
point orbits, captures or even returning to a previously visited moon, since the gravity assist sequence is affected. As a result, the technique is sometimes adjusted on a case-by-case basis.

The focus of the present investigation is an alternative general methodology for transfer design between moons applicable to any given system; in particular, the use of dynamical structures in the three-body problem is examined. 
Using an analytical formulation from the 2BP, this strategy enables construction of transfers in the coupled spatial CR3BP between moons in different planes, owing to the exploration of the problem in space. Also, for any transfer between moons where the analytical formulation
is valid and, for any given angle of departure from the departure moon, a potential transfer configuration is produced. For the resulting transfer configuration, the relative locations of the moons in their respective planes are obtained, as well as the required total $\Delta v_{tot}$ and transfer time-of-flight. This methodology is denoted the Moon-to-Moon Analytical Transfer method: the MMAT Method. Additional insights are then possible when designing tours. For example, potential new options may emerge, such as transfers between different periodic
orbits or alternatives for captures.  The MMAT scheme is applied to different case studies in the Jovian and Uranian systems (see Table \ref{Table:DataEuropaGanymede} for systems data), where transfers between spatial and planar periodic orbits are produced. Given the geometrical nature of the analytical relationships, the method is applicable to any type of direct transfer between moons. The transition of the transfers into higher-fidelity ephemeris models confirms the validity of the technique as a fast strategy to design transfers between moons that are viable for real applications.
 
\begin{table}[htbp!]
\caption{Orbital data for Europa, Ganymede, Titania and Oberon obtained from the SPICE database and referred to the Ecliptic J2000.0 reference frame \citep{SPICE}. Last accessed 08/05/2020. }
\label{Table:DataEuropaGanymede}
\centering 
\begin{tabular}{lclclclclclclc}
\hline\noalign{\smallskip}
& &  & &  &   & Longitude \\
      &		Semi-major 	     &	Orbital	 &CR3BP   &  & & ascending\\
&	axis     & period & mass ratio & Eccentricity  & Inclination& node\\
&	 {[}$10^{5}$ km{]}    &  {[}day{]}& {[}$10^{-5}${]}   &{[}$10^{-3}${]}  & {[}degree{]} &{[}degree{]}\\
\noalign{\smallskip}\hline\noalign{\smallskip}
	Europa & $\phantom{0}6.713$& \phantom{0}3.554 & 2.528 & 9.170 & \phantom{0}2.150 & 331.361\\
\noalign{\smallskip}\hline
	Ganymede & $10.706$ & \phantom{0}7.158 & 7.804 & 2.542 & \phantom{0}2.208 & 340.274\\
\noalign{\smallskip}\hline\noalign{\smallskip}
    	Titania &$4.363$ & \phantom{0}8.708 & 3.917 & 1.871 & 97.829 & 167.627\\
\noalign{\smallskip}\hline
	Oberon & $5.836$& 13.471 & 3.544 & 1.170 & 97.853 & 167.720\\
\noalign{\smallskip}\hline
\end{tabular}
\end{table}

The dynamical models employed in this investigation are introduced in Section \ref{sec:DynamicalModels}. The purpose of Section \ref{sec:coupledCR3BPMoonTransfers} is to discuss some issues that arise when designing moon-to-moon transfers using the coupled spatial CR3BP. Such challenges are addressed in Section \ref{sec:MMATMethod} where the MMAT method and its features are illustrated. Here, the analytical approach offers a useful methodology for moon-to-moon transfers leveraging CR3BP structures. The methodology is demonstrated in a sample scenario with transfers between Ganymede and Europa. Results are compared between a coplanar assumption for the moon orbits and moon motion modeled in terms of the actual inclined planes. The extension of the MMAT method to the spatial application is illustrated via transfers between Titania and Oberon. The transfer results from the MMAT method are then validated in a higher-fidelity ephemeris model in Section \ref{sec:ephemerisValidation}. Finally, some concluding remarks, as well as a summary of the technique, are presented in Section \ref{sec:Conclusions}. 

\section{\label{sec:DynamicalModels}Dynamical Models}
In a multi-body environment with one planet and multiple moons, a spacecraft (s/c) trajectory is subject to  the gravitational accelerations of several bodies simultaneously. Therefore, these gravitational accelerations must be incorporated for accurate design of arcs between moons. For the purpose of this investigation, two variants employing the CR3BP are adopted to construct transfers between moons orbiting a common planet: a spatial 2BP-CR3BP patched model and a coupled spatial CR3BP. Finally, the resulting moon-to-moon transfers in the coupled spatial CR3BP are transitioned to a higher-fidelity ephemeris model to validate the moon-to-moon transfers obtained in the spatial 2BP-CR3BP patched model.

\subsection{\label{subsec:CR3BP}The Circular Restricted Three-Body Problem}
When analyzing the motion in the vicinity of a moon, \citet{poincare1892} introduced the CR3BP, a simplified model that offers useful insight into trajectories and the general dynamical flow
for many different applications. Designs constructed
in such a framework are useful for preliminary analysis and usually straightforwardly transitioned to higher-fidelity
ephemeris models. 

\begin{figure}
\hfill{}\centering\includegraphics[width=10cm]{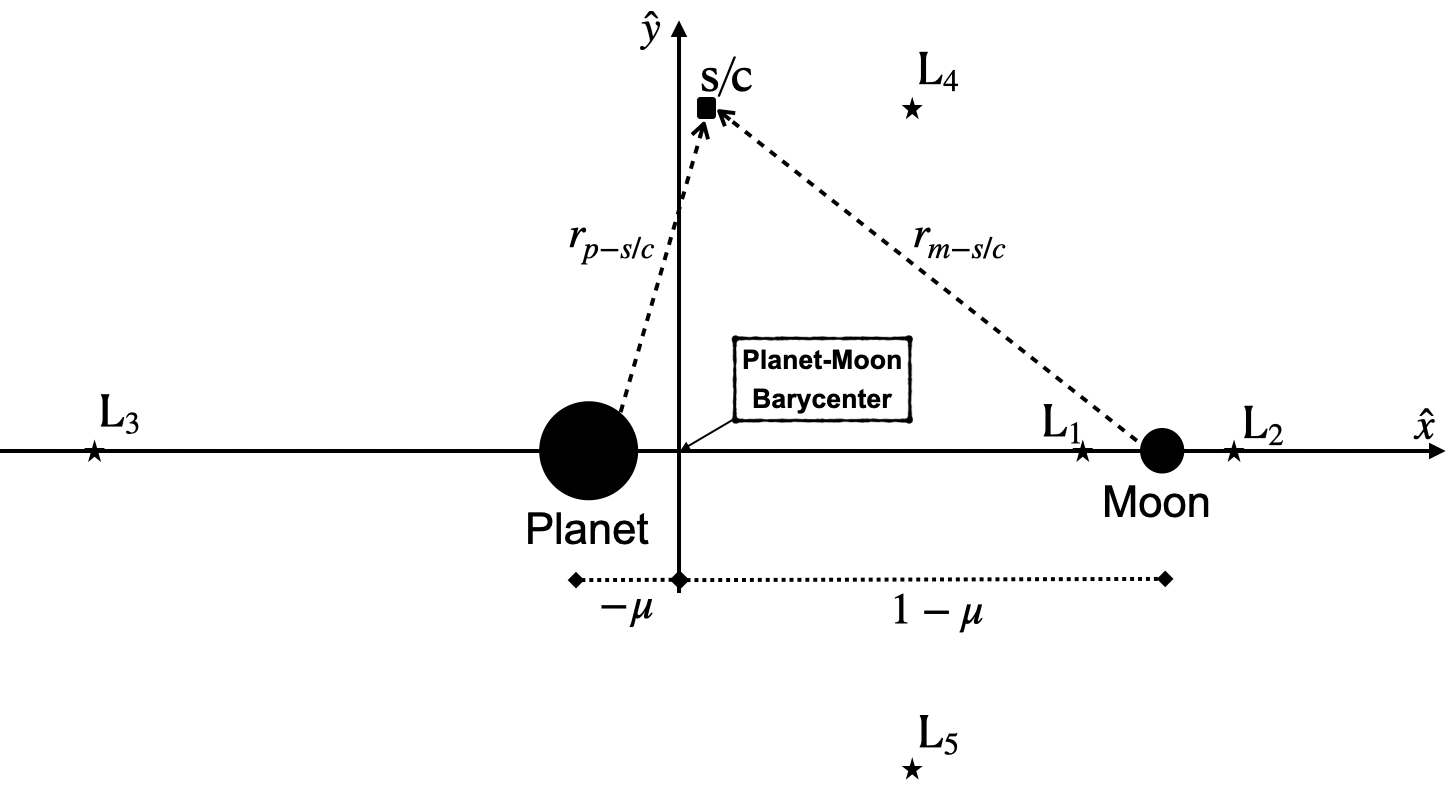}\hfill{}
\caption{\label{fig:schematicCR3BP}CR3BP: the barycentric rotating reference frame with the primaries, the third body and the 5 equilibrium points. }
\end{figure}

For
the purposes of this investigation, the CR3BP describes the
motion of a s/c subject to the gravitational forces of a larger (planet) and a smaller (moon) primary,  
assumed to move in circular orbits about the center of mass of the system. Most moons in the solar system actually 
possess very small orbital eccentricities ($e$); the Earth-Moon system is characterized by one
of the largest eccentricities at a value of 0.055. To model the problem, a system of differential equations in the CR3BP is written in dimensionless form 
 such that the characteristic distance is 
defined as the constant distance between the planet and the moon; the characteristic time is selected
 to guarantee a dimensionless mean motion
of the primaries with a value equal to unity. Note that, henceforth, dimensionless units are denoted as [nondim] in figures and tables. The mass ratio $\mu=m_{m}/(m_{m}+m_{p})$
is defined as the characteristic mass parameter of the system, with $m_{m}$ and $m_{p}$  being the masses of the moon and the planet, respectively. A barycentric rotating frame
is defined with the $\hat{x}$-axis directed from the planet-moon barycenter to the moon, 
and the $\hat{z}$-axis from the barycenter in the direction of the system angular momentum vector. Figure \ref{fig:schematicCR3BP} is a schematic representation of the model. The planet and the moon are located at positions $\bar{r}_{p}=[-\mu,0,0]^T$ and $\bar{r}_{m}=[1-\mu,0,0]^T$, respectively. Note that overbars denote vectors whereas the superscript '$T$' indicates the vector transpose. The evolution of the s/c position $\bar{r}_{rot}=[x,y,z]^T$ and velocity $\dot{\bar{r}}_{rot}=[\dot{x},\dot{y},\dot{z}]^T$ is governed by the following equations of motion:
\begin{align} 
\label{eq:cr3bpEOMS}
\displaystyle
\ddot{x}-2\dot{y}&=\frac{\partial U^{*}}{\partial x}\\
\displaystyle \ddot{y}+2\dot{x}&=\frac{\partial U^{*}}{\partial y}\\
\label{eq:cr3bpEOMS2}
\displaystyle \ddot{z}&=\frac{\partial U^{*}}{\partial z}
\end{align}
where dots indicate derivatives with respect to dimensionless time. Consequently, $x$, $y$, $z$, $\dot{x}$, $\dot{y}$, $\dot{z}$, $\ddot{x}$, $\ddot{y}$, and $\ddot{z}$ are dimensionless. Henceforth, the position states $x$, $y$, $z$ are denoted as X [nondim], Y [nondim] and Z  [nondim] (respectively) in the figures. Then, 
\begin{equation}
U^{*}=\frac{1-\mu}{r_{p-s/c}}+\frac{\mu}{r_{m-s/c}}+\frac{1}{2}(x^2+y^2)
\end{equation}
represents the pseudo-potential function for the system of differential equations; $r_{p-s/c}$
and $r_{m-s/c}$ are the distances of the s/c to the planet and the moon, respectively. The Jacobi constant ($JC$)
is the only scalar integral for the given system, i.e., 
\begin{equation}
JC=2U^{*}-(\dot{x}^{2}+\dot{y}^{2}+\dot{z}^{2}).
\end{equation} 
Note that those locations where the velocity and acceleration terms in Eq. \eqref{eq:cr3bpEOMS}-\eqref{eq:cr3bpEOMS2} are null define the zero velocity curves (ZVCs) in the CR3BP system, associated with a specified value of $JC$. Finally, note that motion exists 
in the vicinity of five equilibrium solutions
in the given formulation. Such equilibrium solutions are denoted as $\text{L}_1$ to $\text{L}_5$ in Fig. \ref{fig:schematicCR3BP}. The motion is frequently categorized by different types of families of periodic and quasi-periodic orbits according to their
geometry and stability, as demonstrated by various authors, e.g., \citet{brouckePOs1968,Campbell1999}. Examples are the planar Lyapunov orbit families and the three-dimensional halo orbit families. These orbits are leveraged for many different types of
mission scenarios. Consistent with their stability properties, hyperbolic invariant manifolds
that emanate from such periodic orbits often serve as
pathways in the vicinity of the moons and to locate connections between
periodic orbits within the same system, as demonstrated in \cite{Haapala2015}.

With one planet and two moons, each planet-moon system
possesses its own characteristic quantities and angular velocities,
and the moons' orbits are not located in the same plane. Given that the CR3BP only incorporates two bodies affecting the dynamics of the s/c, it is challenging to construct valid preliminary trajectories to travel from one moon
to another even within the context of the CR3BP. Hence, it is useful to adopt some simplifications. First, the dynamics of the s/c in the vicinity of the moons is approximated with
the CR3BP. Consequently, both moons are assumed to be moving in circular obits around their common planet. Additionally, the planes of the moon orbits are defined by the appropriate epoch in the Ecliptic J2000 frame, but the moons move in their respective circular orbits in their recognized orbital plane. Three models are introduced to address the transfer design between distinct moons: (a) a two-body/three-body patched model (2BP-CR3BP patched model), (b) a coupled three-body formulation (coupled spatial CR3BP) and (c) a higher-fidelity ephemeris model.

\subsection{\label{subsec:2BP-CR3BP-patched-model}The 2BP-CR3BP patched model}
\begin{figure}
\hfill{}\centering\includegraphics[width=14cm]{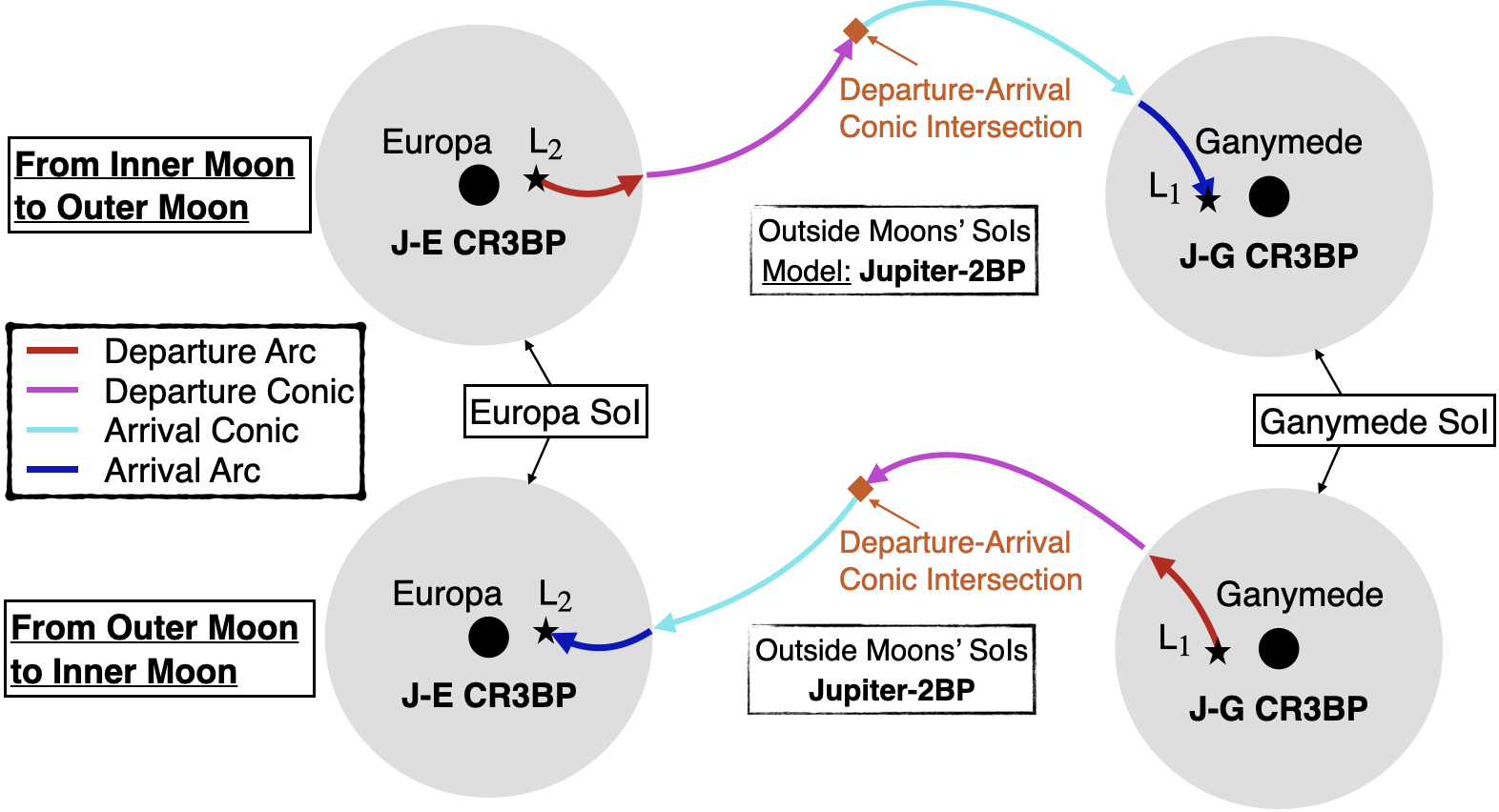}\hfill{}
\caption{\label{fig:patchedScheme}The blending of two different systems in the Ecliptic J2000.0 planet-centered inertial frame according to the 2BP-CR3BP patched model. }
\end{figure}

The 2BP-CR3BP patched model, as detailed in \citet{Fantino2016}, approximates trajectories with either the 2BP or the CR3BP depending on the location of
the s/c within the system. Trajectories in the vicinity of a moon
are modelled with the CR3BP. When these trajectories
reach a certain distance from the moon, the motion is considered
to be Keplerian with a focus at the larger primary and uniquely determined by the osculating orbital elements, namely the semi-major
axis ($a$), the eccentricity ($e$), the right ascension of the
ascending node ($\Omega$), the inclination ($i$), and the argument of
periapsis ($\omega$). The true anomaly ($\theta$) for the s/c orbit is also retrieved. Trajectories in the CR3BP are propagated up to the border of the sphere of influence (SoI) associated with the moon, where the states are determined and used to define the osculating Keplerian orbits relative to the planet. Similarly, CR3BP trajectories arriving
at a moon are propagated backwards in time towards the corresponding 
SoI, where the states are also determined and used to define the osculating back-propagated Keplerian orbits around the planet. In this way, both departure and arrival systems are blended from their respective rotating frames into a common, planet-centered frame with fixed axes; in this case, the Ecliptic J2000.0 planet-centered inertial frame. Hence, an analytical exploration of the intersection between orbits is now possible. The transformation from the rotating frame to the Ecliptic J2000.0 planet-centered inertial frame is described in Appendix \ref{appendixRotToIner}. In this blending procedure, the trajectories computed in the two systems are scaled using the same characteristic quantities. A schematic of the model appears in Fig. \ref{fig:patchedScheme}, where a sample transfer from Ganymede to Europa is employed to illustrate the concept. 

\begin{figure}
\hfill{}\centering\includegraphics[width=7.5cm]{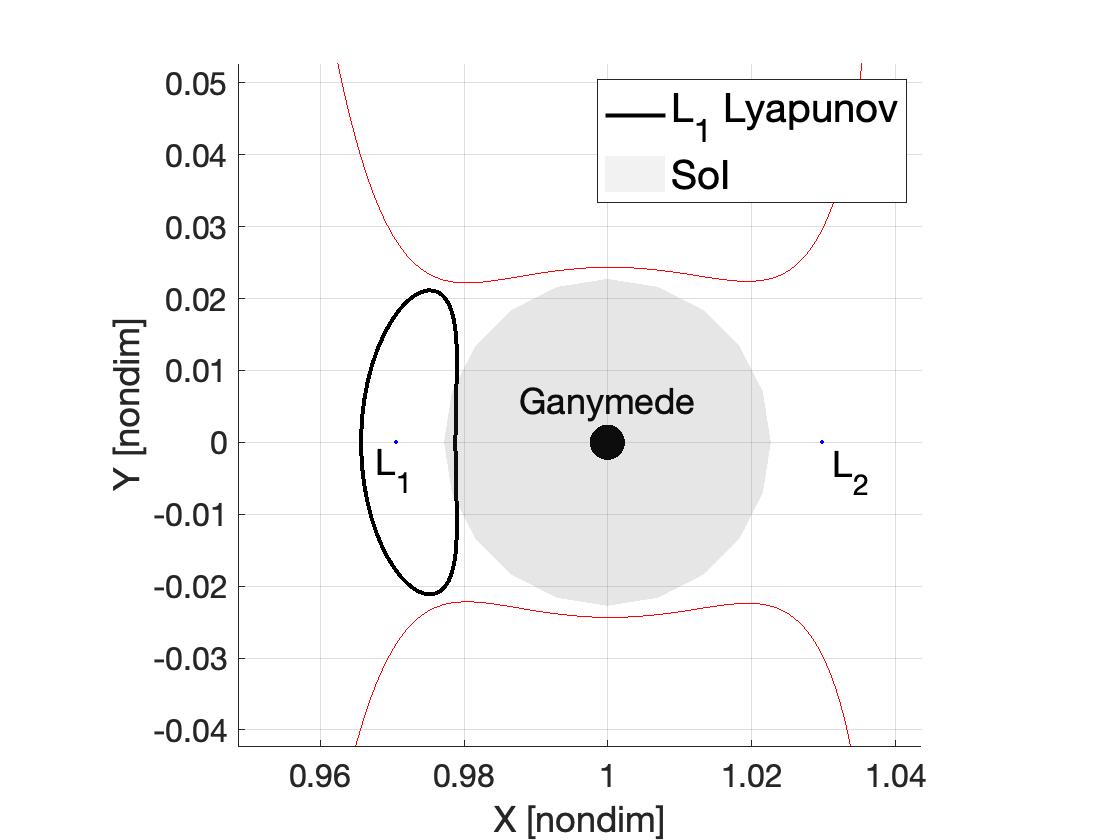}\hfill{}
\caption{\label{fig:RoyDefinition}Representation of the SoI of Ganymede (grey circle) according to the definition of \citet{roy1988} and a planar L$_1$ Lyapunov orbit with $JC=3.0061$ (Jupiter-Ganymede CR3BP and barycentric rotating reference frame).}
\end{figure}

\begin{figure}
\hfill{}\centering\includegraphics[width=7.5cm]{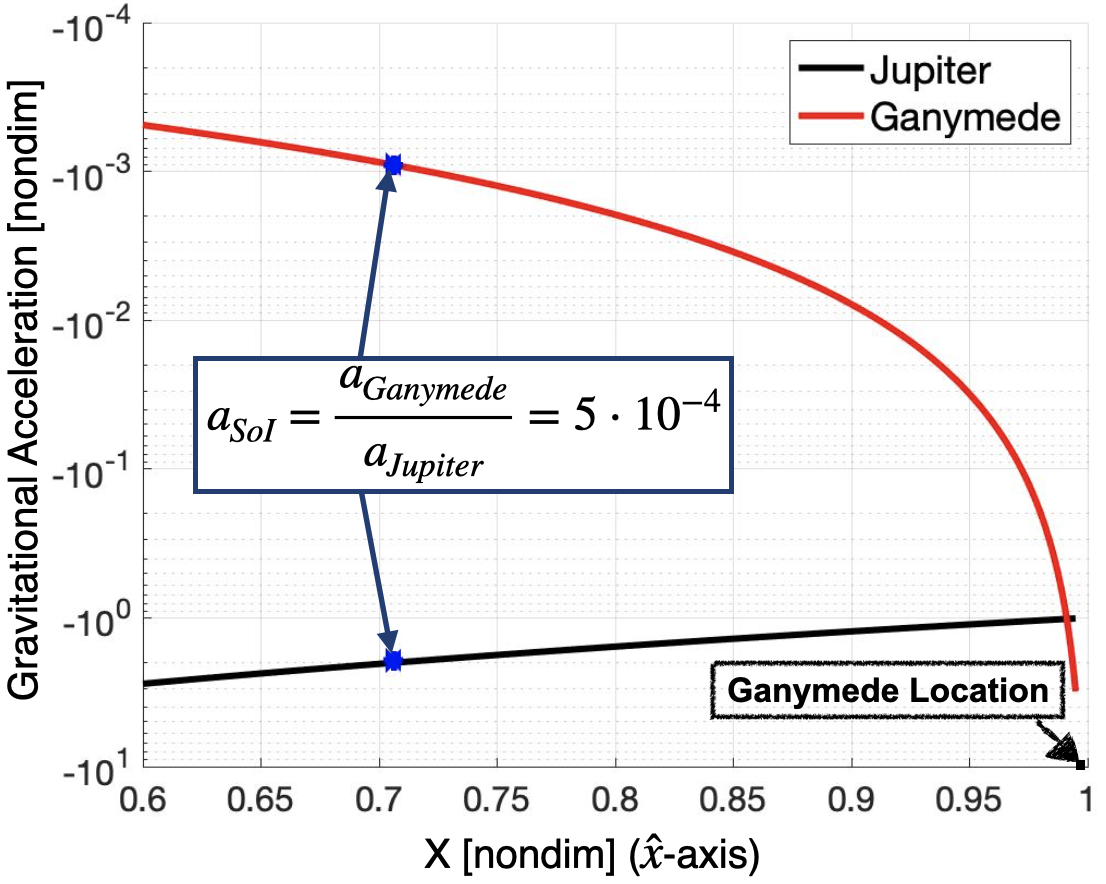}\hfill{}
\caption{\label{fig:gravitationalInfluence}Gravitational acceleration
due to Jupiter and Ganymede along the $\hat{x}$-axis. }
\end{figure}

\begin{figure}
\hfill{}\centering%
\begin{minipage}[b][1\totalheight][t]{0.5\columnwidth}%
\subfigure[A portion of a branch of the hyperbolic unstable manifold of the planar L$_1$ Lyapunov orbit with $JC=3.0061$ in the Jupiter-Ganymede CR3BP.]{\label{UnstableManifoldGanymede}{\includegraphics[width=6.5cm]{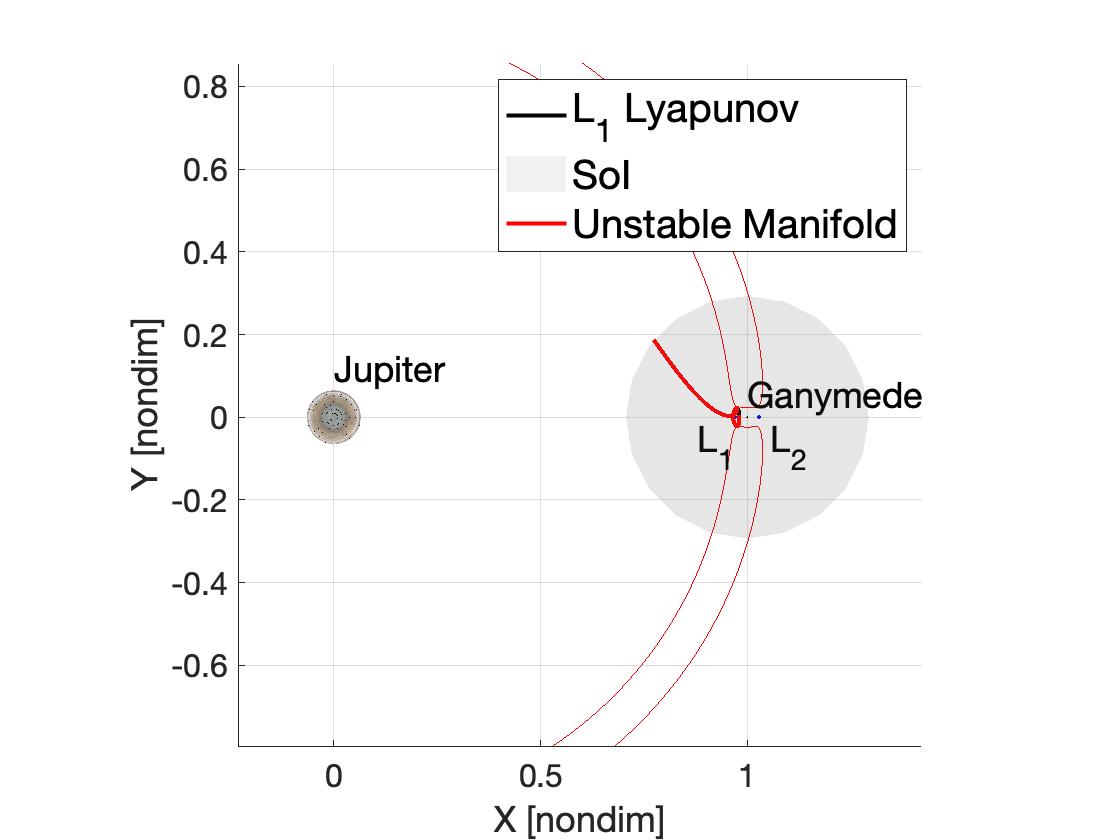}}}%
\end{minipage}\hfill{}%
\begin{minipage}[b][1\totalheight][t]{0.33\columnwidth}%
\subfigure[Osculating Keplerian orbit obtained at the crossing with Ganymede's SoI from the unstable manifold of plot a) (Ecliptic J2000.0 Jupiter-centered inertial frame).]{{\includegraphics[width=6cm]{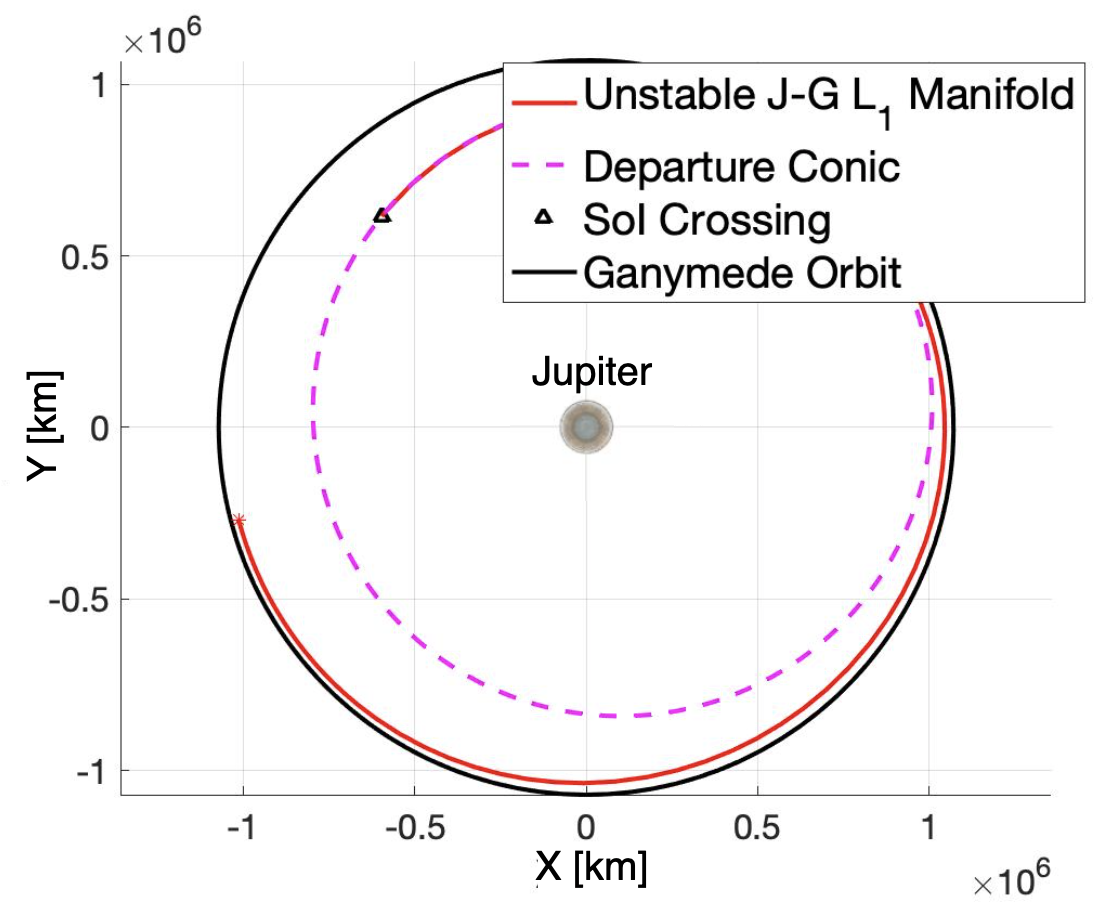}}\label{Patched2BPCR3BP}}%
\end{minipage}\hfill{}
\label{Fig:PatchedModelExample}
\caption{Patched 2BP-CR3BP model for the Jupiter-Ganymede system.}
\end{figure}

Designing transfers in the 2BP-CR3BP patched model depend on how the SoI of the moons is defined. To define the radius ($R_{SoI}$) of the SoI, the most commonly used approximation in astrodynamics is the one provided by \citet{roy1988}, 
\begin{equation}
\label{eq:classicSoI}
R_{SoI}=R_{p-m}(\frac{m_{m}}{m_{p}})^{2/5},
\end{equation}
where $R_{p-m}$ is the orbital radius of the moon. However, if CR3BP periodic orbit families are continued in the vicinity of L$_1$ or L$_2$, most of these orbits extend beyond the SoI defined by Eq. \eqref{eq:classicSoI} (see Fig. \ref{fig:RoyDefinition} for a selected L$_1$ Lyapunov orbit computed in the Jupiter-Ganymede CR3BP). Consequently, a SoI that includes the CR3BP periodic orbits of interest and their associated manifolds is required. In this investigation, the $R_{SoI}$ is defined as the distance from the moon along the $\hat{x}$-axis to the point for which the ratio $a_{SoI}=\frac{a_{m}}{a_{p}}$ is equal to a certain small quantity ($a_m$ and $a_p$ are the gravitational accelerations of the moon and the planet, respectively). An example for Jupiter and Ganymede of such a variation in gravitational influence is plotted in Fig. \ref{fig:gravitationalInfluence}. The value for $a_{SoI}$ is a free parameter for the modeling in the 2BP-CR3BP patched model. For this investigation, $a_{SoI}=5\cdot10^{-4}$ is selected because it is a sufficiently low moon gravitational acceleration for the motion of the s/c to be approximated with an elliptical orbit. The sensitivity of the moon-to-moon transfer trajectory design to the value of $a_{SoI}$ in the 2BP-CR3BP patched model is addressed later. To illustrate the 2BP-CR3BP patched model, Fig. \ref{UnstableManifoldGanymede} shows a single trajectory along the hyperbolic unstable manifold of a planar L$_1$ Lyapunov orbit in the Jupiter-Ganymede CR3BP. When it crosses Ganymede's SoI, it is approximated in the Jupiter-s/c 2BP in the Jupiter-centered inertial frame (Fig. \ref{Patched2BPCR3BP}).


\subsection{\label{subsec:coupledModel}The coupled spatial CR3BP model}
\begin{figure}
\hfill{}\centering\includegraphics[width=12cm]{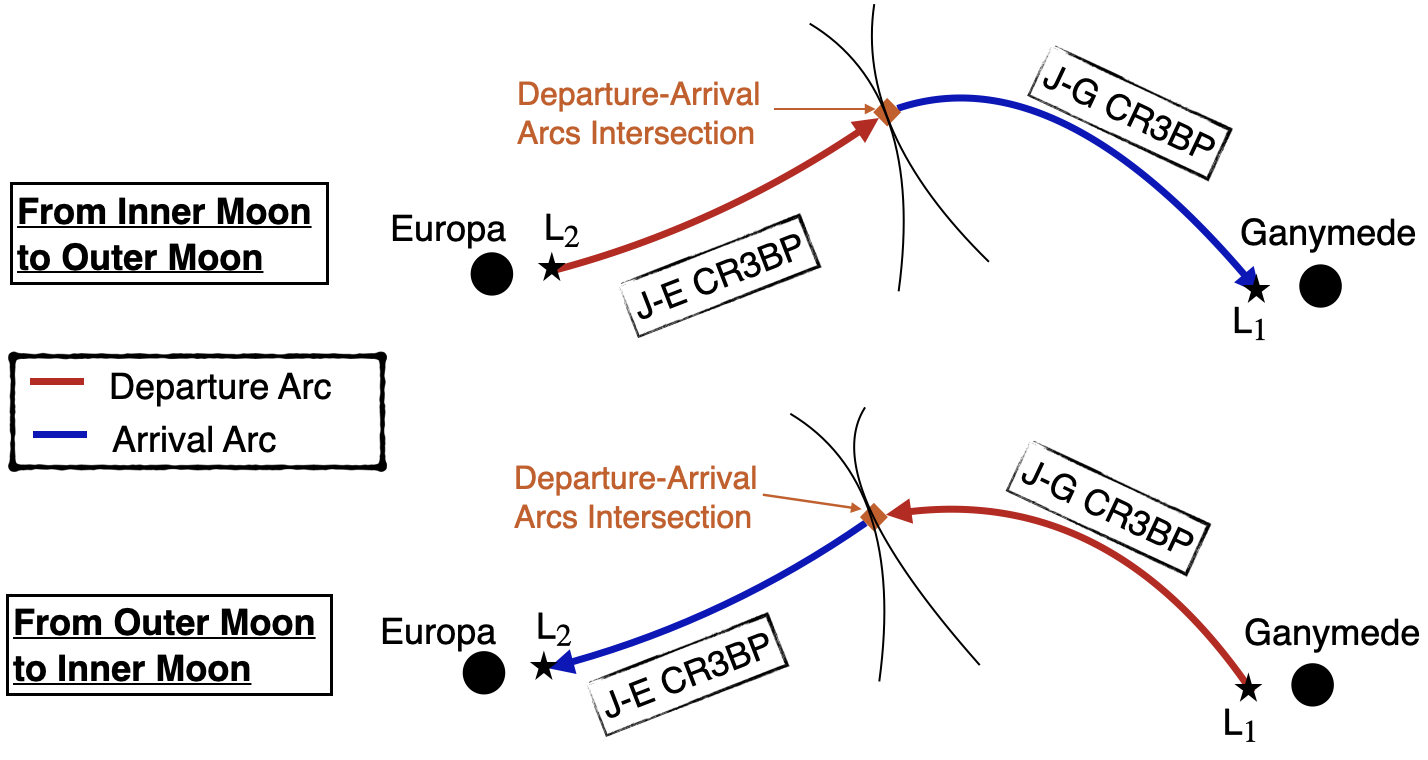}\hfill{}
\caption{\label{fig:coupledCR3BPScheme}The coupled spatial CR3BP. }
\end{figure}

\begin{figure}[h]
\hfill{}\centering\includegraphics[width=8.2cm]{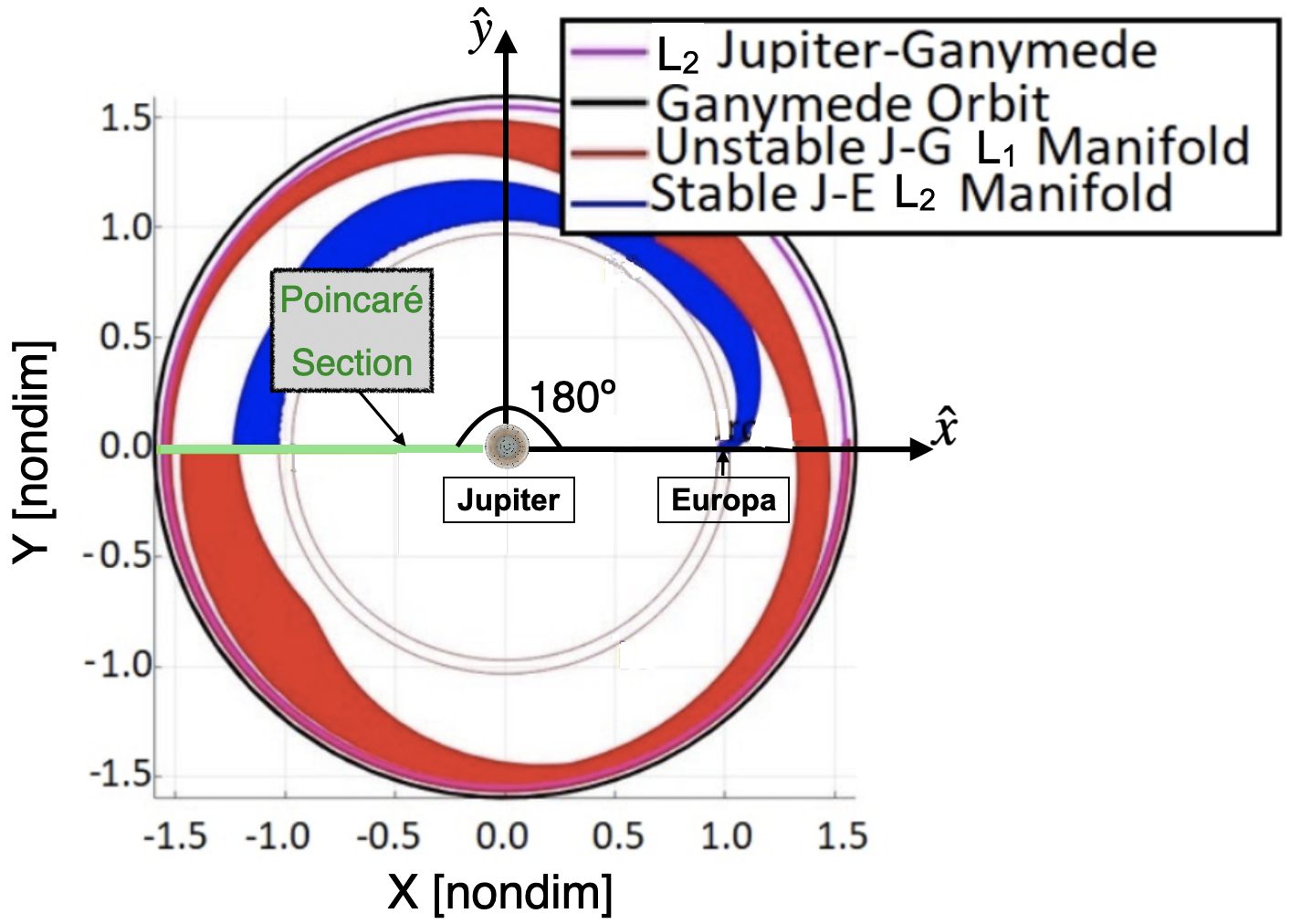}\hfill{}
\caption{\label{fig:coupledCR3BPExample} Intersection at a Poincar\'e section at 180\si{\degree} between unstable manifolds from L$_1$ Lyapunov orbit in the Jupiter-Ganymede CR3BP ($JC_{d}=3.0061$) and stable manifolds from L$_2$ Lyapunov orbit in the Jupiter-Europa CR3BP ($JC_{a}=3.0028$) represented in the Jupiter-Europa rotating frame.}
\end{figure}

The coupled spatial CR3BP represents two CR3BPs with a common primary (Jupiter in the case of a transfer from Ganymede to Europa) and allows the merging of spatial trajectories, each from one of the CR3BP, within the same reference frame. It is not a dynamical four-body problem, but rather a kinematical blending. Departure arcs are propagated in the departure planet-moon CR3BP and arrival arcs in the arrival planet-moon CR3BP. For the purposes of this investigation, departure and arrival arcs are assumed to be unstable and stable manifolds, respectively, that emanate from periodic orbits around L$_1$ or L$_2$. Since each problem is associated with its own reference frame, the following procedure is adopted to reach a common representation:
CR3BP trajectories departing one moon are transformed from the rotating
frame to the Ecliptic J2000.0 planet-centered inertial (see Appendix \ref{appendixRotToIner}); then, the resulting states are rotated into the arrival planet-moon rotating frame (see Appendix \ref{appendixInerToRot}) to seek a link
between both the departure and arrival trajectories. Note that the coupled planar CR3BP is a special case in which the orbital element $i$ is zero and $\Omega$ is undefined.
An illustration of this model appears in Fig. \ref{fig:coupledCR3BPScheme} reflecting sample between Ganymede and Europa. Numerical strategies are employed to approximate the intersections between trajectories from the two CR3BP in the common arrival rotating frame. In this investigation, differential corrections strategies, based on multi-variable Newton methods, are applied to solve boundary value problems that enable feasible spatial transfers between the moons vicinities. The differential corrections process is implemented by means of a multiple-shooter scheme inspired by the $\tau$-$\alpha$ method introduced by \citet{Haapala2015}. For the details of the algorithm, the reader is referred to Appendix \ref{appendix:coupledCorrections}.
Figure \ref{fig:coupledCR3BPExample} recreates an example of blending trajectories in the coupled spatial CR3BP between the Jupiter-Ganymede CR3BP and the Jupiter-Europa CR3BP in the arrival moon rotating frame. Here, a Poincar\'e section
at a specified angle with respect to the rotating Jupiter-Europa $\hat{x}$-axis aids in locating connecting trajectories between Ganymede and Europa.

\subsection{\label{subsec:ephemeris}The higher-fidelity ephemeris model}
\begin{figure}
\hfill{}\centering\includegraphics[width=9cm]{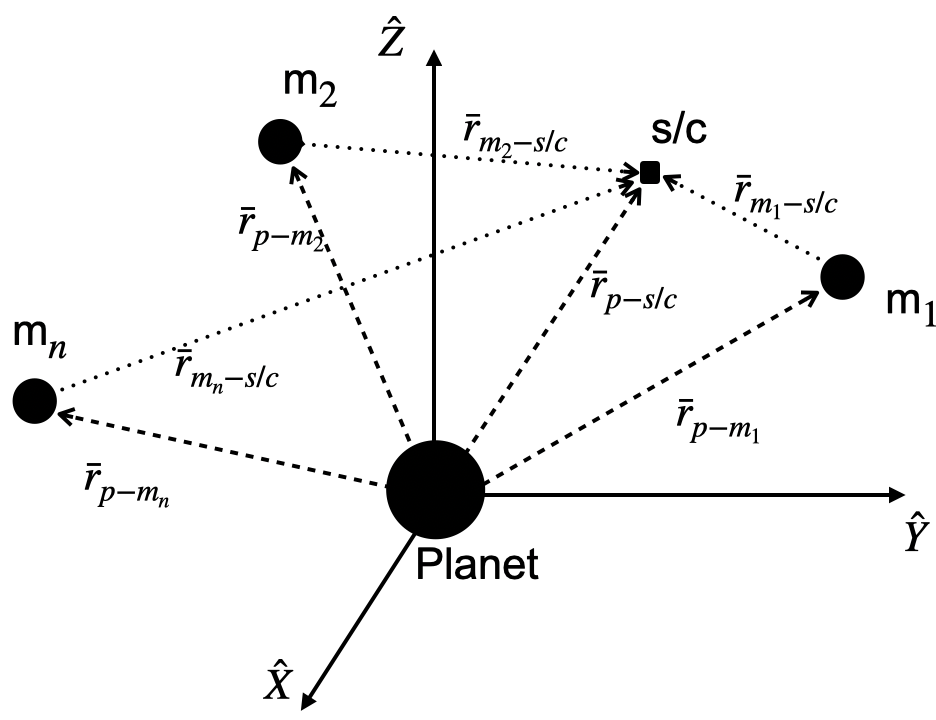}\hfill{}
\caption{\label{fig:schematicEphemeris}The higher-fidelity ephemeris model in the Ecliptic J2000.0 planet-centered inertial frame. }
\end{figure}

The higher-fidelity ephemeris model provides a good representation of the motion of a s/c subject to multiple gravitational accelerations. As noted in Sect. \ref{subsec:CR3BP}, it is possible to produce a good initial guess for this higher-fidelity model from the CR3BP. The motion of a s/c with mass $m_{s/c}$ is modeled relative to a central body of mass $m_{p}$. A number of perturbing bodies located relative to the central body, with masses $m_i$ for $i = 1, \dots, n$, are also included. For the purposes of this investigation, and  consistent with the moon plane definitions in Sect. \ref{subsec:CR3BP}, such a model is represented in the Ecliptic J2000.0 planet-centered reference inertial frame, represented by the axes $\left\{ \hat{X}, ~ \hat{Y}, ~\hat{Z} \right\}$ as seen in Fig. \ref{fig:schematicEphemeris}. This implementation relies on the SPICE libraries \citep{SPICE} supplied by the Ancillary Data Services of NASA's Navigation and Ancillary Information Facility (NAIF), from which precise positions and velocities of most bodies in the solar system are retrieved.
The net acceleration of the s/c in a model with several perturbing bodies is expressed as:
\begin{equation}
	\ddot{\bar{r}}_{p-s/c} = \frac{-G (m_{s/c} + m_{p})}{r_{p-s/c}^3} \bar{r}_{p-s/c} + G \sum\limits_{i=1}^n m_i \left(  \frac{\bar{r}_{s/c-m_i}}{r_{s/c-m_i}^3} - \frac{\bar{r}_{p-m_i}}{r_{p-m_i}^3}\right),
\end{equation}
where the first term is the s/c gravitational acceleration due to the central body, and the summation incorporates the gravitational interactions between the perturbing bodies and the s/c as well as the interactions between the perturbing bodies and the central body. 
Note that, in contrast to the CR3BP, the moons do not move on circular orbits. Therefore, the position and velocity states from the CR3BP deviate with respect to the original path in this model. To accommodate such deviations, it is useful to express components of position and velocity vectors variously in terms of rotating and inertial frames (see Appendix \ref{appendixRotationsEphemeris}). Trajectories originally computed in the coupled spatial CR3BP are corrected in the ephemeris model using a multiple shooting algorithm \citep{pavlak2012strategy} for position and epoch continuity along the entire transfer. Velocity continuity is ensured along most of the trajectory, but two impulsive $\Delta v_{tot}$s are allowed to enter/depart the periodic orbits and another to transfer from the unstable to the stable manifold trajectories. Note that the scheme is similar to the one illustrated in Appendix \ref{appendix:coupledCorrections}, but without the $\tau$ component. Therefore, the multiple shooting algorithm is merely a multi-segment corrections process but using the analytical partials computed from propagating segments in the higher-fidelity ephemeris model. Additionally, to represent periodic orbits in such a model, multiple revolutions of the CR3BP periodic orbit are stacked, one on top of the other, and are corrected for position and velocity continuity \citep{pavlakJournal2012}. The fidelity of the model is enhanced by adding the effects of a large number of celestial bodies as perturbing bodies that depend on the multi-moon system.

\section{\label{sec:coupledCR3BPMoonTransfers}Moon-to-moon transfers relying solely on the coupled spatial CR3BP}
Consider a s/c located in an L$_1$ Lyapunov orbit in the Jupiter-Ganymede system with a Jacobi Constant value $JC_{d}=3.0061$. Let
 the target destination be an L$_2$ Lyapunov orbit ($JC_{a}=3.0024$)
in the Jupiter-Europa system. Determining an intersection between unstable and stable manifolds from these periodic orbits depends mainly upon two factors: (a) the relative position between the moons at the initial time and (b) the location where the intersection is expected to occur. To assess (b), Poincar\'e sections are employed (as introduced in Sect. \ref{subsec:coupledModel}). Here, the problem is explored at two levels: first, assuming that the moons move in coplanar orbits and, then, adding the real inclinations of their orbital planes.

\subsection{\label{subsec:coplanar} Moons in coplanar orbits}
\begin{figure}
\hfill{}\centering\includegraphics[width=7.2cm]{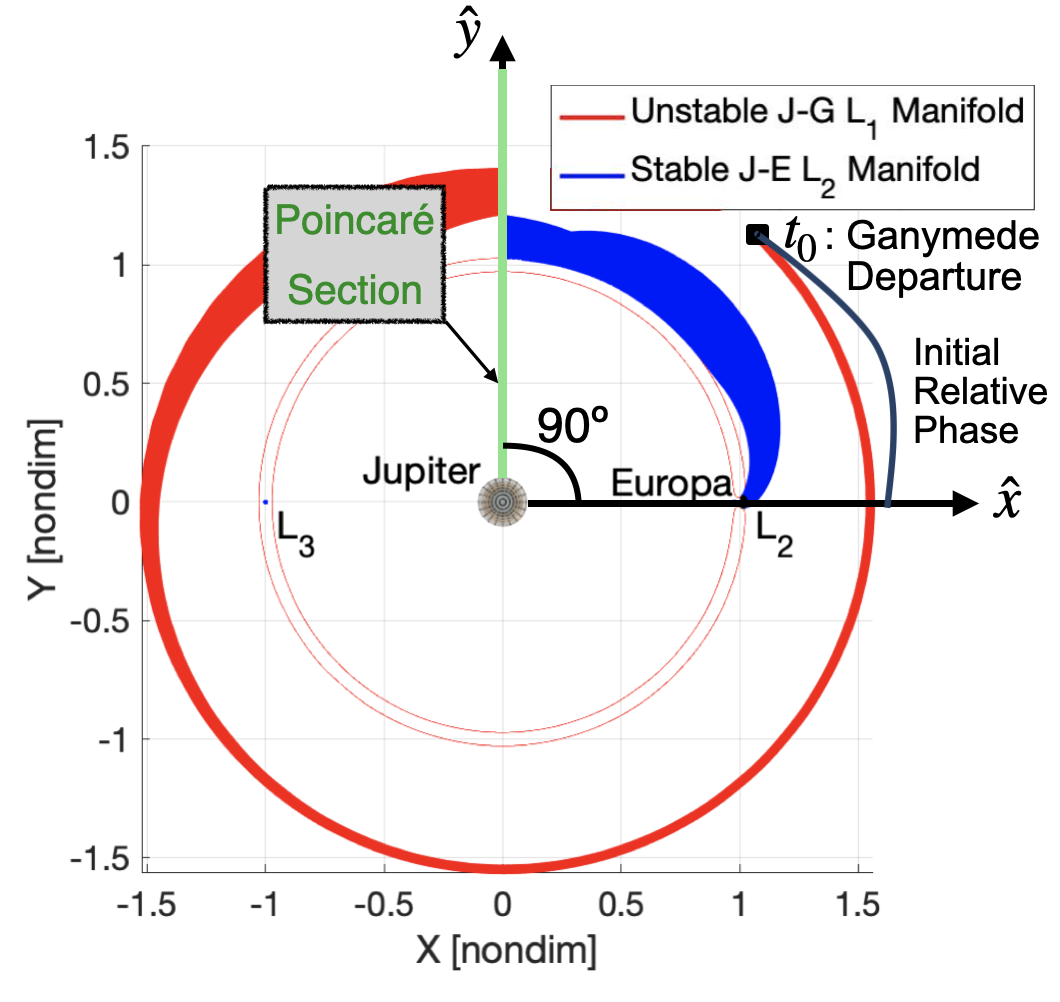}\hfill{}
\caption{\label{fig:Section90Deg45Deg0}Intersection at a Poincar\'e section at 90\si{\degree} between unstable manifolds from L$_1$ Lyapunov orbit in the Jupiter-Ganymede CR3BP ($JC_{d}=3.0061$) and stable manifolds from L$_2$ Lyapunov orbit in the Jupiter-Europa CR3BP ($JC_{a}=3.0028$) represented in the Jupiter-Europa rotating frame (coupled planar CR3BP).}
\end{figure}

\begin{figure}
\hfill{}\centering%
\begin{minipage}[b][1\totalheight][t]{0.38\columnwidth}%
\subfigure[$y$-$\dot{x}$ Poincar\'e section.]{\label{YXDOTSection90Deg45Deg0}{\includegraphics[width=7cm]{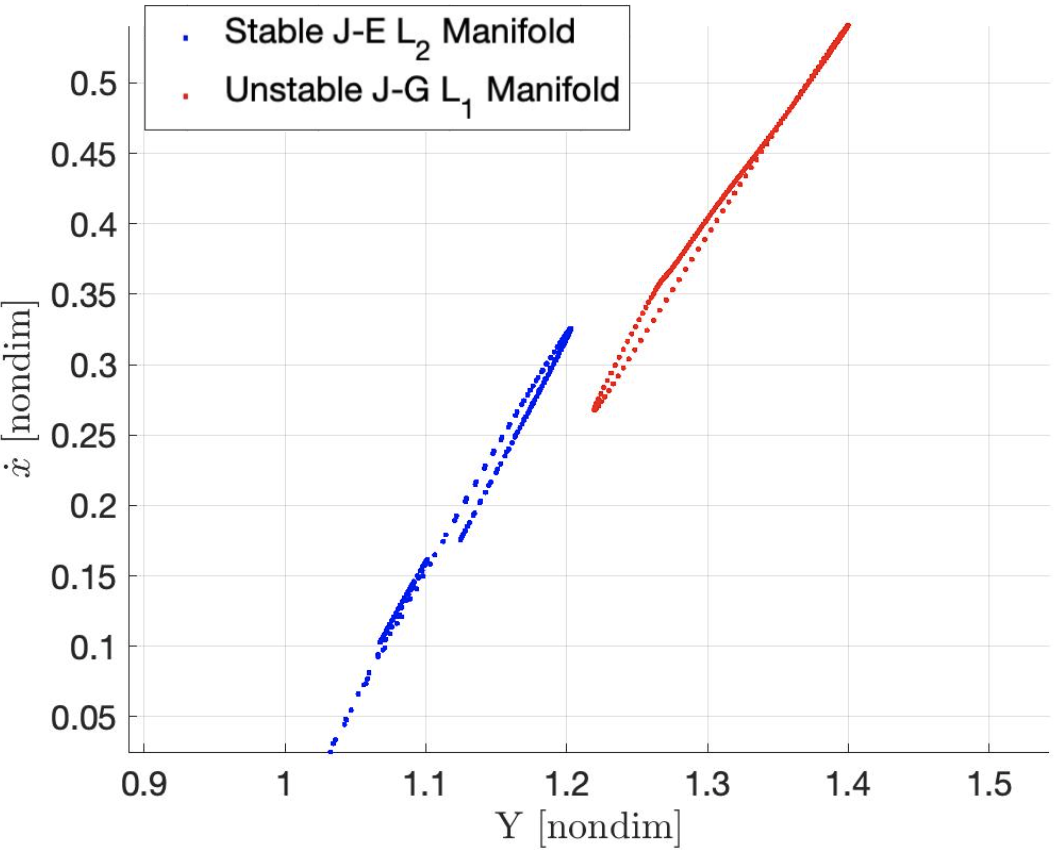}}}%
\end{minipage}\hfill{}%
\begin{minipage}[b][1\totalheight][t]{0.33\columnwidth}%
\subfigure[$y$-$\dot{y}$ Poincar\'e section.]{{\includegraphics[width=7cm]{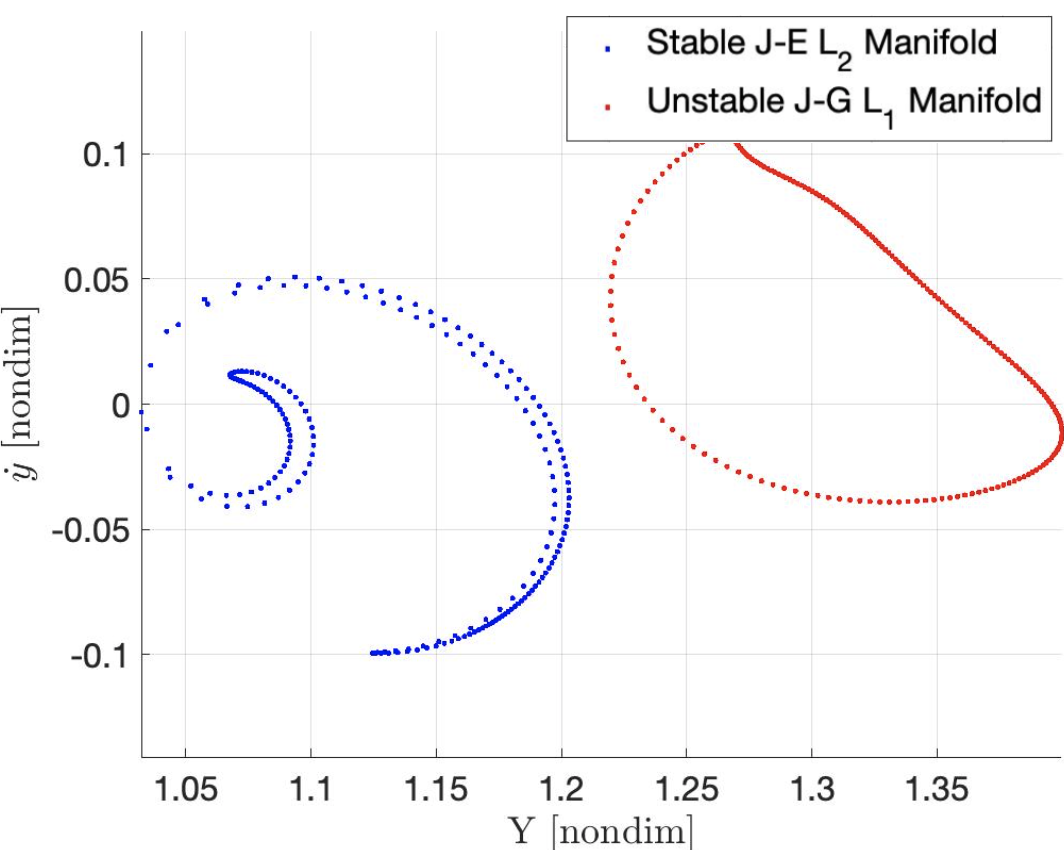}}\label{YYDOTSection90Deg45Deg0}}%
\end{minipage}\hfill{}
\caption{\label{fig:section90DegreeComp}State components of both manifold trajectories at the Poincar\'e section at 90\si{\degree} (coupled planar CR3BP).}
\end{figure}

\begin{figure}
\hfill{}\centering\includegraphics[width=7.2cm]{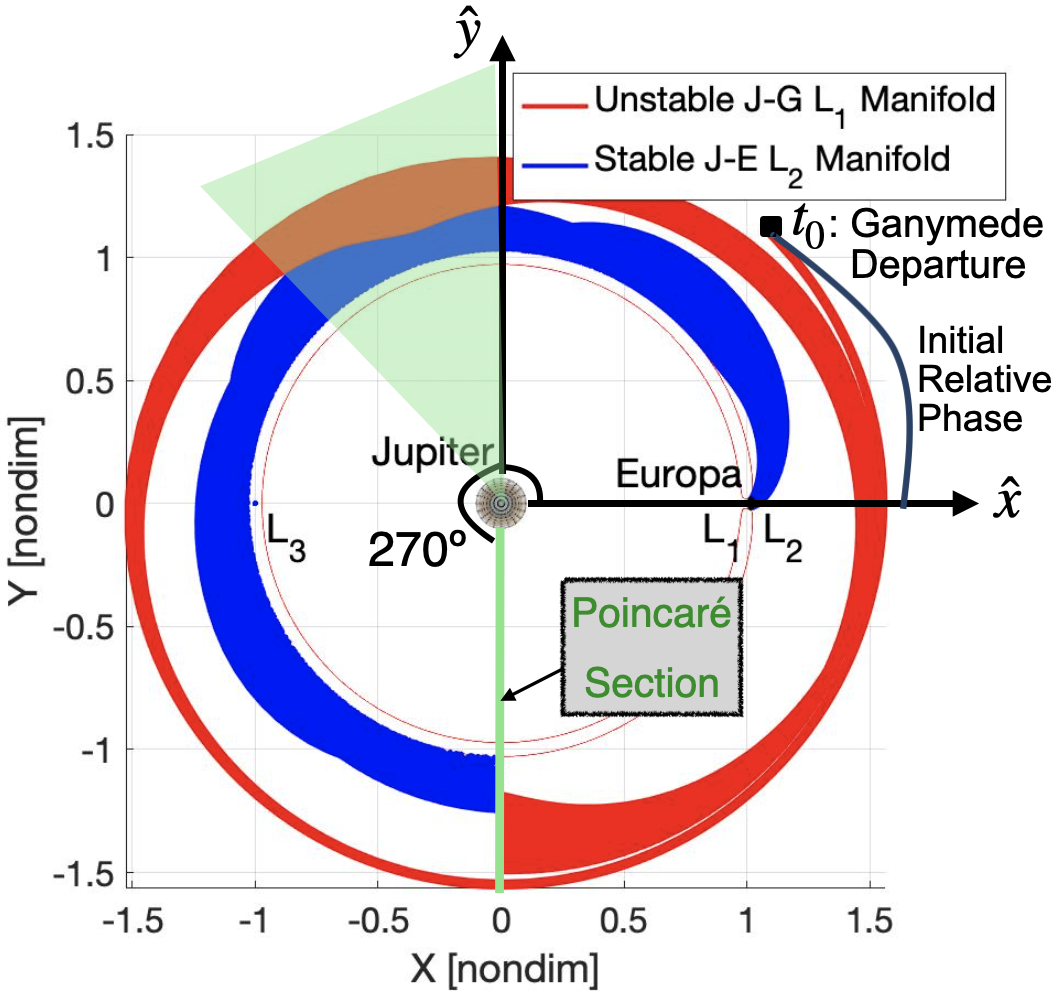}\hfill{}
\caption{\label{fig:Section270Deg45Deg0}Intersection at a Poincar\'e section at 270\si{\degree} between unstable manifolds from L$_1$ Lyapunov orbit in the Jupiter-Ganymede CR3BP ($JC_{d}=3.0061$) and stable manifolds from L$_2$ Lyapunov orbit in the Jupiter-Europa CR3BP ($JC_{a}=3.0028$) represented in the Jupiter-Europa rotating frame (coupled planar CR3BP). }
\end{figure}

\begin{figure}
\hfill{}\centering%
\begin{minipage}[b][1\totalheight][t]{0.47\columnwidth}%
\subfigure[$y$-$\dot{x}$ Poincar\'e section.]{\label{YXDOTSection270Deg45Deg0}{\includegraphics[width=6.5cm]{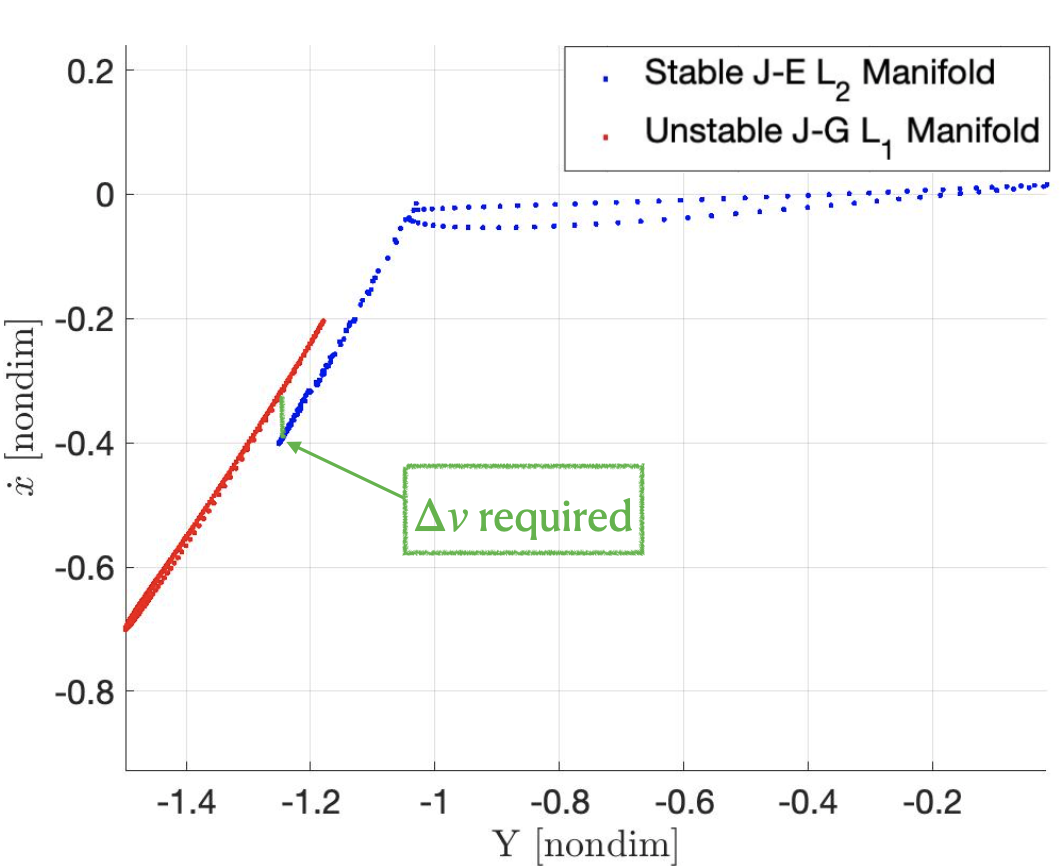}}}%
\end{minipage}\hfill{}%
\begin{minipage}[b][1\totalheight][t]{0.33\columnwidth}%
\subfigure[$y$-$\dot{y}$ Poincar\'e section.]{{\includegraphics[width=6.5cm]{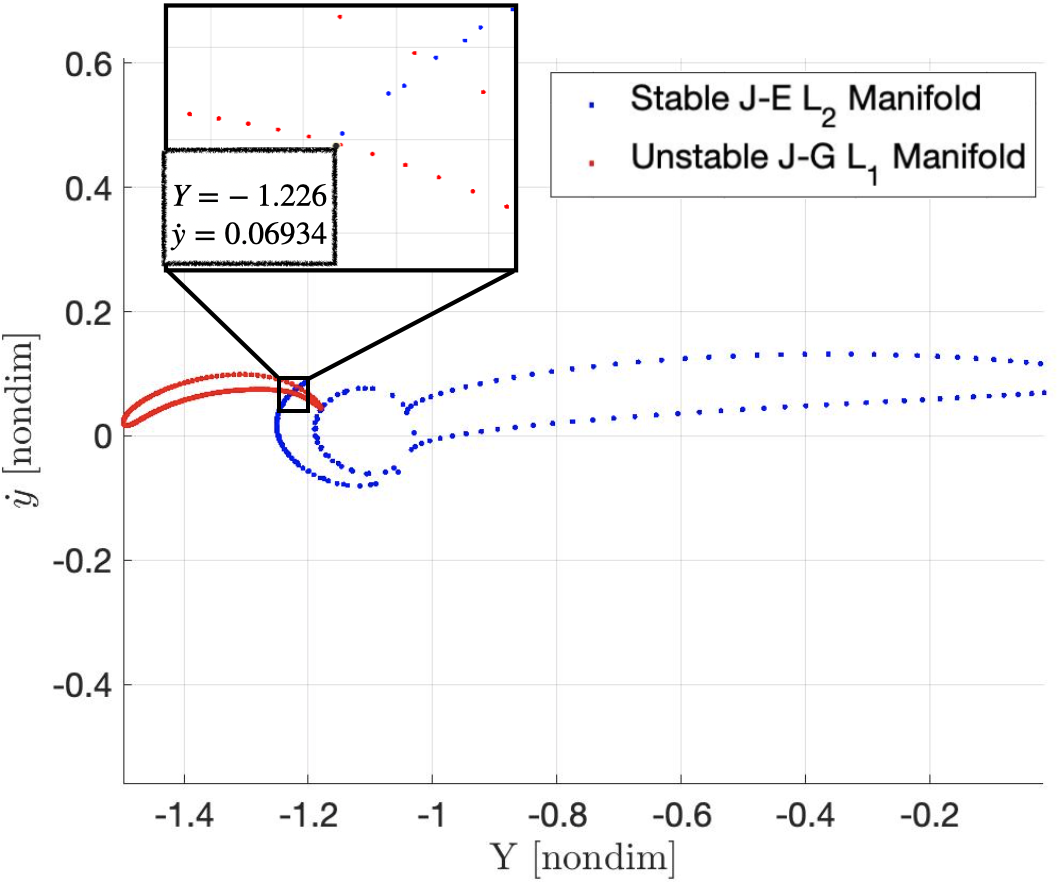}}\label{YYDOTSection270Deg45Deg0}}%
\end{minipage}\hfill{}
\label{Fig:PoincareSection90Deg}
\caption{State components of both manifold trajectories at the Poincar\'e section at 270\si{\degree} (coupled planar CR3BP).}
\end{figure}

\begin{figure}
\hfill{}\centering%
\begin{minipage}[b][1\totalheight][t]{0.55\columnwidth}%
\subfigure[Ecliptic J2000.0 Jupiter-centered inertial frame.]{\label{fig:sectionInertial}{\includegraphics[width=7cm]{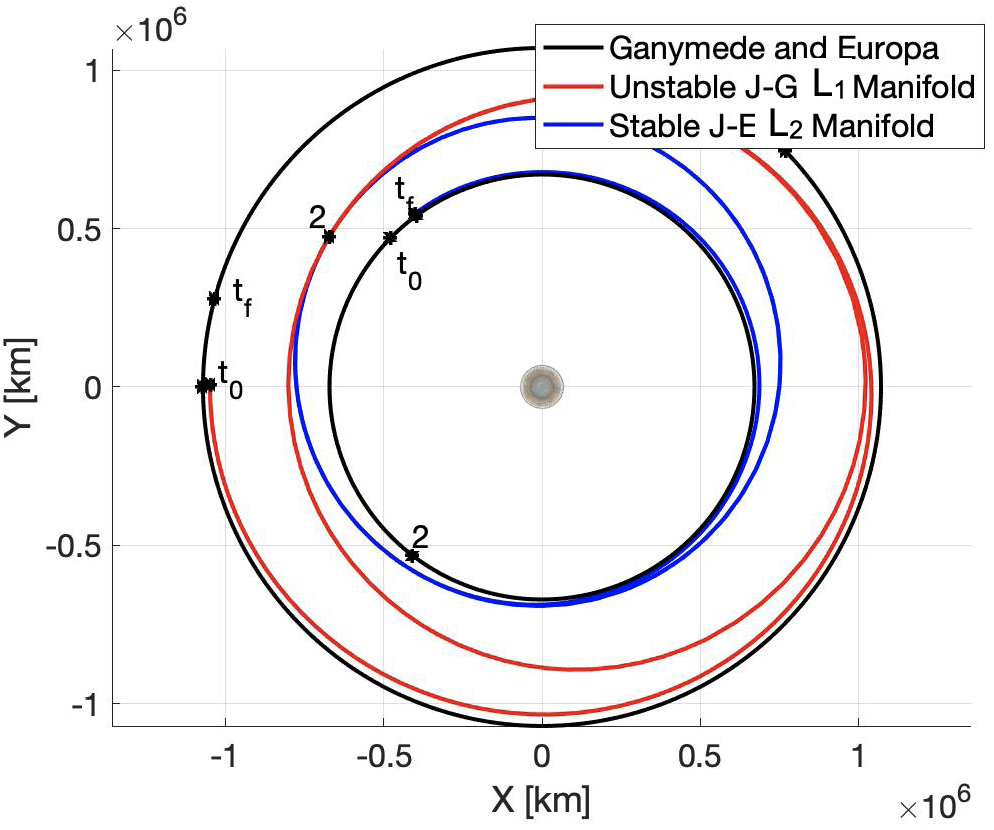}}}%
\end{minipage}\hfill{}%
\begin{minipage}[b][1\totalheight][t]{0.33\columnwidth}%
\subfigure[Jupiter-Europa rotating frame.]{\label{fig:sectionRotating}{\includegraphics[width=8cm]{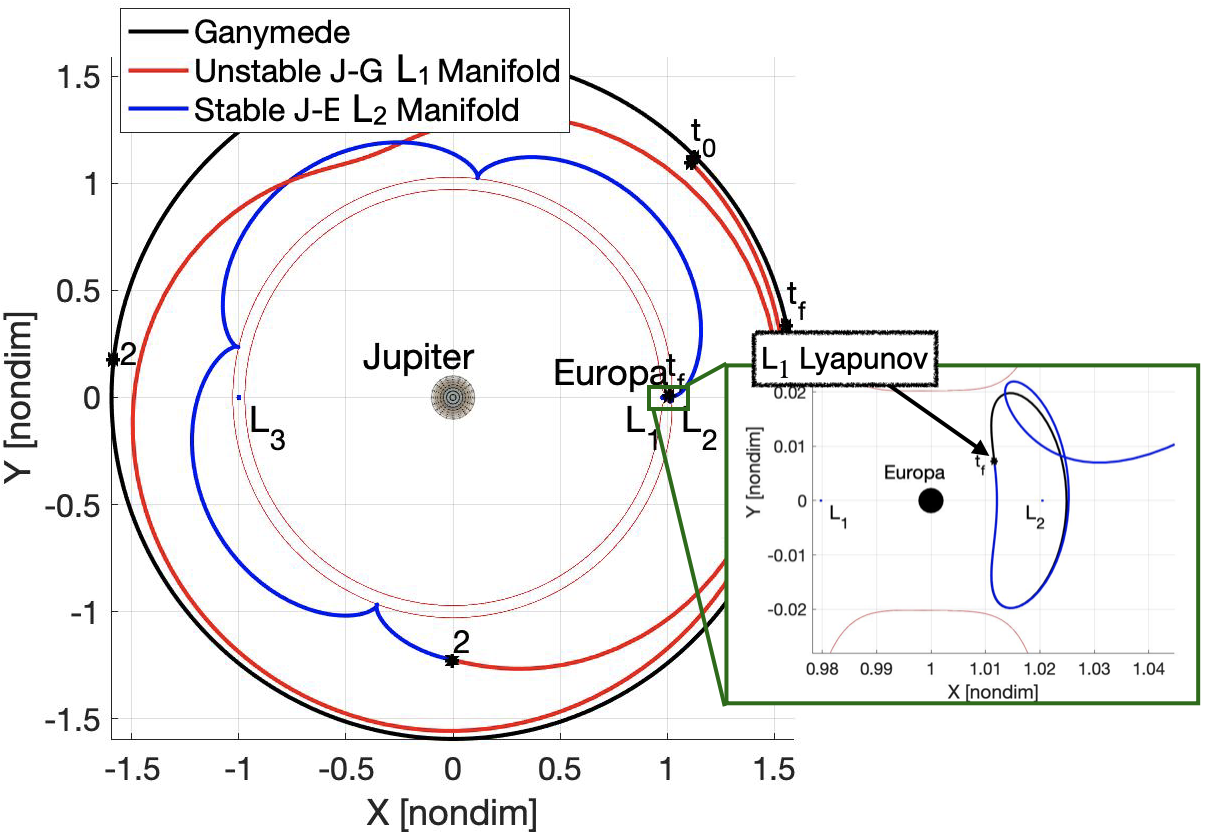}}}%
\end{minipage}\hfill{}
\caption{\label{fig:PoincareSectionCoupledTransfer}Converged solution in the coupled
planar CR3BP for a transfer from an L$_1$ Lyapunov orbit of
Jupiter-Ganymede system, to an L$_2$ Lyapunov orbit of Jupiter-Europa
system (coupled planar CR3BP): $\Delta v_{tot}=1.0192$ km/s and $t_{tot}=28.31$ days.}
\end{figure}

Let Ganymede be located at an orbital phase of 45\si{\degree} with respect to Europa at the initial time for the transfer, $t_0$. Both unstable and stable manifold trajectories are propagated towards a Poincar\'e section located at an angle of 90\si{\degree} from the $\hat{x}$-axis as defined in the Jupiter-Europa rotating frame, and illustrated in Fig. \ref{fig:Section90Deg45Deg0}. Since the $x$-coordinate for all the states is zero at the intersection on the section, the velocity components need only be plotted against the $y$-component of the position (see Fig. \ref{fig:section90DegreeComp}). In this case, no intersections occur between the curves in either section; hence the direct transfer between manifolds is not possible. However, if the intersection between unstable and stable manifold trajectories occurs at a Poincar\'e section located at 270\si{\degree} (Fig. \ref{fig:Section270Deg45Deg0}), it is possible to deliver a crossing between the unstable and stable manifolds as observed in the $y$-$\dot{y}$ Poincar\'e section (see Fig. \ref{YYDOTSection270Deg45Deg0}). Nevertheless, since there is no intersection at this $y$ location in the $y$-$\dot{x}$ Poincar\'e section in Fig. \ref{YXDOTSection270Deg45Deg0}, then a $\Delta v_{tot}$ is required to transfer from the unstable J-G L$_1$ manifold to the stable J-E L$_2$ manifold. Note that if a crossing also occurs at the $y$-$\dot{x}$ section, a heteroclinic connection is produced as explained in \cite{Haapala2015}. 
Once a good initial guess is generated by means of Poincar\'e sections, the closest unstable and stable manifold trajectories to the intersection point are selected by means of a nearest neighbor algorithm. However, the trajectory is discontinuous. To produce a continuous transfer in the coupled spatial CR3BP, a multiple shooting scheme serves as the basis for the differential corrections algorithm. The resulting converged trajectory is plotted in Fig. \ref{fig:PoincareSectionCoupledTransfer}. One maneuver in the $\hat{x}$-direction is required to complete the transfer (Fig. \ref{YXDOTSection270Deg45Deg0}). Nevertheless, as observed in Fig. \ref{fig:Section270Deg45Deg0}, potential connections between manifolds exist if the Poincar\'e section is located inside the green area. All these possibilities offer options in a search for smaller values of $\Delta v_{tot}$ and $t_{tot}$, but the process becomes computationally expensive and time consuming.

\subsection{\label{subsec:coupledPoincareSectionsTruePlanes} Moons in their true orbital planes}
\begin{figure}
\hfill{}\centering\includegraphics[width=7cm]{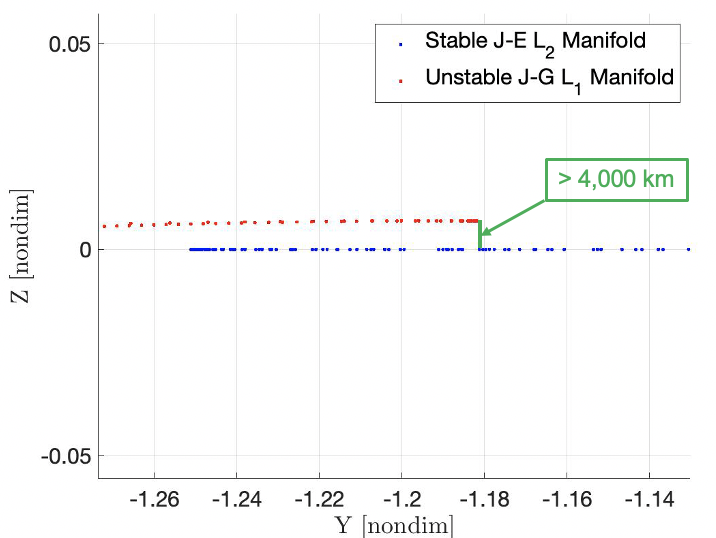}\hfill{}
\caption{\label{fig:YZCoupledSpatialSection}$y$-$z$ Poincar\'e section of both manifold trajectories at 270\si{\degree} in the Jupiter-Europa rotating frame (coupled spatial CR3BP). }
\end{figure}

Assume that both Europa and Ganymede are located in their true orbital planes with orbital properties as defined in Table \ref{Table:DataEuropaGanymede}. An intersection between unstable and stable manifold trajectories is sought at a Poincar\'e section oriented 270\si{\degree} from the $\hat{x}$-axis, similar to Fig. \ref{fig:Section270Deg45Deg0}, whereas the phase of Ganymede relative to Europa at the time of departure is 45\si{\degree}. Now, since the unstable manifolds in the J-G system are propagated along the Ganymede plane, which is different from the Europa plane, a $z$-component appears in the motion when rotated into the J-E rotating frame and, consequently, a $\dot{z}$-velocity component. As apparent in in Fig. \ref{fig:YZCoupledSpatialSection}, the unstable and stable manifold trajectories for both systems no longer intersect in space due to the difference in the Z-component. As a result, a trajectory cannot be converged for these conditions.

\subsection{Discussion}
The analysis of Sects. \ref{subsec:coplanar} and \ref{subsec:coupledPoincareSectionsTruePlanes} demonstrate the challenges when designing moon-to-moon transfers in the coupled spatial CR3BP. Notable barriers include:
\begin{itemize}
\item It becomes computationally expensive to locate intersections between departure and arrival arcs with a nearest neighbor algorithm in the Poincar\'e section at arbitrary orientations of the Poincar\'e section with respect to the $\hat{x}$-axis in the arrival moon rotating frame.
\item Similarly, framing the search over different initial relative positions between moons to determine lower $\Delta v_{tot}$s and times-of-flight is also time consuming.
\item An intersection in the coupled planar CR3BP may not transition to the coupled spatial CR3BP. This aspect complicates the search for ideal intersections. Furthermore, this fact becomes more challenging with a wider difference between the departure and arrival moon planes. Notably, construction of transfers between spatial periodic orbits is complex.
\end{itemize}
It is, thus, apparent that simplifications may efficiently narrow the search for the relative phases and locations for intersections in the coupled spatial CR3BP. As a result, the search for optimal $\Delta v_{tot}$s and transfer times is then more straightforward.

\section{\label{sec:MMATMethod}The Moon-to-Moon Analytical Transfer Method}
\begin{figure}
\hfill{}\centering\includegraphics[width=13cm]{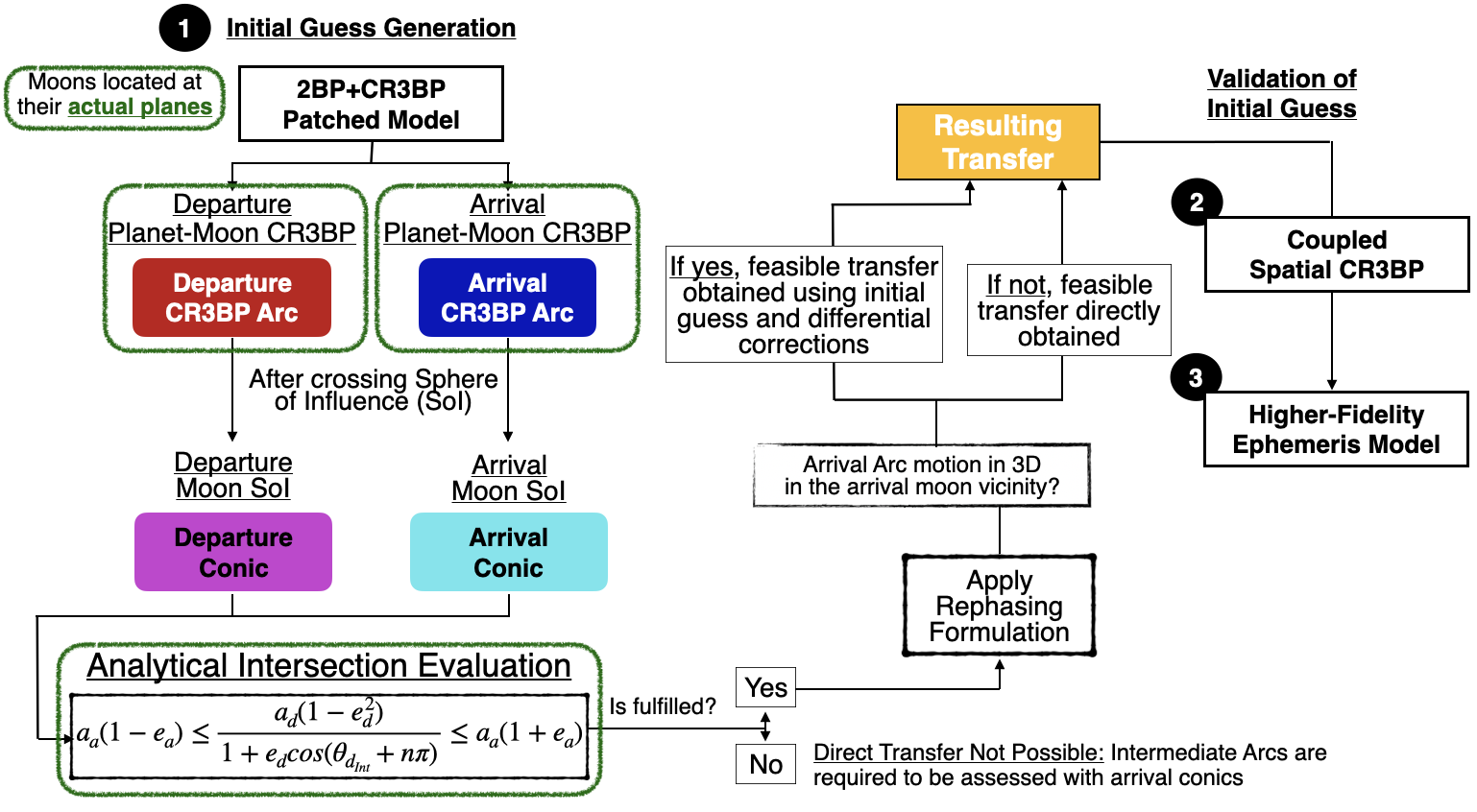}\hfill{}
\caption{\label{fig:MMATmethod}MMAT method scheme.}
\end{figure}

Consider the transfer from Ganymede to Europa as discussed in Sect. \ref{sec:coupledCR3BPMoonTransfers}. It is previously demonstrated that relying solely on the coupled spatial CR3BP to determine suitable transfers between periodic orbits in two different planet-moon systems brings many complications. An alternative strategy, 'the MMAT method', is introduced that leverages some simplifications to produce lower costs and shorter times-of-flight assuming that both moon orbits are in their true orbital planes. A brief schematic of the MMAT method appears in Fig. \ref{fig:MMATmethod}. First, the 2BP-CR3BP
patched model is used to approximate CR3BP trajectories as arcs of conic sections. Thus, it is possible to analytically explore promising trajectories
and configurations between the moons. 
For a given angle of departure from one moon,
if the geometrical properties between departure and arrival conics satisfy a given condition, an orbital phase for the arrival moon is produced implementing a rephasing formulation. Consequently, the adequate relative phasing between the moons at their respective planes yields a unique
natural transfer between CR3BP arcs with a single $\Delta v_{tot}$. Once a promising
transfer is uncovered in the 2BP+CR3BP patched model, it serves as an initial guess
for the coupled spatial CR3BP. Finally, results in the coupled spatial CR3BP are transitioned to the higher-fidelity ephemeris model.

To demonstrate the methodology and, for the sake of comparison, the problem is first explored assuming the orbits of the moons are coplanar and, then, in different planes. Note that, in this section, the following definitions hold: instant 0 denotes the beginning of the transfer from the departure moon; instant 1 denotes the time at which the departure arc reaches the departure moon SoI, where it is approximated by a conic section; instant 2 corresponds to the intersection between the departure and arrival conics (or arcs in the coupled spatial CR3BP); instant 3 matches the moment when the arrival conic reaches the arrival moon SoI; finally, instant 4 labels the end of the transfer. Thus, a feasible end-to-end transfer is constructed.

\subsection{Moons on coplanar orbits}
\subsubsection{Analytical constraint for a successful transfer}
Assume, for now, that the orbits of Ganymede and Europa are in the same plane.
The 2BP-CR3BP patched model is employed to construct analytical connections
between trajectories in the different systems. Trajectories departing Ganymede's
vicinity are propagated towards the Ganymede SoI, where a departure conic with orbital elements dependent on the initial
Ganymede epoch angle $\theta_{0_{Gan}}$ is produced. Desirable trajectories for arrival in the vicinity of Europa are also back-propagated
to the Europa SoI, where an arrival conic is produced with orbital elements dependent
on the Europa arrival epoch angle $\theta_{4_{Eur}}$. Recall that the subscripts '0' and '4' represent the initial and arrival instants, repectively. To determine a potential moon-to-moon transfer leveraging the departure and arrival conics, the following theorem is presented: 
\begin{theopargself}
\begin{theorem}\label{theorem:planar}
If the geometrical properties for two coplanar, confocal conics fulfill the inequality
constraint represented by 
\begin{equation}
2a_{d}a_{a}(1+e_{d}e_{a})\ge b_{d}^{2}+b_{a}^{2}\ge2a_{d}a_{a}(1-e_{d}e_{a}),\label{eq:planarConstraint}
\end{equation}
either one of the two conics can be reoriented such that they tangentially intersect. Consequently, the ideal phase of the arrival moon at arrival, $\theta_{4_{Eur}}$, for the intersection to occur is produced considering that the departure epoch, $\theta_{0_{Gan}}$, is fixed.
\end{theorem}
\end{theopargself}

\noindent {\it \textbf{Proof}}\\
Similar to \cite{doi:10.2514/8.9010}, the objective is the determination of the geometrical condition that both departure and arrival conics must possess for intersection.
Two coplanar confocal ellipses allow up to two common points (excluding the degenerate case in which the two curves coincide) at a distance, $r_{int}$, as long as 
\begin{equation}
r_{int}=\frac{p_{d}}{1+e_{d}\cos\theta_{d_{Int}}}=\frac{p_{a}}{1+e_{a}\cos\theta_{a_{Int}}},
\end{equation}
where $a_{d}$ and $a_{a}$ are the semi-major axes, $e_{d}$ and
$e_{a}$ are the eccentricities, $\theta_{d_{Int}}$ and $\theta_{a_{Int}}$ the
true anomalies at the intersection point along the departure and arrival conics,
respectively, and $p_{d}=a_{d}(1-e_{d}^{2})$ and $p_{a}=a_{a}(1-e_{a}^{2})$.
Due to the fact that the moons are assumed to be in coplanar orbits, the angle $\theta_{a_{Int}}$ is defined as $\theta_{a_{Int}}=\theta_{d_{Int}}-\Delta\omega$,
where $\Delta\omega=\omega_{a}-\omega_{d}$. Given the trigonometric
property $\cos(\theta_{d_{Int}}-\Delta\omega)=\cos\theta_{d_{Int}}\cos\Delta\omega+\sin\theta_{d_{Int}}\sin\Delta\omega$,
the following expression is produced:
\begin{equation}
p_{d}-p_{a}+(p_{d}e_{a}\cos\Delta\omega-p_{a}e_{d})\cos\theta_{d_{Int}}=-p_{d}e_{a}\sin\Delta\omega\sin\theta_{d_{Int}}.\label{eq:squaring}
\end{equation}
Squaring both sides in Eq. \eqref{eq:squaring} yields the quadratic
equation 
\begin{equation}
k_{1}\cos^{2}\theta_{d_{Int}}+2k_{2}\cos\theta_{d_{Int}}+k_{3}=0
\end{equation} with
the following solution:
\begin{equation}
\cos\theta_{d_{Int}}=\frac{-k_{2}\pm\sqrt{k_{2}^{2}-k_{1}k_{3}}}{k_{1}},\label{eq:quadratic}
\end{equation}
where 
\begin{equation}
  \begin{split}
k_{1}&=(p_{d}e_{a}\cos\Delta\omega-p_{a}e_{d})^{2}+(-p_{d}e_{a}\sin\Delta\omega)^{2}, \\
k_{2}&=(p_{d}-p_{a})(p_{d}e_{a}\cos\Delta\omega-p_{a}e_{d}), \\
k_{3}&=(p_{d}-p_{a})^{2}-(-p_{d}e_{a}\sin\Delta\omega)^{2}.
  \end{split}
\end{equation}
Examining Eq. \eqref{eq:quadratic}, only one solution is available when $k_{2}^{2}-k_{1}k_{3}=0$.
This limiting case corresponds to a tangential intersection between
the two conics. As reflected in \cite{Fantino2016},
trajectories that intersect tangentially frequently 
correspond to a theoretical minimum-$\Delta v_{tot}$, certainly 
in terms of the minimum energy for the given maneuver at the tangential intersection between ellipses.
From $k_{2}^{2}-k_{1}k_{3}=0$, an ideal phase for the arrival moon
is delivered such that a tangent configuration is guaranteed by isolating $\Delta\omega$. The appropriate 
relative argument of periapsis between the arrival and departure conics for a tangent configuration to occur is defined by:
\begin{equation}
\cos\Delta\omega=\frac{2a_{a}a_{d}-a_{a}^{2}(1-e_{a}^{2})-a_{d}^{2}(1-e_{d}^{2})}{2a_{a}a_{d}e_{a}e_{d}}.\label{eq:RephasingFormula}
\end{equation}
Therefore, by fixing $\omega_d$, the ideal argument of periapsis for the arrival conic $\omega_a$ is obtained,  eventually leading to the ideal phase for the arrival moon at the arrival epoch to produce a tangential (hence, minimum cost) transfer. The departure conic true anomaly at intersection with the arrival conic, $\theta_{d_{Int}}$, is then:
\begin{equation}
\begin{array}{cc}
\cos \theta_{d_{Int}}=-\frac{k_2}{k_1}; & \sin \theta_{d_{Int}}=\frac{p_{d}-p_{a}+(p_{d}e_{a}\cos\Delta\omega-p_{a}e_{d})\cos\theta_{d_{Int}}}{-p_{d}e_{a}\sin\Delta\omega}.
\end{array}
\end{equation}
Finally, the angle $\theta_{a_{Int}}$ is evaluated from $\theta_{a_{Int}}=\theta_{d_{Int}}-\Delta\omega$. To determine the feasibility of a transfer, focus on the right side of Eq. \eqref{eq:RephasingFormula}.
The expression 
\begin{equation}
2a_{d}a_{a}(1+e_{d}e_{a})<b_{d}^{2}+b_{a}^{2}
\end{equation}
is produced when 
\begin{equation}
\frac{2a_{a}a_{d}-a_{a}^{2}(1-e_{a}^{2})-a_{d}^{2}(1-e_{d}^{2})}{2a_{a}a_{d}e_{a}e_{d}}<-1,
\end{equation} where $b_{d}$ and $b_{a}$ are the semi-minor axis of
the departure and arrival conics, respectively. This conditions implies that the departure and
arrival conics never intersect. When both sides
equate, the minimum condition for tangency is produced,
which occurs between the apogee of the inner conic and the perigee
of the outer conic. Similarly, when 
\begin{equation}
\frac{2a_{a}a_{d}-a_{a}^{2}(1-e_{a}^{2})-a_{d}^{2}(1-e_{d}^{2})}{2a_{a}a_{d}e_{a}e_{d}}>1,
\end{equation}
the inequality constraint yields 
\begin{equation}
2a_{d}a_{a}(1-e_{d}e_{a})>b_{d}^{2}+b_{a}^{2},
\end{equation}
a result that implies  two intersections between the conics are available. Therefore, when $2a_{d}a_{a}(1-e_{d}e_{a})=b_{d}^{2}+b_{a}^{2}$, the maximum
limiting geometrical relationship between the ellipses emerges, one such that a tangent
configuration occurs: an apogee-to-apogee or perigee-to-perigee configuration, depending on the properties of both ellipses. \textbf{QED}

\begin{figure}[h]
\hfill{}\centering\includegraphics[width=13cm]{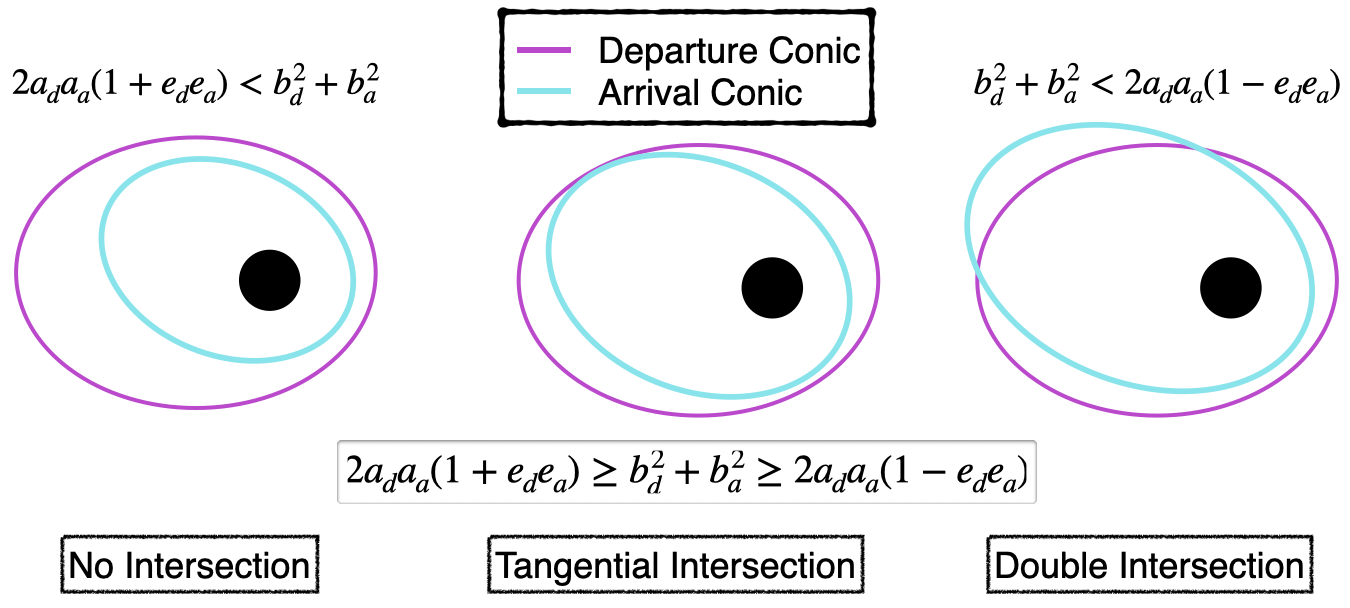}\hfill{}
\caption{\label{fig:threeConditions} Possible intersections between two coplanar, confocal ellipses with given semi-major axes and eccentricities, according to Theorem \ref{theorem:planar}.}
\end{figure}

As a result, Theorem \ref{theorem:planar} is proved and Figure \ref{fig:threeConditions} represents the three possible intersections between two coplanar, confocal ellipses depending on $a$ and $e$. 

\subsubsection{\label{subsec:Arrival-moon's-rephasing}Rephasing the arrival moon}
\begin{figure}[h]
\hfill{}\centering\includegraphics[width=12.5cm]{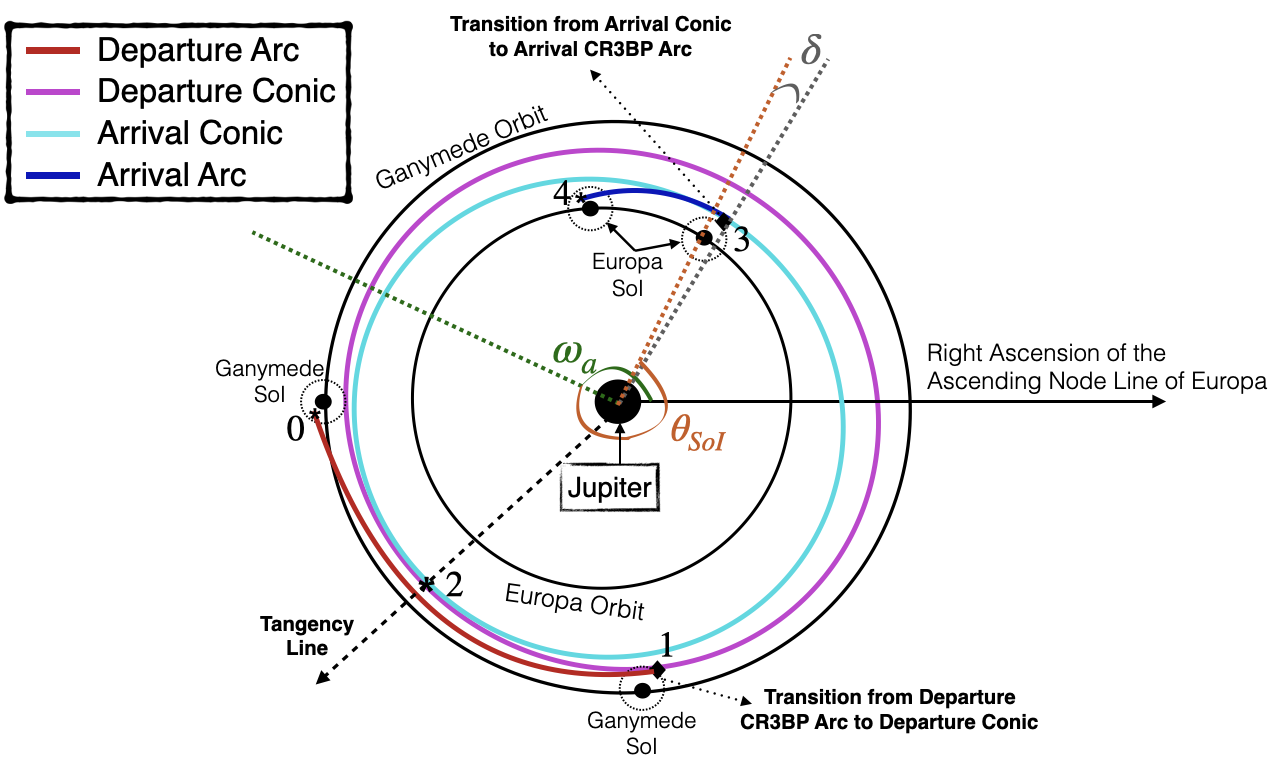}\hfill{}
\caption{\label{fig:reorientationGraph} Representation of the rephasing of
the arrival moon for a spatial transfer configuration (all angles projected upon the arrival moon plane). }
\end{figure}

Given departure and arrival conics that satisfy Theorem \ref{theorem:planar}, it is then possible to construct the ideal arrival conic argument of periapsis $\omega_a$ such that a tangential intersection with the departure conic occurs. Therefore, the angle $\sigma$ at which the arrival conic intersects
the arrival moon SoI is produced: $\sigma=\omega_{a}+\theta_{SoI}$. Note that the angle $\sigma$ is defined as the projection onto the arrival moon plane, which is measured from its right ascension
of the ascending node line in the planet-centered Ecliptic J2000.0 frame; in the case where the inclination of the arrival
moon plane is zero, it is measured from the $\hat{X}$-axis in the Ecliptic J2000.0 Jupiter
centered inertial frame (Fig. \ref{fig:reorientationGraph}). Let
$\delta=\tan ^{-1}(\frac{y_{SoI}}{|x_{SoI}|})$ be the angle between the intersection of the arrival arc with
the arrival moon SoI and the $\hat{x}$-axis in the Jupiter-Europa rotating frame (arrival moon rotating frame). Then,
the angle of the moon in its orbit at the end of the transfer, $\theta_{4_{Eur}}$,
is evaluated in as done by \cite{cite-keyviale}:
\begin{equation}
\theta_{4_{Eur}}=\sigma-\delta+\frac{2\pi\Delta t_{a}}{P_{m}},\label{eq:rephasing}
\end{equation}
where $\Delta t_{a}$ is the total time along the arrival CR3BP trajectory
in the arrival moon vicinity, and $P_{m}$ is the period of the
arrival moon in its orbit.

\subsubsection{\label{subsec:coplanarApplication}Ganymede to Europa transfer application}
\begin{figure}
\hfill{}\centering%
\begin{minipage}[b][1\totalheight][t]{0.5\columnwidth}%
\subfigure[L$_1$ Lyapunov orbit's unstable manifolds towards Ganymede SoI ($JC_d=3.0061$) in the J-G rotating frame.]{{\includegraphics[width=6.5cm]{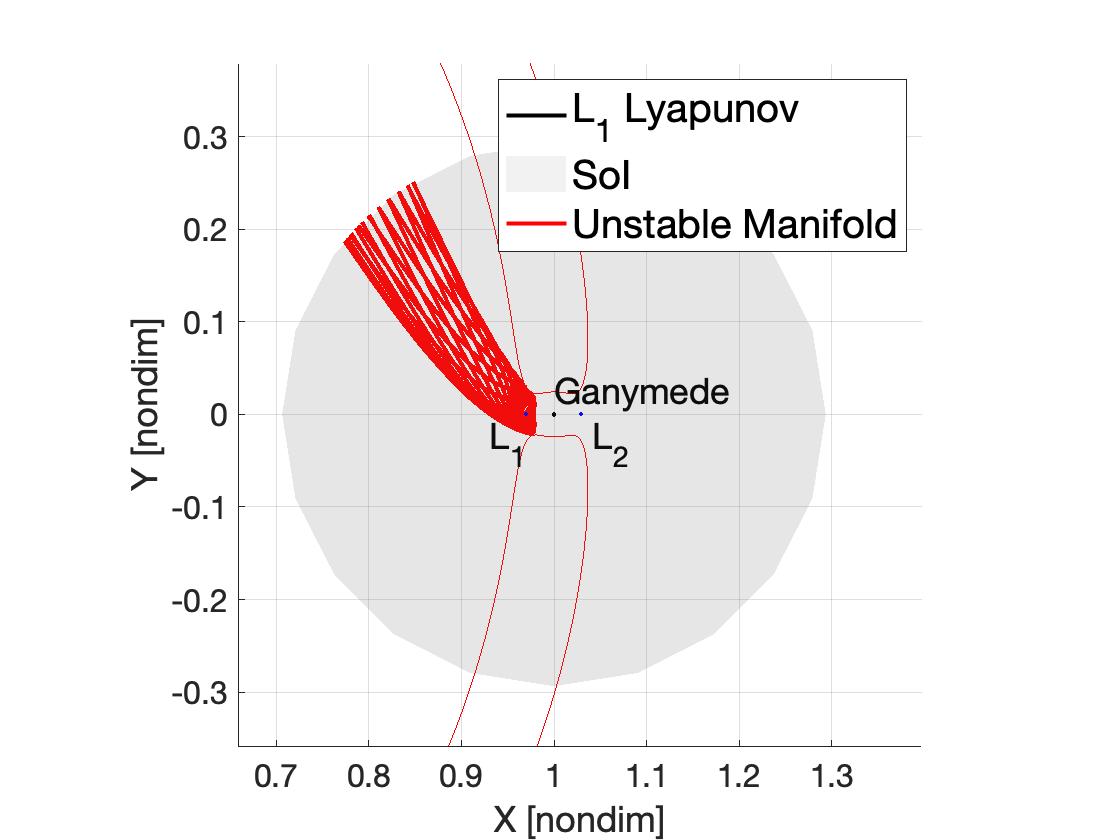}}}%
\end{minipage}\hfill{}%
\begin{minipage}[b][1\totalheight][t]{0.33\columnwidth}%
\subfigure[L$_2$ Lyapunov orbit's stable manifolds towards Europa SoI ($JC_a=3.0024$)  in the J-E rotating frame.]{{\includegraphics[width=6.5cm]{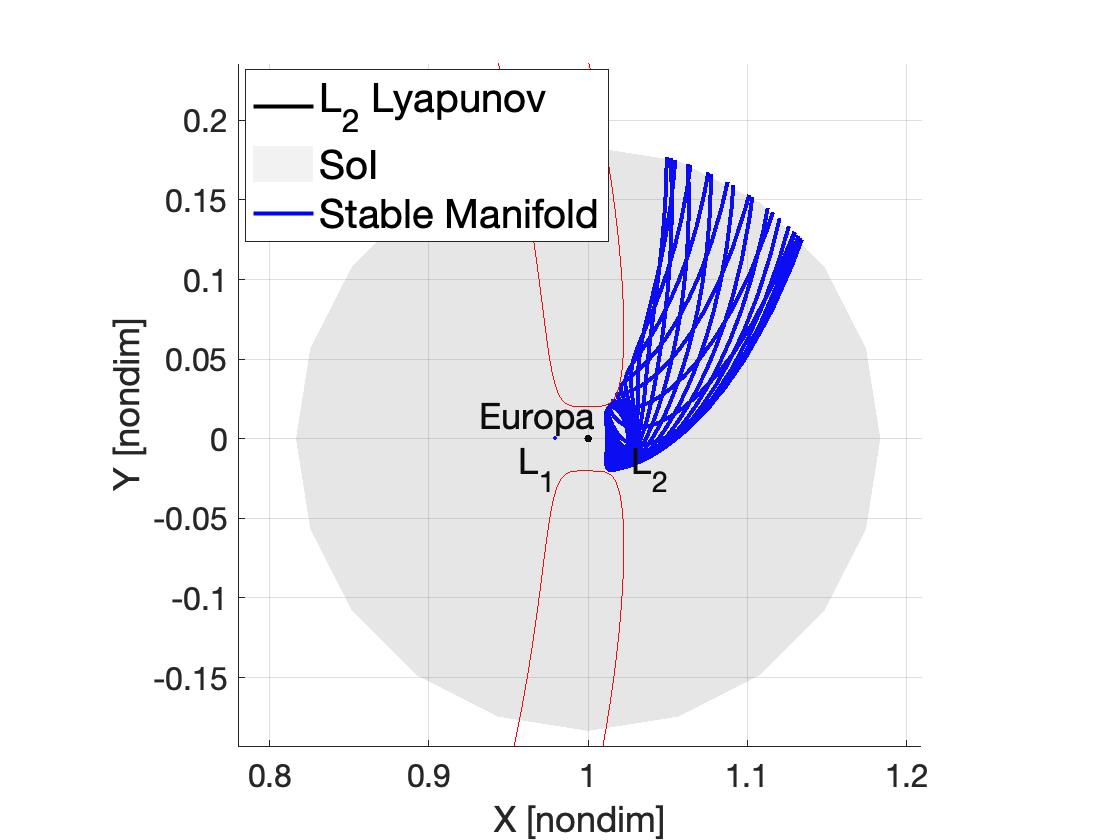}}\label{fig:PlanarInitialGuess-1}}%
\end{minipage}\hfill{}
\caption{{\label{fig:manifoldTowardsSoi}{Departure
trajectories from the Ganymede vicinity (left) and arrival trajectories
at the Europa vicinity (right).}}}
\end{figure}

\begin{figure}
\hfill{}\centering%
\begin{minipage}[b][1\totalheight][t]{0.55\columnwidth}%
\subfigure[All minimum-$\Delta v_{tot}$ tangent configurations. Angle projected onto the x-y plane and measured counter-clockwise from the $\hat{x}$-axis of the rotating frame.]{\label{fig:minDvPlanar}{\includegraphics[width=6.5cm]{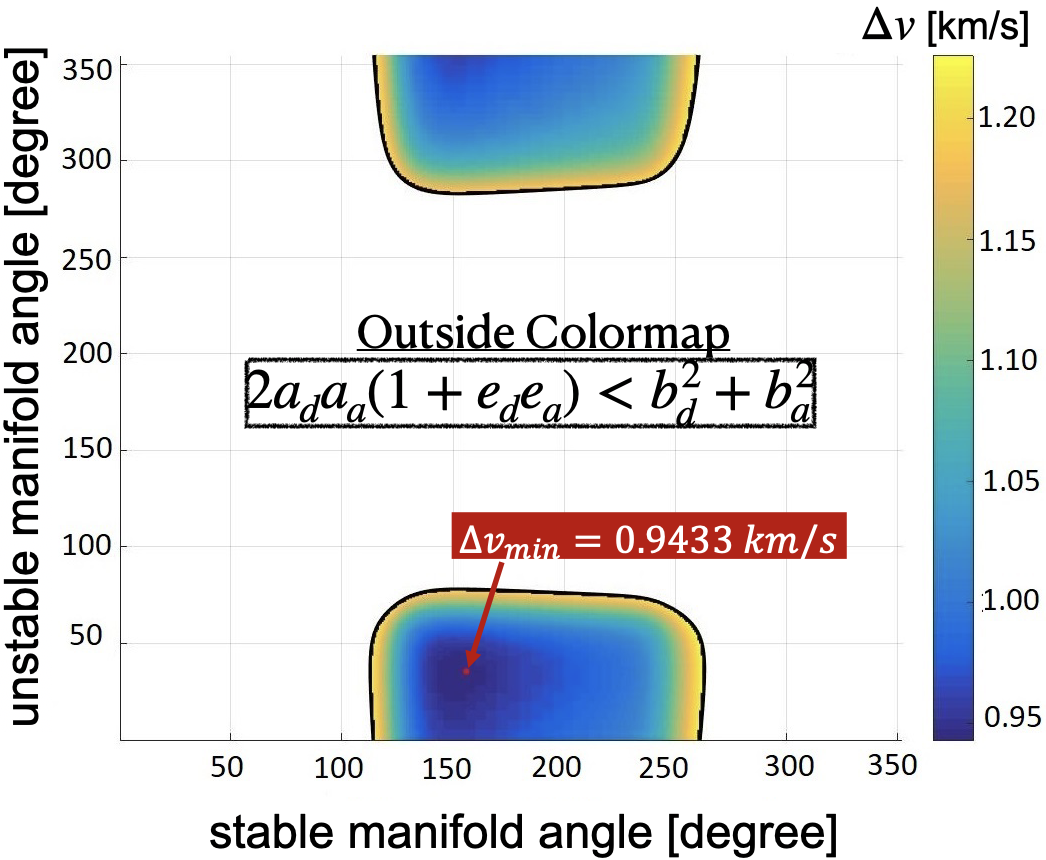}}}%
\end{minipage}\hfill{}%
\begin{minipage}[b][1\totalheight][t]{0.33\columnwidth}%
\subfigure[Minimum-$\Delta v_{tot}$ trajectory in the patched model CR3BP + 2BP in the Ecliptic J2000.0 Jupiter-centered inertial frame: $\Delta v_{tot}=0.9433$ km /s and $t_{tot}=9.47$ days.]{{\includegraphics[width=6.3cm]{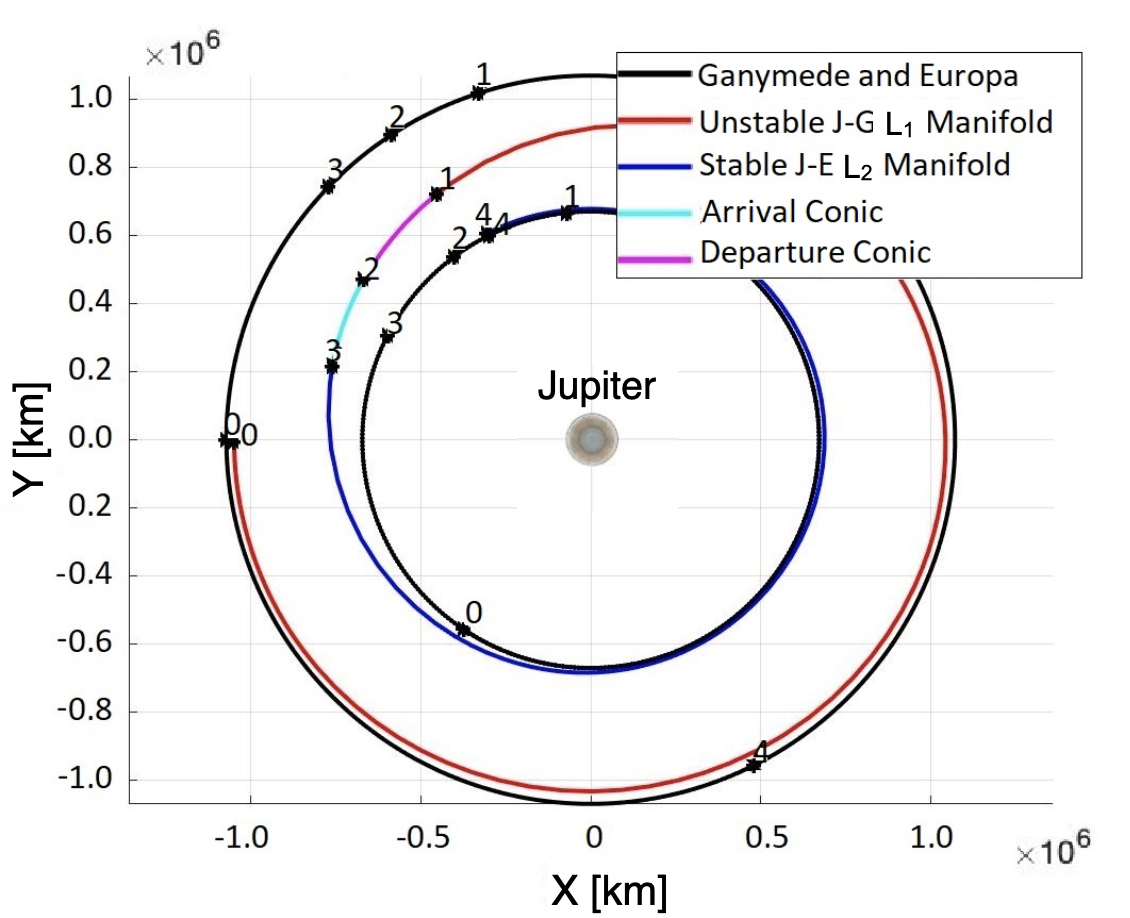}}\label{fig:PlanarInitialGuess}}%
\end{minipage}\hfill{}
\caption{{\label{fig:InitialGuess}{Transfer from unstable
manifolds leaving from L$_1$ in the J-G vicinity ($JC_{d}=3.0061$ ) to stable manifolds
arriving to L$_2$ in the J-E vicinity ($JC_a=3.0024$), assuming moons are coplanar (2BP-CR3BP patched model).}}}
\end{figure}

Unstable manifolds are propagated from the L$_1$ Lyapunov orbit in the
J-G system towards the Ganymede SoI (with a radius $R_{SoI_{Gan}}\approx 3\cdot10^5$ km), where they transition into departure
conics. Similarly, the stable manifolds arriving into the L$_2$
Lyapunov orbit in the J-E system are propagated from the Europa SoI (with a radius $R_{SoI_{Eur}}\approx 1.2\cdot10^5$ km), where they become
arrival conics in backwards time (Fig. \ref{fig:manifoldTowardsSoi}). Then,  Theorem \ref{theorem:planar} is evaluated for all permutations of unstable and stable manifold trajectories
(Fig. \ref{fig:minDvPlanar}). If the selected unstable manifold and stable manifold trajectories lead to departure and arrival conics, respectively, that fulfill Eq. \eqref{eq:planarConstraint}, then it is possible to obtain $\omega_a$ such that a tangential intersection occurs, as well as the arrival moon location at the final epoch along the transfer ($\theta_{4_{Eur}}$) from Eq. \eqref{eq:rephasing}.
The black line in Fig. \ref{fig:minDvPlanar}
bounds permutations of departure and arrival conics that satisfy Theorem \ref{theorem:planar} with those where the lower boundary reflected in Eq. \eqref{eq:planarConstraint} is not satisfied;
i.e., outside the colormap, all the departure conics are too large
for any arrival conics to intersect tangentially. The
angle in Fig. \ref{fig:minDvPlanar} corresponds to the
location of the departure/arrival arc on the manifold along the periodic orbit,
measured from the $\hat{x}$-axis regardless of the periodic orbits being planar or spatial. The manifold
trajectory leading to a tangential $\Delta v_{tot}$ connection between the
moons is then constructed, similar to \cite{Fantino2016} (Fig. \ref{fig:InitialGuess}).
Since the orbits of the moons are coplanar, the total $\Delta v_{tot}$, $t_{tot}$ and the transfer
configuration remain the same regardless of the angle of departure $\theta_{0_{Gan}}$ (i.e., the departure epoch in the Ganymede orbit). 

\begin{figure}
\hfill{}\centering%
\begin{minipage}[b][1\totalheight][t]{0.5\columnwidth}%
\subfigure[Jupiter-Europa Rotating frame.]{\label{fig:minDvPlanar-1}{\includegraphics[width=6cm]{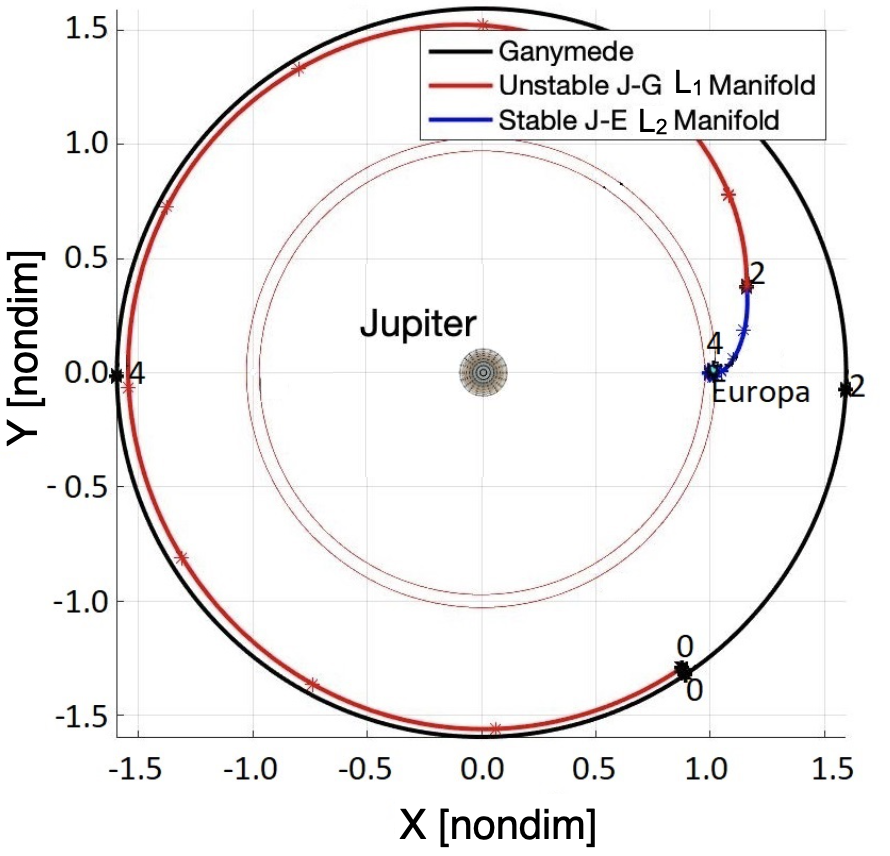}}}%
\end{minipage}\hfill{}%
\begin{minipage}[b][1\totalheight][t]{0.33\columnwidth}%
\subfigure[Ecliptic J2000.0 Jupiter-centered inertial frame.]{{\includegraphics[width=7.2cm]{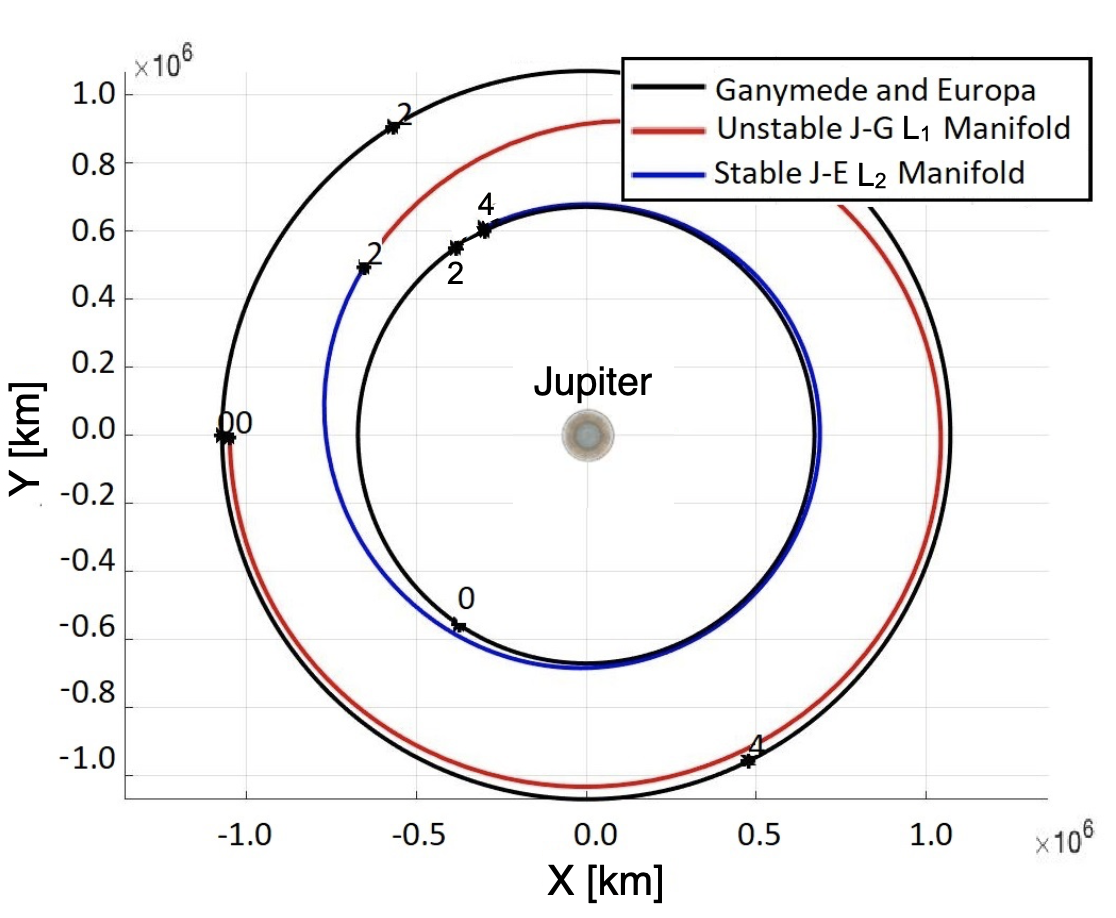}}\label{fig:Inertial}}%
\end{minipage}\hfill{}
\caption{\label{fig:Converged-solution-in-Planar}Transfer from unstable
manifolds leaving from L$_1$ in the J-G vicinity ($JC_{d}=3.0061$ ) to stable manifolds
arriving to L$_2$ in the J-E vicinity ($JC_a=3.0024$), assuming moons are coplanar (coupled planar CR3BP): $\Delta v_{tot}$=0.9456
km/s and $t_{tot}$=9.47 days.}
\end{figure}

Selection of the manifold corresponding to the minimum-$\Delta v_{tot}$ in such a simplified
model enables connections in the coupled planar CR3BP. To identify such links, the following angles from Fig.
\ref{fig:PlanarInitialGuess} are necessary: (a) the initial phase between the moons is computed measuring the location of
Ganymede with respect to the Europa location at instant 0; (b) a time-of-flight is determined for both the unstable and stable manifolds at instant 2 (intersection between departure and arrival conics in Fig. \ref{fig:PlanarInitialGuess}). By leveraging the result from the 2BP-CR3BP patched model as the initial guess, the differential corrections scheme in Appendix \ref{appendix:coupledCorrections} delivers the transfer in the coupled planar CR3BP. A final converged connection between the moons
then appears in Fig. \ref{fig:Converged-solution-in-Planar}. Furthermore,
the resulting trajectory possesses a very similar $t_{tot}$ and $\Delta v_{tot}$
as compared to the patched conic model. The initial guess from the
simplified model is reasonably efficient as compared to analyzing all possible scenarios
with Poincar\'e sections. Consequently, the 2BP-CR3BP patched model
offers a good initial guess to construct a transfer in the coupled planar CR3BP problem. The importance of Eq. \eqref{eq:planarConstraint} is significant: that is, a necessary condition that must be fulfilled to locate tangential connections between planar trajectories linking two coupled coplanar CR3BP models.
Also, Eq. \eqref{eq:planarConstraint} is leveraged as a constraint to produce feasible transfers
in the CR3BP where the motion of the s/c is mostly governed by one
primary and the trajectories are planar.

\subsection{\label{subsec:Moons-in-different-planes}Moons on non-coplanar orbits}
\subsubsection{Analytical constraint for a successful transfer}
The next level of complexity is a strategy to produce a successful transfer
considering that both moon orbits are in different planes, with different
values of $i$ and $\Omega$. The design
procedure to deliver a suitable transfer between the planes also employs
the 2BP-CR3BP patched model, given the advantage of initially exploring
the problem analytically. Given that moon orbits are in their true orbital plane, departure and arrival conics possess different values of  $\Omega$ and $i$. The objective is the determination of the geometrical
condition necessary for both departure and arrival conics
to intersect, at some point in space, given a uniquely determined
phase between the moons. Similar to the example for coplanar moon orbits, the arrival epoch of the arrival moon is assumed free with the aim of rephasing the arrival moon in its orbit such that an intersection between
departure and arrival conics is accomplished. Consequently, the following theorem is delivered: 
\begin{theopargself}
\begin{theorem}
\label{theorem:spatial}
As long as the geometrical properties of two conics located in different planes fulfill the inequality
constraint represented by 
\begin{equation}
a_{a}(1-e_{a})\le\frac{a_{d}(1-e_{d}^{2})}{1+e_{d}\cos(\theta_{d_{Int}}+n\pi)}\le a_{a}(1+e_{a}),\ \text{being }n=0,\ 1.\label{eq:3dConstraint}
\end{equation}
either one of the two conics can be reoriented such that they intersect in space. Consequently, the ideal phase of the arrival moon at arrival, $\theta_{4_{Eur}}$, for the moon-to-moon transfer to occur is obtained considering that the departure epoch, $\theta_{0_{Gan}}$, is fixed.
\end{theorem}
\end{theopargself}

\begin{figure}
\hfill{}\centering\includegraphics[scale=0.35]{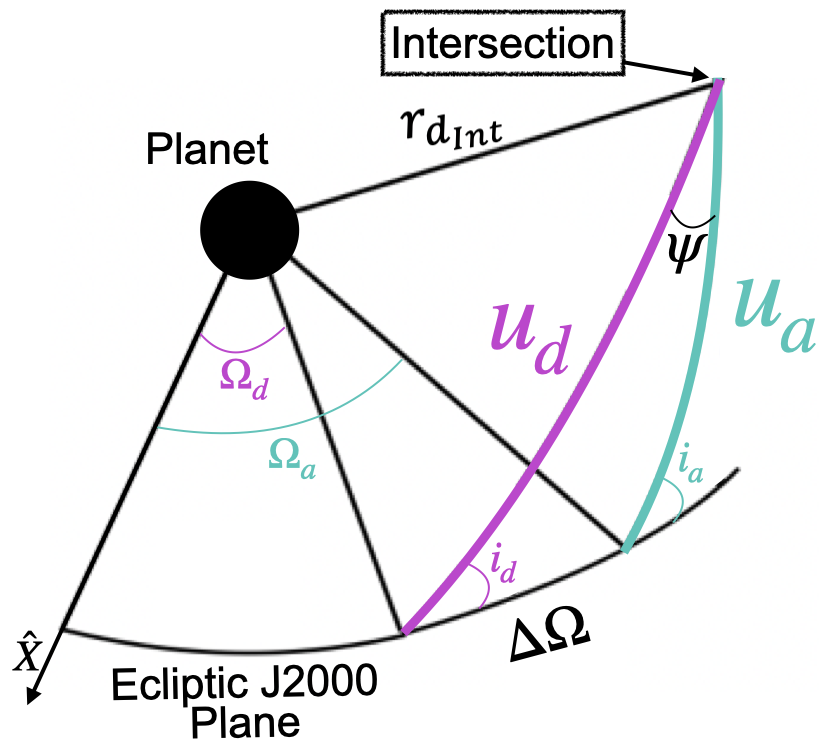}\hfill{}
\caption{\label{fig:sphericalTrigonometry} Intersection between two planes
using spherical trigonometry \citep{Vallado}.}
\end{figure}

\noindent {\it \textbf{Proof}}\\
By applying spherical trigonometry (see schematic in Fig. \ref{fig:sphericalTrigonometry}),
it is possible to determine the angle of intersection for the departure and arrival conics with respect
to their node line (right ascension of the ascending node), $u_{d}$ and $u_{a}$ respectively, between the two orbital
planes for both moon orbits. These two angles are obtained given the difference in the right ascension of the ascending nodes of the two conics, $\Delta\Omega=\Omega_{d}-\Omega_{a}$, and the inclinations
of the two planes, $i_{d}$ and $i_{a}$:
\begin{align} 
\displaystyle
\sin u_{d}&=\frac{\sin(i_{a})\sin(\Delta\Omega)}{\sin(\psi)}\\
\displaystyle 
\cos u_{d}&=\frac{\frac{\sin(i_{a})\cos(\Delta\Omega)}{\cos(i_{d})}-\cos(\psi)\tan(i_{d})}{\sin(\psi)}\\
\displaystyle 
\sin u_{a}&=\frac{\sin(i_{d})\sin(\Delta\Omega)}{\sin(\psi)}\\
\displaystyle 
\cos u_{a}&=\cos(\Delta\Omega)\cos(u_{d})+\sin(\Delta\Omega)\sin(u_{d}),
\end{align}
where $\psi=\cos^{-1}(\cos(i_{a})\cos(i_{d})+\sin(i_{a})\sin(i_{d})\cos(\Delta\Omega)).$
As a result, the true anomaly at which the departure conic intersects
the plane of the arrival conic is evaluated as $\theta_{d_{Int}}$ or $\theta_{d_{Int}}+\pi$,
depending on the argument of periapsis for the departure conic, $\omega_{d}$:
$\theta_{d_{Int}}=u_{d}-\omega_{d}$. The spatial intersection
of the departure conic occurs at:
\begin{equation}
r_{d_{Int}}=\frac{a_{d}(1-e_{d}^{2})}{1+e_{d}\cos(\theta_{d_{Int}}+n\pi)},\ \text{being }n=0,\ 1.
\end{equation}
Note that the angle $\theta_{d_{Int}}$ or $\theta_{d_{Int}}+\pi$
is generally defined by the $i$ and $\Omega$ values for the departure and arrival
planes, as apparent in Fig. \ref{fig:sphericalTrigonometry}, as well as by $\omega_{d}$, that is strictly dependent on the departure epoch from the departure moon. For a successful spatial connection between the departure
and arrival conic, the following relationship must hold for the arrival conic:
\begin{equation}
r_{d_{Int}}=\frac{a_{a}(1-e_{a}^{2})}{1+e_{a}\cos(\theta_{a_{Int}}+n\pi)},\ \text{being }n=0,\ 1.
\end{equation}
As a result, the true anomaly for the intersection of the arrival conic
at $\theta_{a_{Int}}$ and $\theta_{a_{Int}}+\pi$
is obtained:
\begin{equation}
\cos(\theta_{a_{Int}}+n\pi)=\frac{\frac{p_{a}}{r_{d_{Int}}}-1}{e_{a}},\ \text{being }n=0,\ 1,\label{eq:thetaArrivalIntersection}
\end{equation}
where $p_{a}=a_{a}(1-e_{a}^{2})$. From Eq. \eqref{eq:thetaArrivalIntersection},
given that $-1\le\cos(\theta_{a_{Int}}+n\pi)\le1$,
Theorem \ref{theorem:spatial} is deduced. \textbf{QED}

\begin{figure}[h!]
\hfill{}\centering\includegraphics[width=14cm]{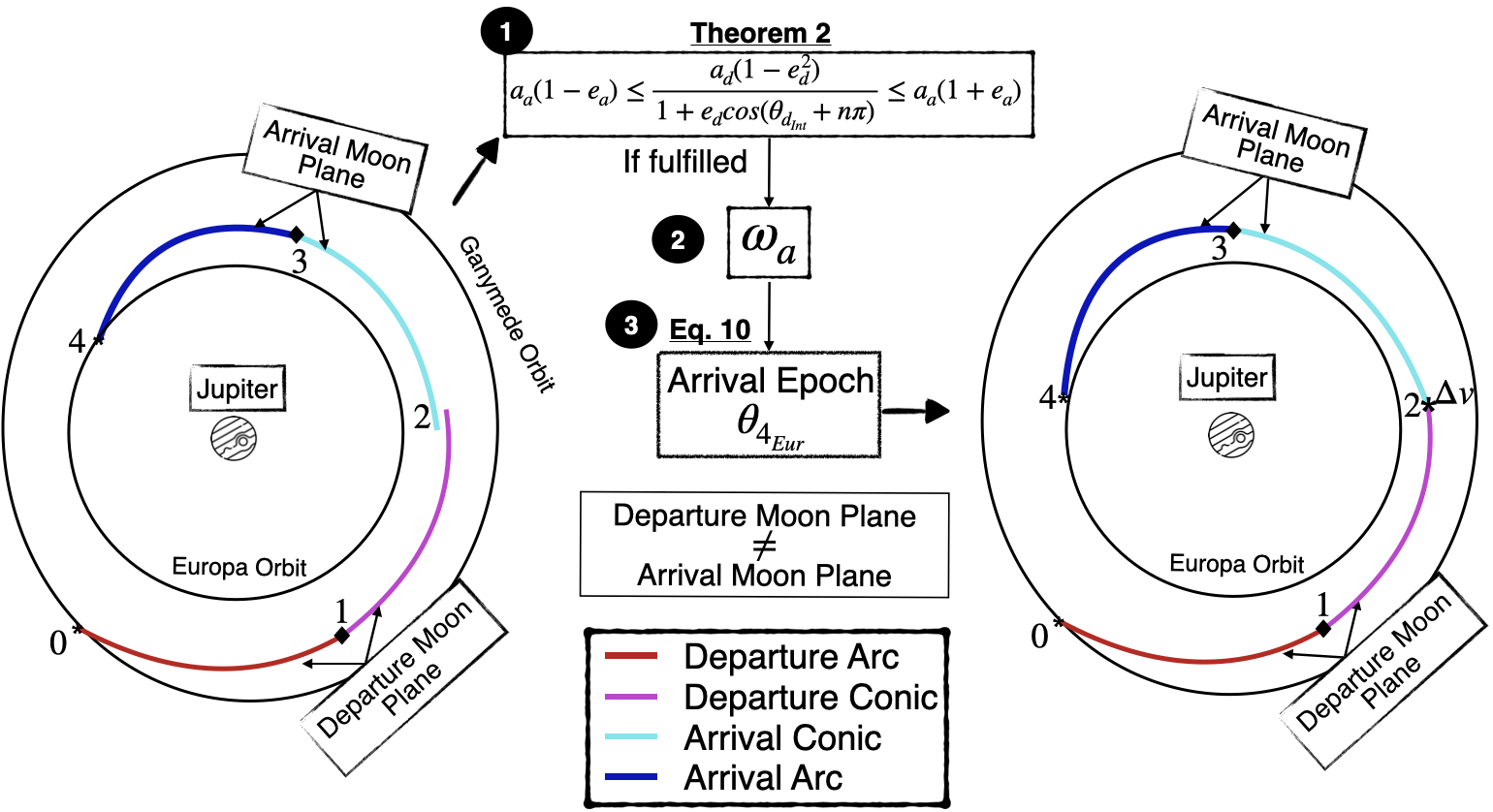}\hfill{}
\caption{\label{fig:reorientationGraph3dProcedure} Representation of the rephasing of
the arrival moon if the geometrical and plane properties of the departure and arrival conics fulfill Theorem \ref{theorem:spatial}.}
\end{figure}

As a result, Theorem \ref{theorem:spatial} is proved. Note that the lower limit of Eq.  \eqref{eq:3dConstraint} corresponds to the periapsis of the arrival conic, $r_{\pi_a}=a_a(1-e_a)$, and that the upper limit to its apoapsis, $r_{\alpha_a}=a_a(1+e_a)$. The lower boundary thus defines an arrival conic
that is too large to connect with the departure conic; the upper limit represents an arrival conic that is too small to link with the departure
conic. If the inequality constraint in Eq. \eqref{eq:3dConstraint}
is fulfilled, the angle of intersection for the arrival conic $\theta_{a_{Int}}$
is determined and, since the intersection occurs at $\theta_{a_{Int}}$
and $\theta_{a_{Int}}+\pi$, $\omega_{a}=u_{a}-\theta_{a_{Int}}$ is obtained. The optimal phase for the
arrival moon to yield such a configuration follows 
the same procedure as detailed in Sect. \ref{subsec:Arrival-moon's-rephasing}. Once $\omega_a$ is delivered, the arrival moon location at the final epoch along the transfer ($\theta_{4_{Eur}}$) is computed with Eq. \eqref{eq:rephasing}. The concept is represented in Fig. \ref{fig:reorientationGraph3dProcedure}. If a transfer is available, it is computed for both the angles $\theta_{a_{Int}}$
and $\theta_{a_{Int}}+\pi$, which provide two distinct values for $\omega_a$ that lead to different values for the $\Delta v_{tot}$ of the transfer. Therefore, the one with the smallest $\Delta v_{tot}$ is selected. 

\subsubsection{\label{subsec:Ganymede-to-Europa}Ganymede to Europa transfer application}
\begin{figure}
\hfill{}\centering%
\begin{minipage}[b][1\totalheight][t]{0.4\columnwidth}%
\subfigure[Evaluation of the constraint Eq. \eqref{eq:3dConstraint} at $\theta_{d_{Int}}$ and $\theta_{d_{Int}}+\pi$.]{\label{fig:constraintVariation}{\includegraphics[width=5.5cm]{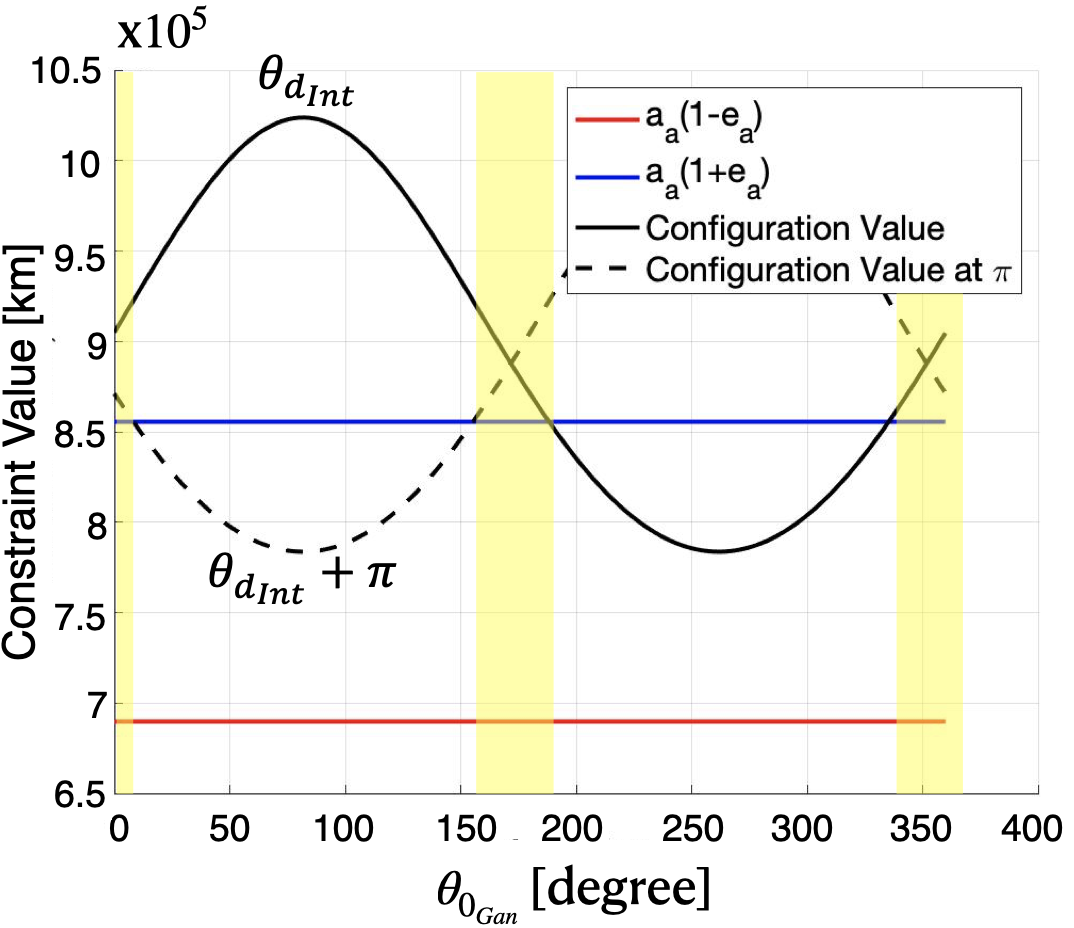}}}%
\end{minipage}\hfill{}%
\begin{minipage}[b][1\totalheight][t]{0.33\columnwidth}%
\subfigure[Phase of Ganymede and Europa with respect to their ascending node direction at $t_0$.]{{\includegraphics[width=5.5cm]{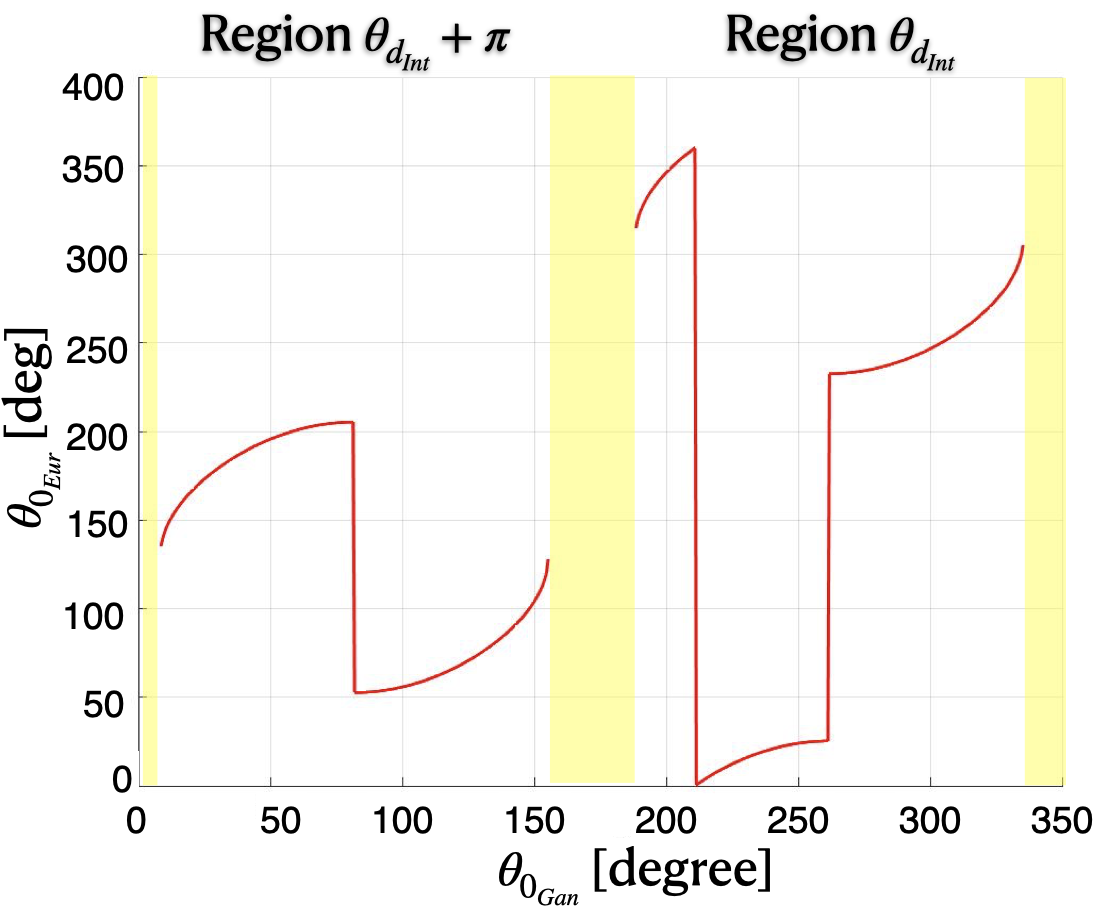}}\label{fig:europaVariation}}%
\end{minipage}\hfill{}
\hfill{}\centering%

\hfill{}\centering%
\begin{minipage}[b][1\totalheight][t]{0.4\columnwidth}%
\subfigure[Transfer $\Delta v_{tot}$ given a $\theta_{0_{Gan}}$.]{\label{fig:dVVariationGan}{\includegraphics[width=5.5cm]{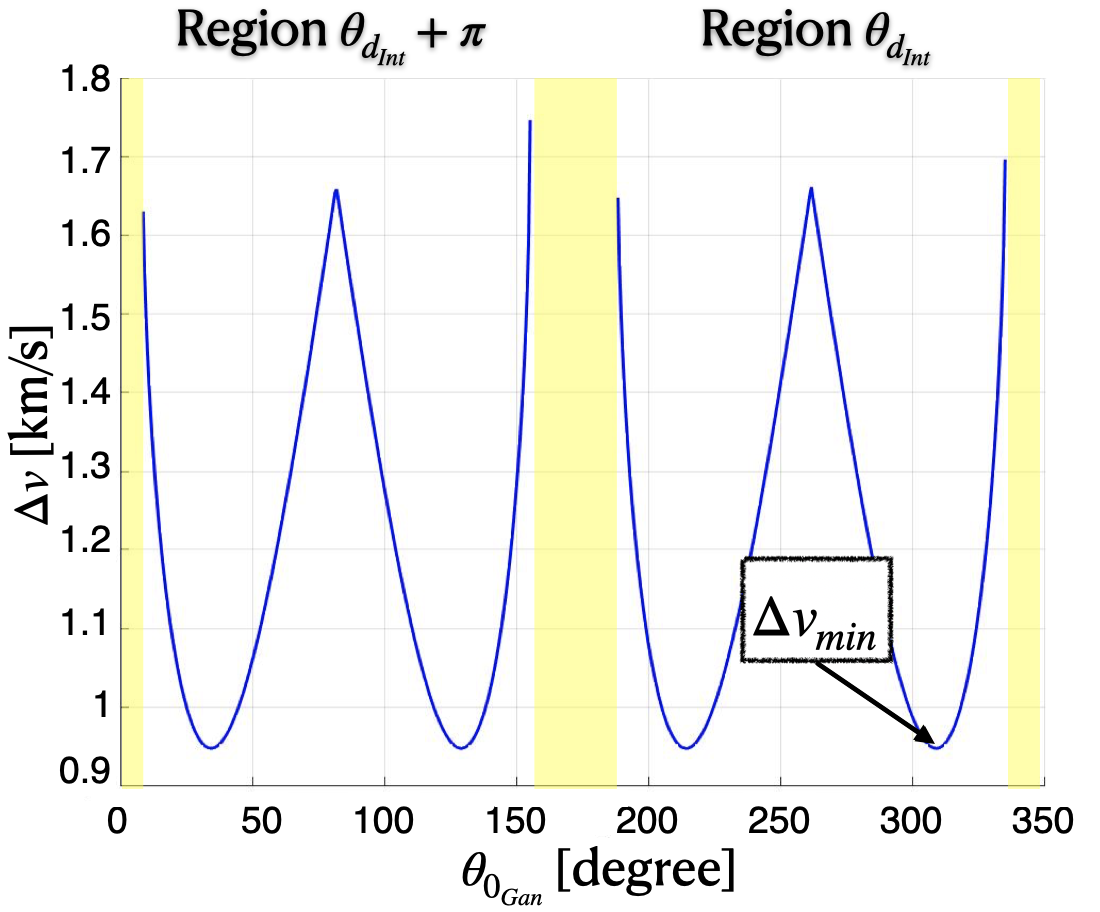}}}%
\end{minipage}\hfill{}%
\begin{minipage}[b][1\totalheight][t]{0.33\columnwidth}%
\subfigure[Transfer $t_{tot}$ given a $\theta_{0_{Gan}}$]{{\includegraphics[width=5.5cm]{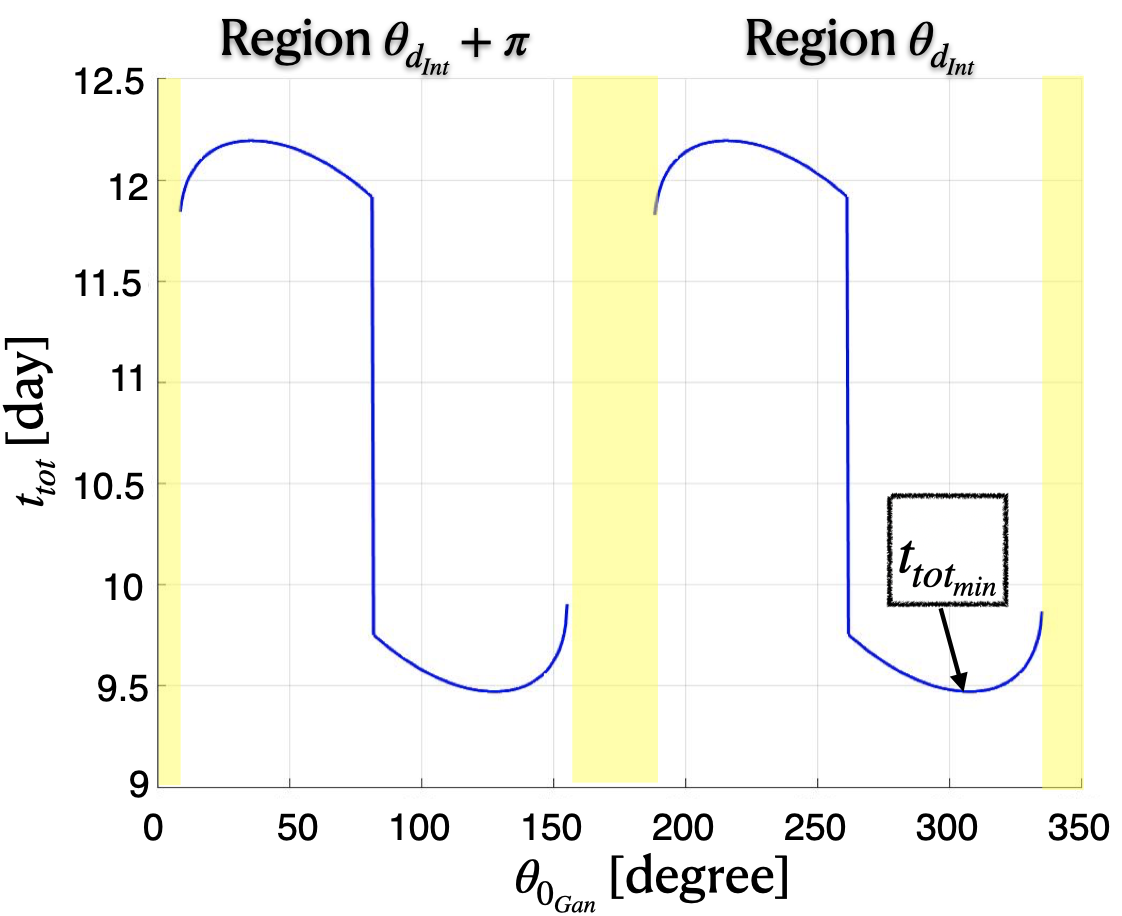}}\label{fig:timeOfFLightVariation}}%
\end{minipage}\hfill{}
\hfill{}\centering%
\centering{}\caption{\label{fig:thetaGanVariation}Available successful configurations for both intersections
at $\theta_{d_{Int}}$ and $\theta_{d_{Int}}+\pi$.}
\end{figure}

\begin{figure}[h!]
\hfill{}\centering%
\begin{minipage}[b][1\totalheight][t]{0.5\columnwidth}%
\subfigure[Minimum-$\Delta v_{tot}$ configuration: $\Delta v_{tot}=0.9448$ km/s and $t_{tot}=9.473$ days.]{\label{fig:minDvSpatialConic}{\includegraphics[width=6.3cm]{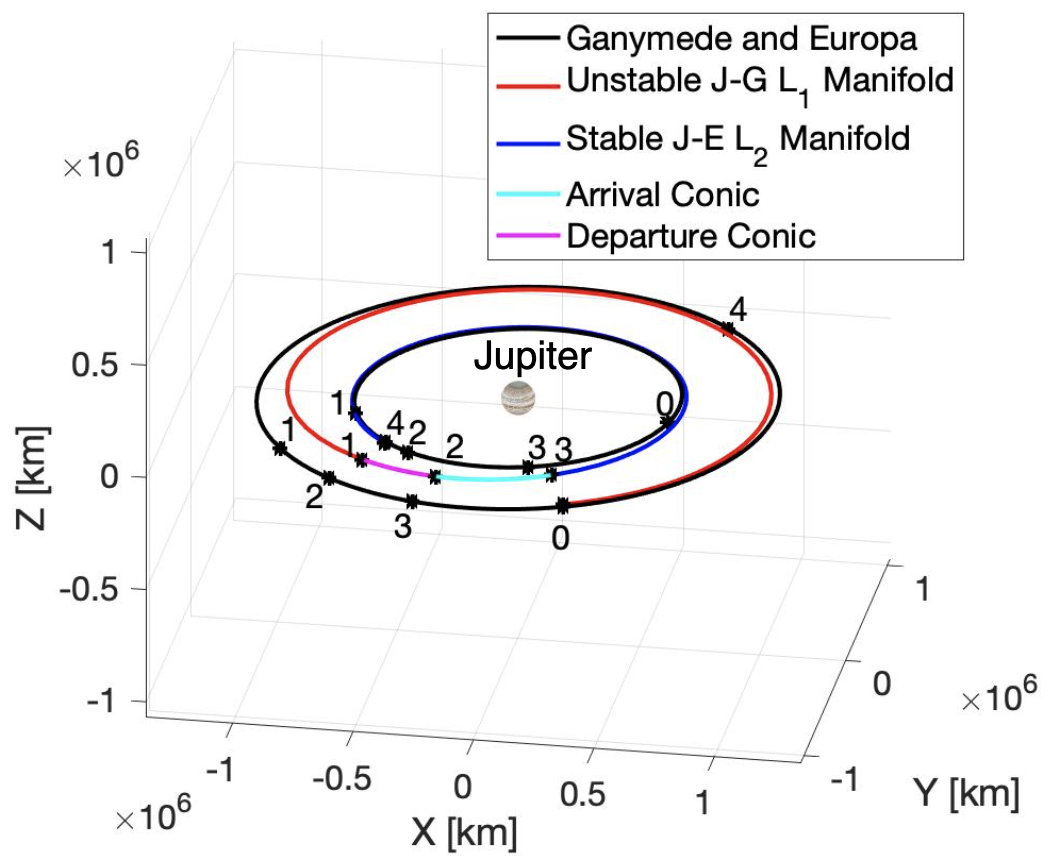}}}%
\end{minipage}\hfill{}%
\begin{minipage}[b][1\totalheight][t]{0.33\columnwidth}%
\subfigure[Minimum-$t_{tot}$  configuration: $\Delta v_{tot}=0.9455$ km/s and $t_{tot}=9.471$ days.]{{\includegraphics[width=6.5cm]{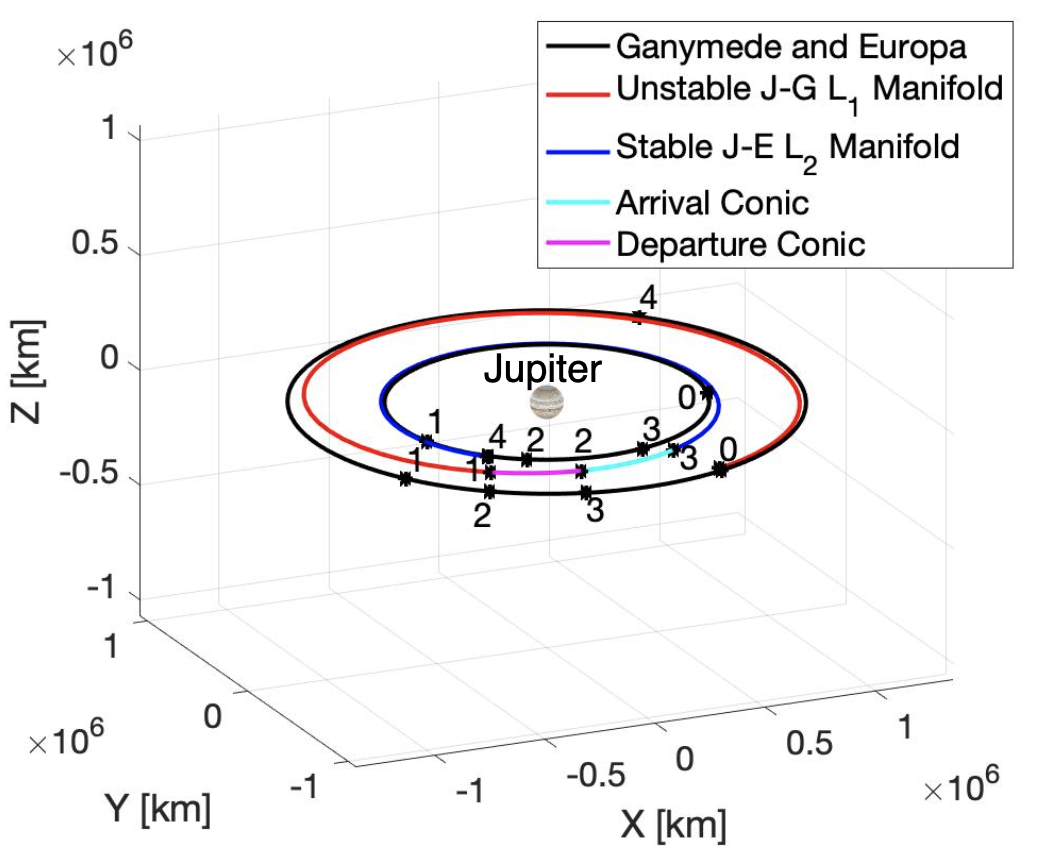}}\label{fig:minimumtofspatialpatched}}%
\end{minipage}\hfill{}
\centering{}\caption{\label{fig:SpatialPatchedModel}Transfer from unstable
manifolds leaving from L$_1$ in the J-G vicinity ($JC_{d}=3.0061$) to stable manifolds
arriving to L$_2$ in the J-E vicinity ($JC_a=3.0024$) in the Ecliptic J2000.0 Jupiter-centered inertial frame (2BP-CR3BP patched model).}
\end{figure}

\begin{figure}[h!]
\hfill{}\centering%
\begin{minipage}[b][1\totalheight][t]{0.5\columnwidth}%
\subfigure[Minimum-$\Delta v_{tot}$ configuration: $\Delta v_{tot}=0.9422$ km/s and $t_{tot}=9.473$ days.]{\label{fig:minDvSpatialCoupled}{\includegraphics[width=6.3cm]{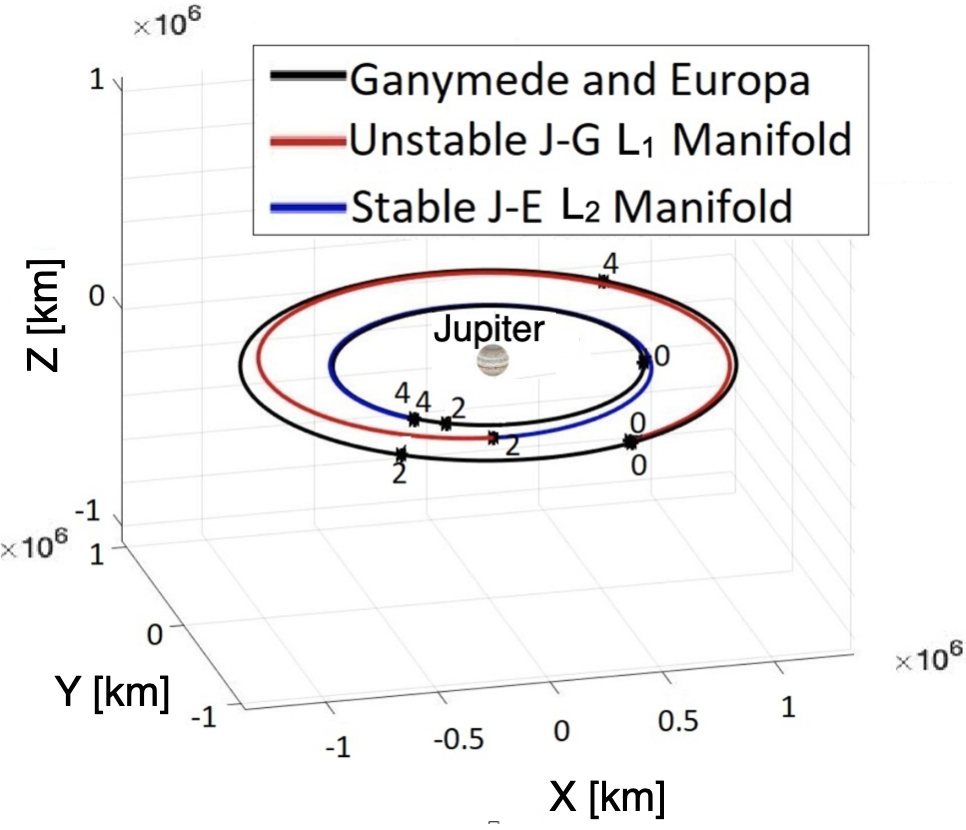}}}%
\end{minipage}\hfill{}%
\begin{minipage}[b][1\totalheight][t]{0.33\columnwidth}%
\subfigure[Minimum-$t_{tot}$ configuration: $\Delta v_{tot}=0.9428$ km/s and $t_{tot}=9.472$ days.]{{\includegraphics[width=6.5cm]{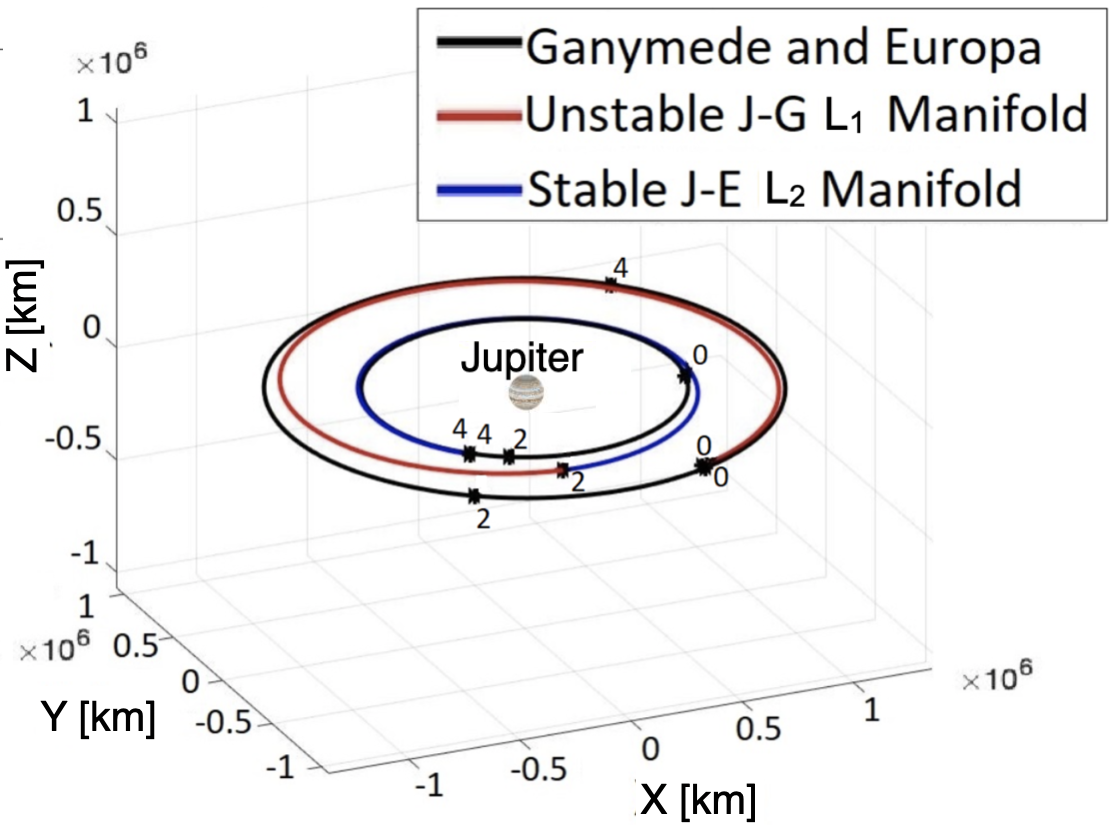}}\label{fig:minimumtofspatialpatched-1-1}}%
\end{minipage}\hfill{}
\centering{}\caption{\label{fig:SpatialCoupledCR3BP-1}Transfer from unstable
manifolds leaving from L$_1$ in the J-G vicinity ($JC_{d}=3.0061$) to stable manifolds
arriving to L$_2$ in the J-E vicinity ($JC_a=3.0024$) in the Ecliptic J2000.0 Jupiter-centered inertial frame (coupled spatial CR3BP).}
\end{figure}

Consider the Ganymede-Europa transfer discussed in Sect. \ref{subsec:coplanarApplication}. For purposes of comparison, the minimum-$\Delta v_{tot}$ trajectory in
the planar case (Fig. \ref{fig:InitialGuess}) is introduced but now assuming that the moons are in their true orbital planes. For a given departure angle from Ganymede
$\theta_{0_{Gan}}$ in its orbit, it is possible to complete a feasibility analysis (Fig.
\ref{fig:thetaGanVariation}) where Theorem \ref{theorem:spatial} is fulfilled. Then, it is possible to determine for which $\theta_{0_{Gan}}$ it is possible to locate a transfer between the two specified moons. The black line in Fig.
\ref{fig:constraintVariation} corresponds to the middle term in Eq. \eqref{eq:3dConstraint}, i.e., the constraint value, for each given configuration between departure and arrival conics assuming an intersection at $\theta_{d_{Int}}$ or $\theta_{d_{Int}}+\pi$. For every $\theta_{0_{Gan}}$, if the constraint value
 is between the red and blue lines (lower and upper boundaries from Eq.
\eqref{eq:3dConstraint}, respectively), the total $\Delta v_{tot}$, $t_{tot}$,
and the position of Europa at the initial epoch $\theta_{0_{Eur}}$ are
noted, as reflected in Fig. \ref{fig:thetaGanVariation}. The light yellow regions in Fig. \ref{fig:thetaGanVariation} correspond to configurations where the inequality constraint represented by Eq.
\eqref{eq:3dConstraint} is not met and, as a result, a direct transfer for such departure epochs at $\theta_{0_{Gan}}$ is not possible.
However, many promising scenarios are available, such as the 
minimum-$\Delta v_{tot}$ (Fig. \ref{fig:minDvSpatialConic}) or minimum
$t_{tot}$ configurations (Fig. \ref{fig:minimumtofspatialpatched}). Note that about 5.6 days is required for the s/c to leave the Ganymede SoI and 2.9 days to arrive to the L$_2$ Lyapunov orbit in the Europa vicinity from the Europa SoI.
Similar to the coplanar moon orbits analysis, these solutions are efficiently transitioned to the coupled spatial CR3BP using the transfers from the 2BP-CR3BP patched model as an initial guess (Fig. \ref{fig:SpatialCoupledCR3BP-1}). It is, thus, interesting to observe that it is possible to implement transfers between the libration points of both Ganymede and Europa in less than 10 days using a single $\Delta v_{tot}$ equal to 942 m/s.  Also, despite the observation that, in this case, the minimum-$\Delta v_{tot}$ and minimum-$t_{tot}$ solutions seem to occur at similar epochs, such a correlation is not guaranteed for other transfers or systems. These results are summarized in Table \ref{Table:resultsMMATJupiter}. In addition, using the analytical method that emerges from simplifying the problem to the 2BP-CR3BP patched model, it is possible to deduce the relative angles between the moons where intersections between departure and arrival conics in different planes occur, as long as Eq. \eqref{eq:3dConstraint} is fulfilled. Therefore, this approach offers an advantage over the Poincar\'e sections as employed in Sect. \ref{subsec:coupledPoincareSectionsTruePlanes}, since it can be challenging to locate sections where the position discontinuity in the $\hat{z}$-axis direction does not occur. Using this method prior to the introduction of a Poincar\'e section aids in the down-selection of relative positions between the departure and arrival moons, as well as possible locations in space where the departure and arrival arcs intersect.
 
\begin{figure}
\hfill{}\centering\includegraphics[width=6.5cm]{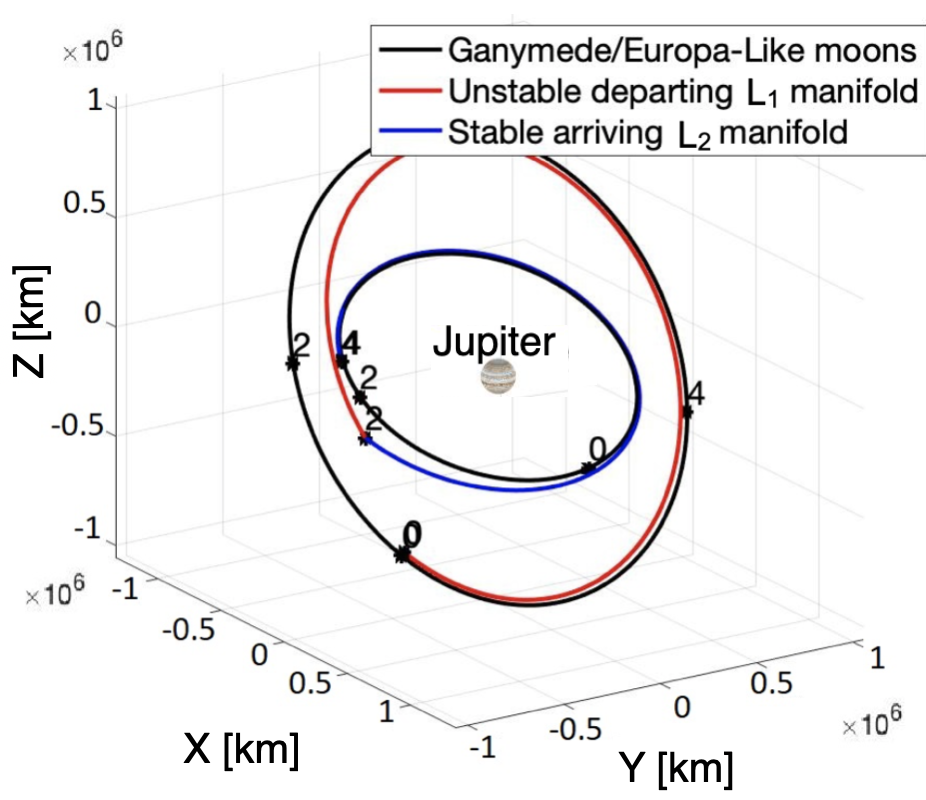}\hfill{}
\caption{\label{fig:minimumtofspatialrandom} Transfer from unstable
manifolds departing  from L$_1$ in the J-G vicinity ($JC_{d}=3.0061$) to stable manifolds
arriving to L$_2$ in the J-E vicinity ($JC_a=3.0024$) in the Ecliptic J2000.0 Jupiter-centered inertial frame, considering that Ganymede and Europa are in different planes from their actual orbits. $\Delta v_{tot}=12.98$ km/s and $t_{tot}=11.83$ days. This figure is uniquely presented to validate the method. Model: coupled spatial CR3BP.}
\end{figure}

 \begin{table}[htbp!]
    \caption{Comparison of resulting transfers assuming moons reside in coplanar orbits and in their true orbital planes in the coupled CR3BP.}
   \label{Table:resultsMMATJupiter}
        \centering 
        \begin{tabular}{llll}
\hline\noalign{\smallskip}
&  $\Delta v_{tot}$  [km/s]&$t_{tot}$ [days] \\
\noalign{\smallskip}\hline\noalign{\smallskip}
	Coplanar moon orbits  & 0.9456& 9.470 \\
\noalign{\smallskip}\hline
	True moon orbits: Minimum-$\Delta v_{tot}$ transfer & 0.9422& 9.473 \\
\noalign{\smallskip}\hline
	True moon orbits: Minimum-$t_{tot}$ transfer & 0.9428& 9.472 \\
\noalign{\smallskip}\hline
\end{tabular}
\end{table}
It is possible to produce a transfer solution between CR3BP arcs connecting two moons as long as the properties of the resulting departure and arrival conics fulfill Eq. \eqref{eq:3dConstraint}. This fact is reflected in Fig. \ref{fig:minimumtofspatialrandom}, where a transfer between two moons with the same properties as Ganymede and Europa in Table
\ref{Table:DataEuropaGanymede}, but located in very different
planes, is constructed. The properties of the planes are as follows: $\Omega_{JG}=100$\si{\degree},\ $i_{JG}=60$\si{\degree},\ $\Omega_{JE}=200$\si{\degree} and $i_{JG}=20$\si{\degree}. Note that the moon planes are arbitrarily selected for demonstration.

\subsection{\label{subsec:Comparison}Comparison between transfers for coplanar and non-coplanar moon orbits}
\begin{figure}
\hfill{}\centering%
\begin{minipage}[b][1\totalheight][t]{0.5\columnwidth}%
\subfigure[Total $\Delta v_{tot}$ depending on $\theta_{0_{Gan}}$.]{\label{fig:comparisonDeltavCoplanarReal}{\includegraphics[width=6.5cm]{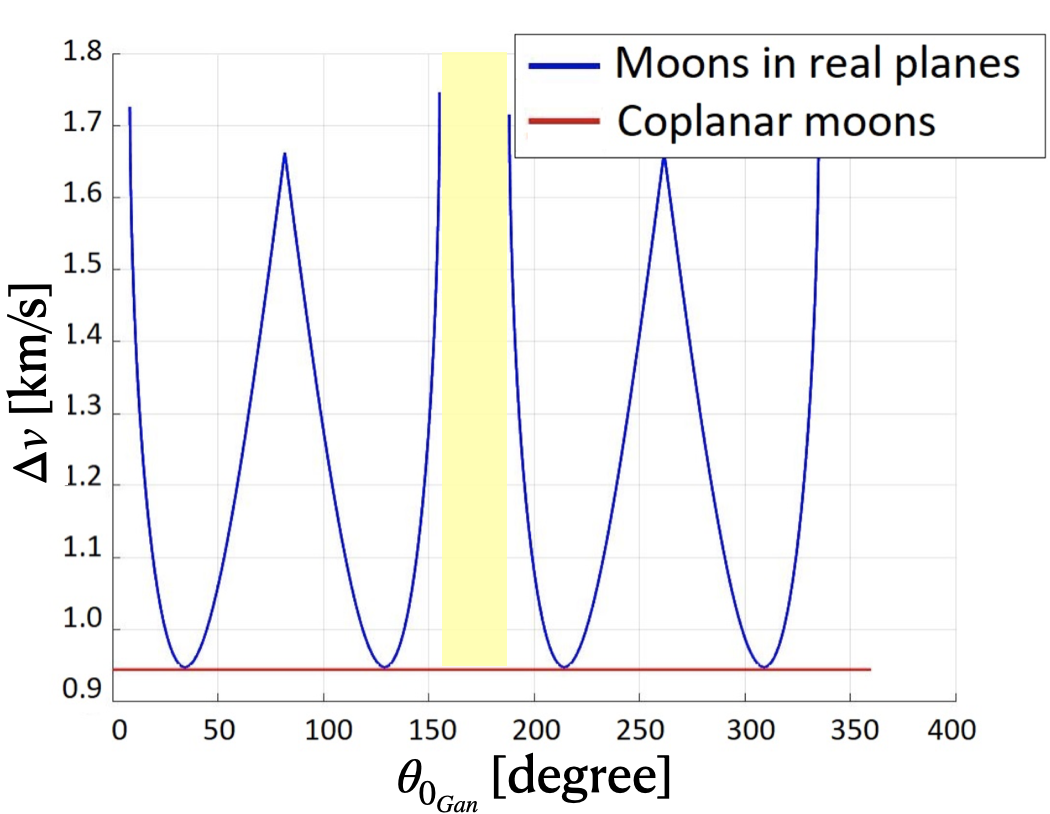}}}%
\end{minipage}\hfill{}%
\begin{minipage}[b][1\totalheight][t]{0.33\columnwidth}%
\subfigure[Final $t_{tot}$ depending on $\theta_{0_{Gan}}$.]{{\includegraphics[width=6.5cm]{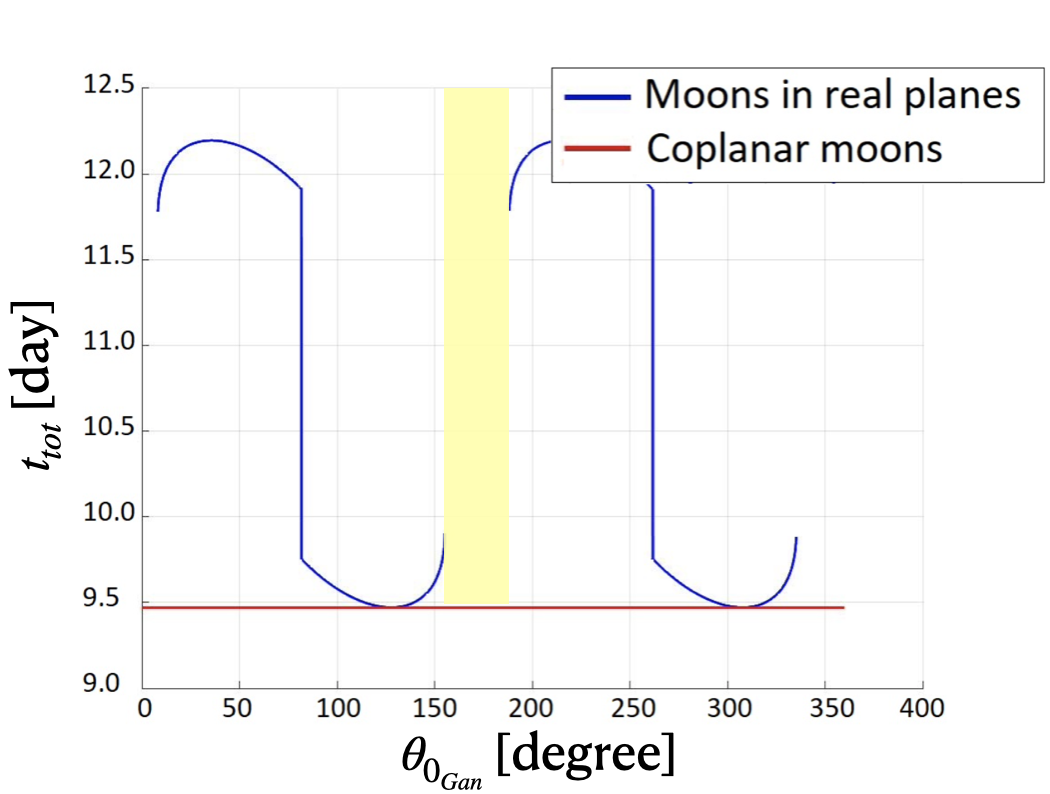}}\label{fig:timecomparisonCoplanar}}%
\end{minipage}\hfill{}
\caption{\label{Comparison-Coplanar-Real}Comparison of the total $\Delta v_{tot}$ and $t_{tot}$
for both (i) assuming that the moons orbits are coplanar (red curve) and (ii) in their real planes (blue line). In
the case of moons moving in their real planes, successful configurations
with intersection at $\theta_{d_{Int}}$ and $\theta_{d_{Int}}+\pi$
are included in the same plot for the sake of comparison. }
\end{figure}

It is possible to compare results between the solutions produced assuming coplanar and non-coplanar
moon orbits. This comparison is accomplished within the context of the example
previously introduced. If the transfer design approach assumes both moon orbits are coplanar, Fig. \ref{Comparison-Coplanar-Real}
illustrates that the total $\Delta v_{tot}$ for the transfer, as well as $t_{tot}$,
remains constant regardless of the initial epoch $\theta_{0_{Gan}}$. Nevertheless, when the moons are incorporated in their true orbital planes, despite
the fact that their relative inclinations are small, and given that there is a difference
between $\Omega_{G}$ and $\Omega_{E}$ , it is apparent in Fig. \ref{fig:comparisonDeltavCoplanarReal} that, depending on $\theta_{0_{Gan}}$, the single $\Delta v_{tot}$ as well as $t_{tot}$ values vary depending on $\theta_{0_{Gan}}$ (Fig. \ref{fig:timecomparisonCoplanar}).
There exist some departure angles, $\theta_{0_{Gan}}$, for which
a connection between the departure and arrival trajectories does not occur. This scenario occurs  due to the fact that Eq. \eqref{eq:3dConstraint} is not satisfied.
Generally, as is well known, the $\Delta v_{tot}$ and
$t_{tot}$ results assuming the moon orbits are coplanar are undervalued with respect to the spatial model, since the results sensibly vary depending
on the position of Ganymede in its plane at departure (Fig. \ref{Comparison-Coplanar-Real}). It is also possible to observe in Table \ref{Table:resultsMMATJupiter} that the coplanar analysis corresponds to an approximate result for the cheapest and lowest time-of-flight corresponding to the transfer between moons in their true orbital planes. Therefore, although the coplanar analysis supplies
preliminary information concerning the transfers between the moons, the spatial technique
supplies more accurate insight to exploit in real applications assuming the goal is a direct transfer. For a transfer between two moons involving a single maneuver, the cost and time are entirely dependent upon the initial epoch, represented in this example as $\theta_{0_{Gan}}$, with some values of the angle providing no access whatsoever.

\subsection{\label{subsec:spatialApplication}Application of the method to a transfer between Titania and Oberon using spatial periodic orbits }
\begin{figure}
\hfill{}\centering%
\begin{minipage}[b][1\totalheight][t]{0.5\columnwidth}%
\subfigure[Unstable manifolds propagated from L$_2$ halo orbit with a $JC_d=3.0035$ towards Titania SoI in the U-T rotating frame.]{{\includegraphics[width=6.6cm]{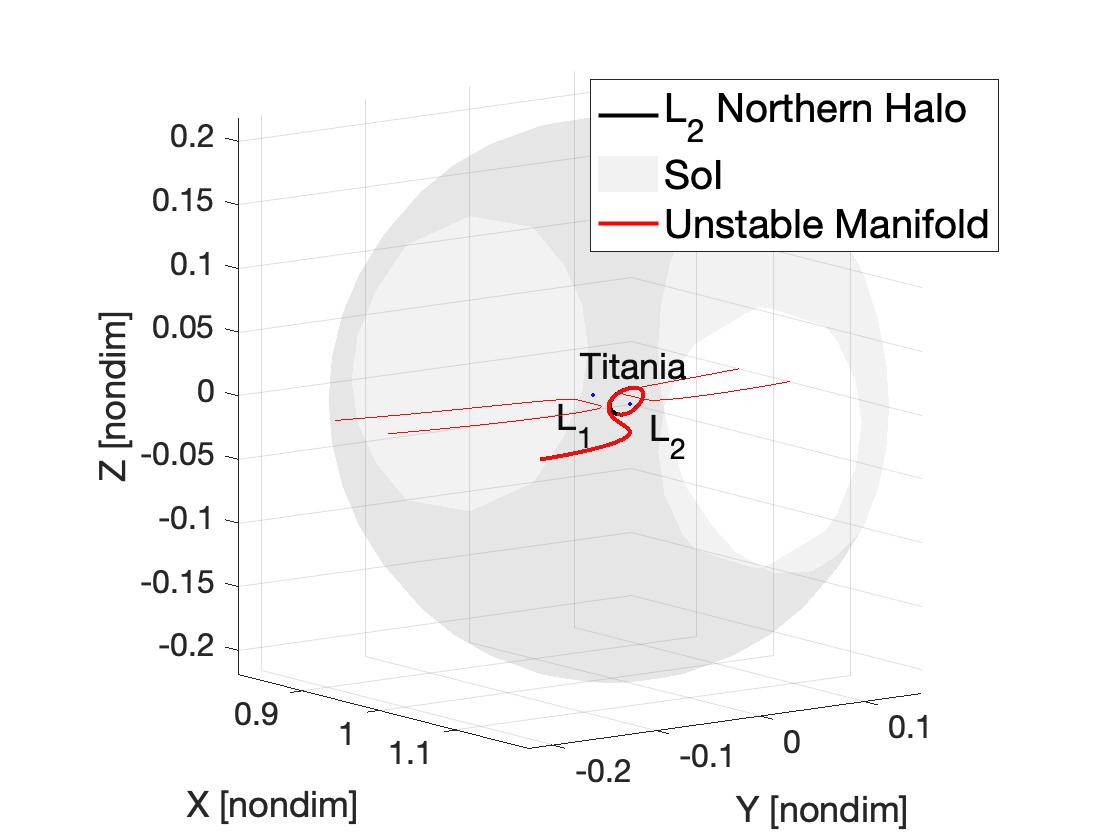}}}%
\end{minipage}\hfill{}%
\begin{minipage}[b][1\totalheight][t]{0.33\columnwidth}%
\subfigure[Stable manifolds propagated towards L$_1$ halo orbit with a $JC_a=3.003$ from Oberon SoI in the U-O rotating frame.]{{\includegraphics[width=6.6cm]{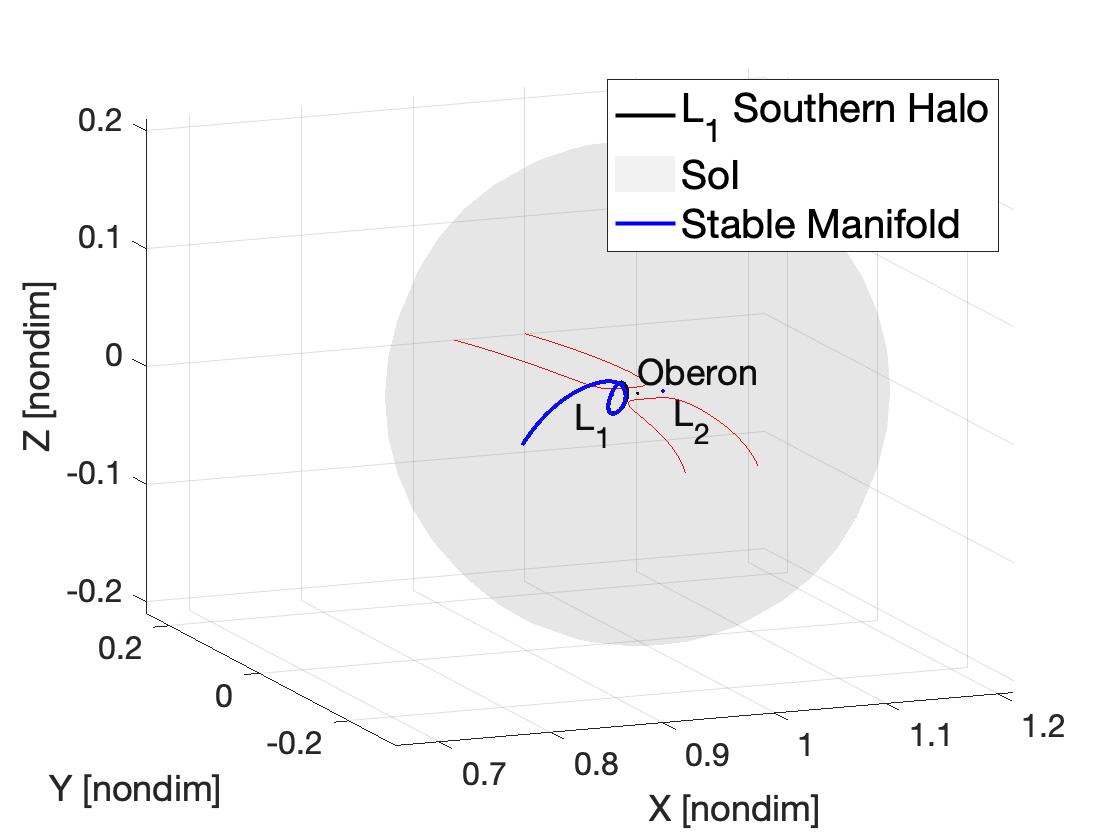}}}%
\end{minipage}\hfill{}
\caption{{\label{fig:manifoldIntersectionUTO}{Departure
manifolds from the Titania vicinity (left) and arrival manifolds to
the local Oberon region (right).}}}
\end{figure}

Consider a s/c located in an L$_2$ northern halo orbit in the Uranus-Titania
(U-T) system with a given Jacobi Constant value $JC_{d}=3.0035$,
and assume that the target location is an L$_1$ southern halo orbit ($JC_{a}=3.003$)
in the Uranus-Oberon (U-O) system (see Table \ref{Table:DataEuropaGanymede}
for system data). Unstable manifolds depart from the L$_2$ halo
orbit in the U-T system and, similarly, stable manifolds
arrive into the L$_1$ halo orbit in the U-O system. For explanatory purposes, assume that the s/c is departing an L$_2$ northern halo orbit in the Titania vicinity along a trajectory on the unstable manifold and, upon arrival, it flows into the L$_1$ southern halo orbit in the Oberon vicinity along a trajectory on the stable manifold. As illustrated in
Fig. \ref{fig:manifoldIntersectionUTO}, the unstable manifold trajectory
is propagated from the L$_2$ northern halo orbit in the U-T system
towards the Titania SoI (with a radius $R_{SoI_{Tit}}\approx 9.54\cdot10^4$ km), where it is evaluated as a departure conic. The
stable manifold trajectory associated with the L$_1$ southern halo orbit
in the U-O system, reaches the Oberon SoI in backwards time (with a radius $R_{SoI_{Obe}}\approx 1.23\cdot10^5$ km), where the state is transformed to an arrival
conic. Numerical algorithms to produce feasible
transfers between spatial orbits assuming coplanar moons are detailed by other authors, such as \citet{FantinoHaloTransfers}. In this investigation, the spatial strategy assumes that the moons are modeled in their
respective true orbital planes and, thus, the spatial transfer is derived directly. To do so,
the 2BP-CR3BP patched model delivers a useful initial guess
to pass to the coupled spatial CR3BP. 

\begin{figure}
\hfill{}\centering%
\begin{minipage}[b][1\totalheight][t]{0.5\columnwidth}%
\subfigure[$i_a$ evolution with respect to $\theta_{4_{Obe}}$.]{{\includegraphics[width=6.7cm]{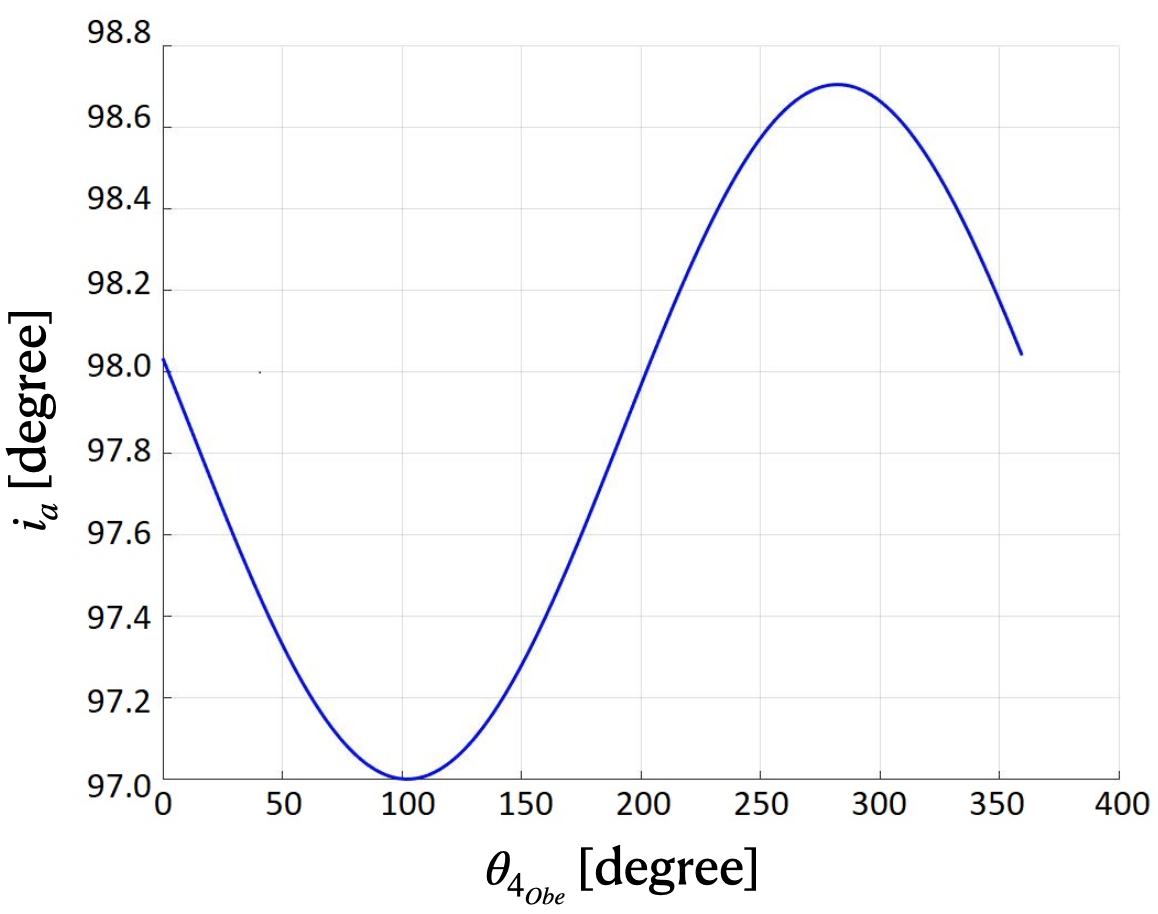}}}
\end{minipage}\hfill{}%
\begin{minipage}[b][1\totalheight][t]{0.33\columnwidth}%
\subfigure[$\Omega_a$ evolution with respect to $\theta_{4_{Obe}}$.]{{\includegraphics[width=6.7cm]{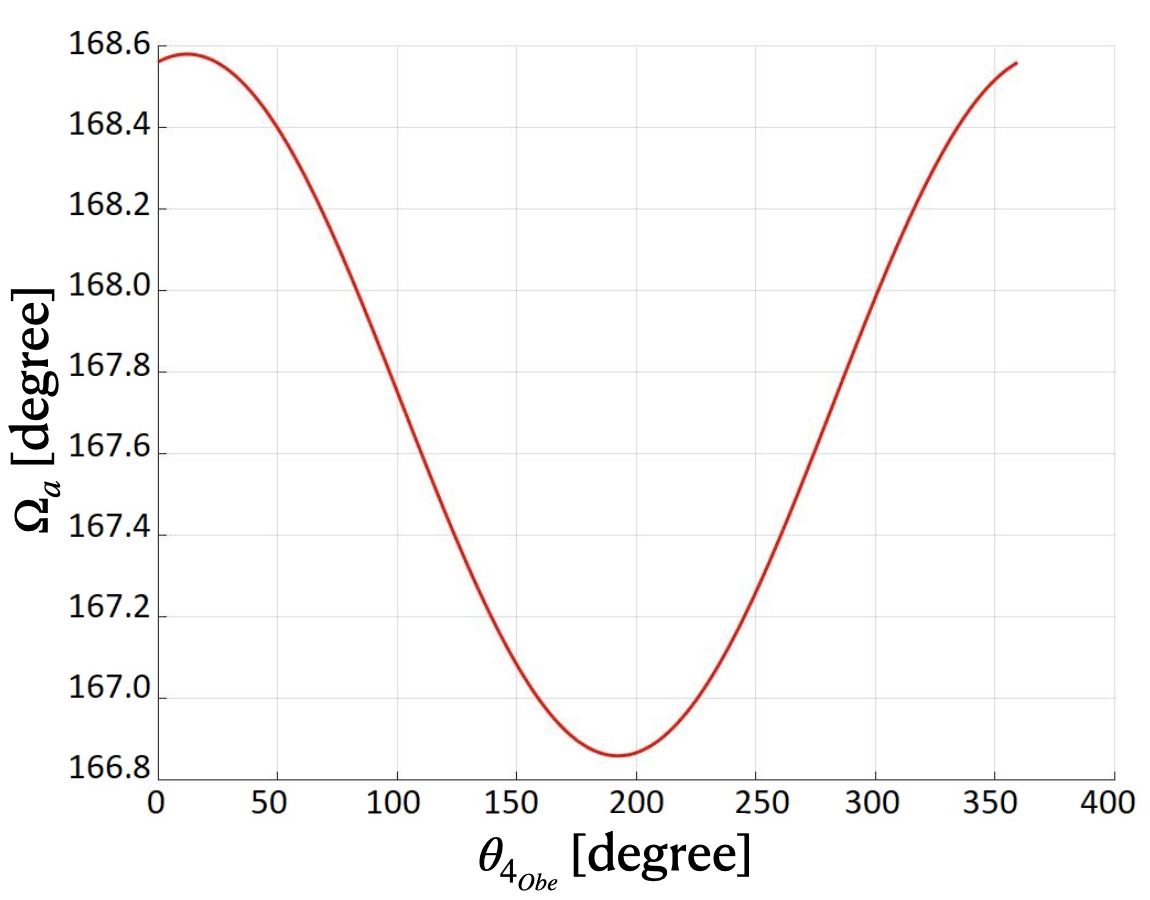}}}%
\end{minipage}\hfill{}
\caption{\label{fig:RAANIncEvolution}Evolution of $\Omega_{a}$ and $i_{a}$
depending on the arrival angle at Oberon. }
\end{figure}

\begin{figure}
\hfill{}\centering\includegraphics[width=14cm]{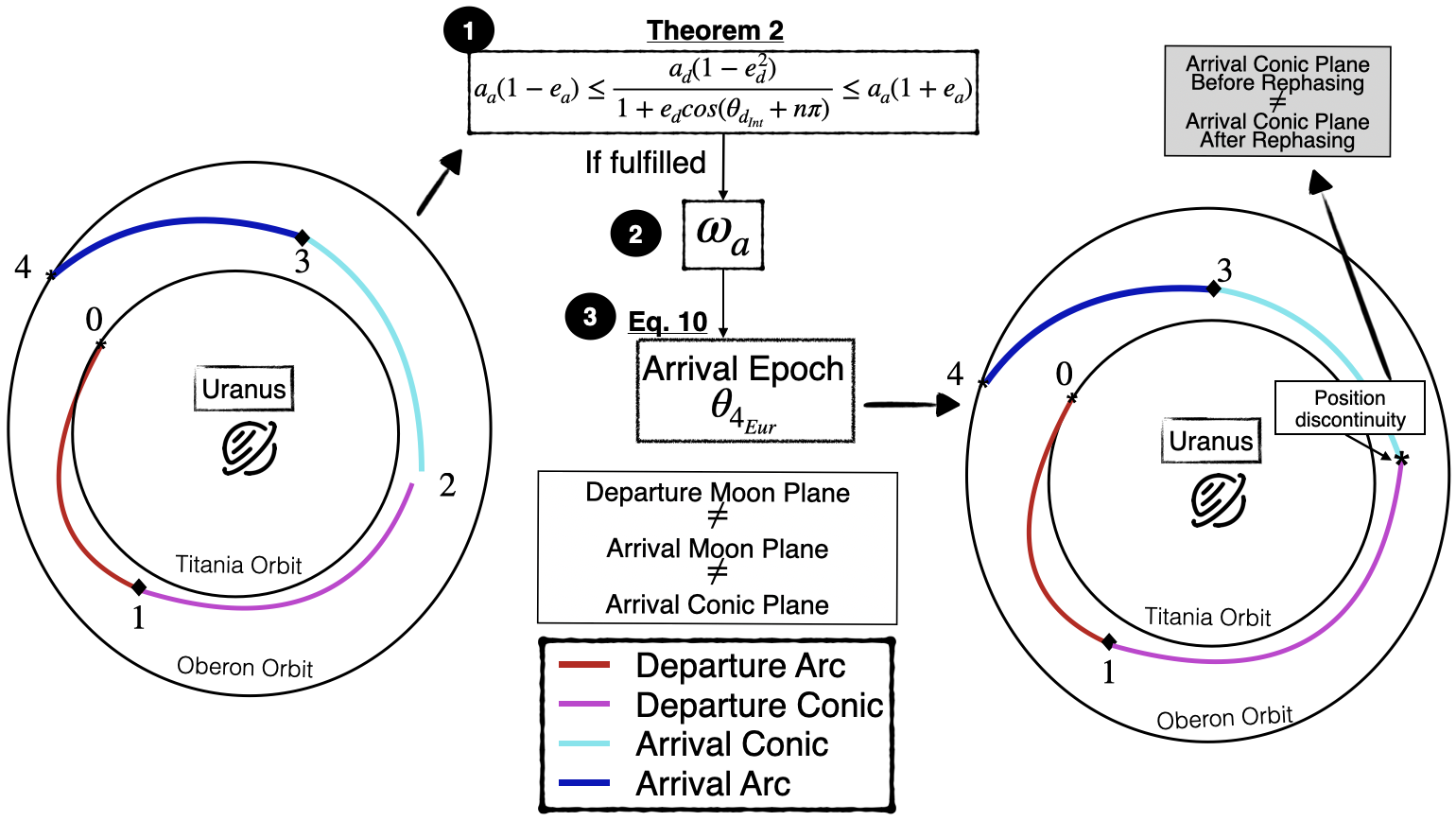}\hfill{}
\caption{\label{fig:reorientationGraph3d} Representation of the rephasing of
the arrival moon for transfers between spatial orbits of two different systems that fulfill Theorem \ref{theorem:spatial}, but that lead to position discontinuity due to a difference in the arrival plane before and after applying Eq. \eqref{eq:rephasing}.}
\end{figure}

\begin{figure}
\hfill{}\centering\includegraphics[width=6.5cm]{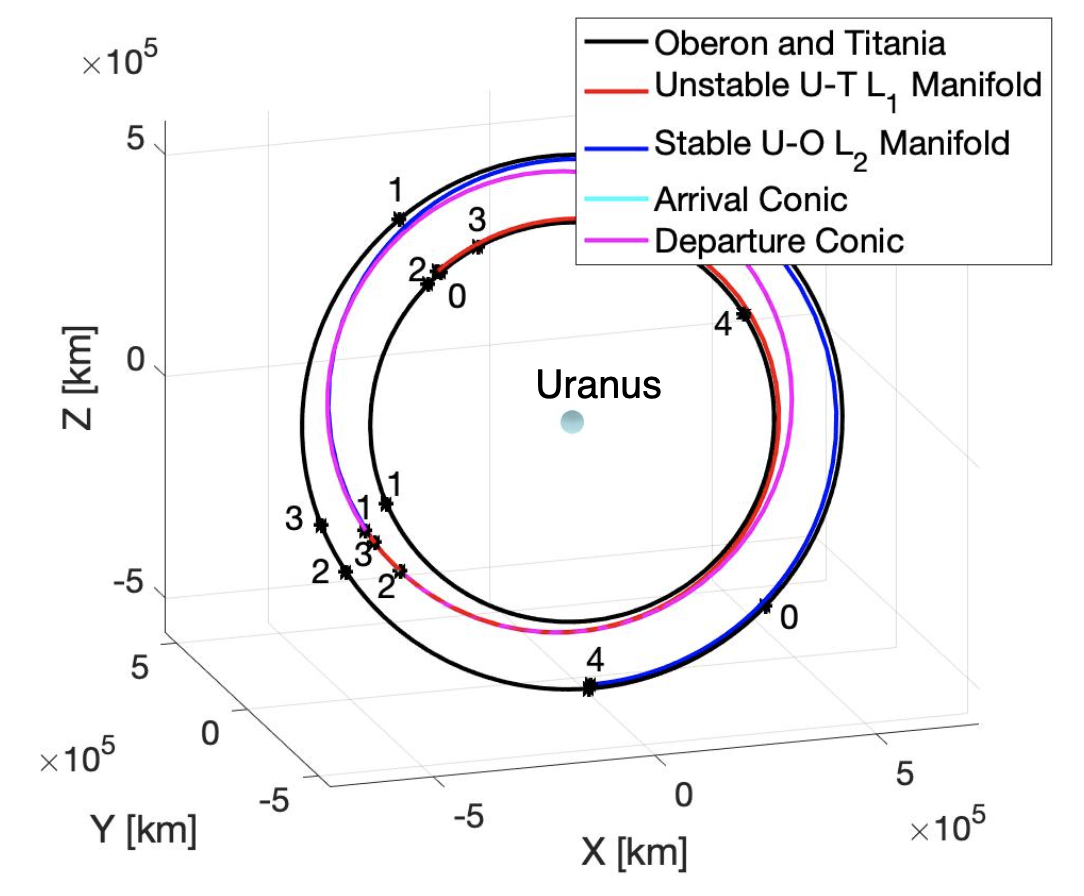}\hfill{}
\caption{\label{fig:mindVHaloOrbit}Transfer from northern halo L$_2$ in the Uranus-Titania
system to southern halo L$_1$ in the Uranus-Oberon system in the Ecliptic
J2000.0 Uranus-centered inertial frame (2BP-CR3BP patched model): $\Delta v_{tot}=62.6$ m/s and $t_{tot}=28.5$ days.}
\end{figure}

\begin{figure}
\hfill{}\centering%
\begin{minipage}[b][1\totalheight][t]{0.45\columnwidth}%
\subfigure[Ecliptic J2000.0 Uranus-centered inertial frame.]{\label{fig:SpatialTransfer}{\includegraphics[width=6cm]{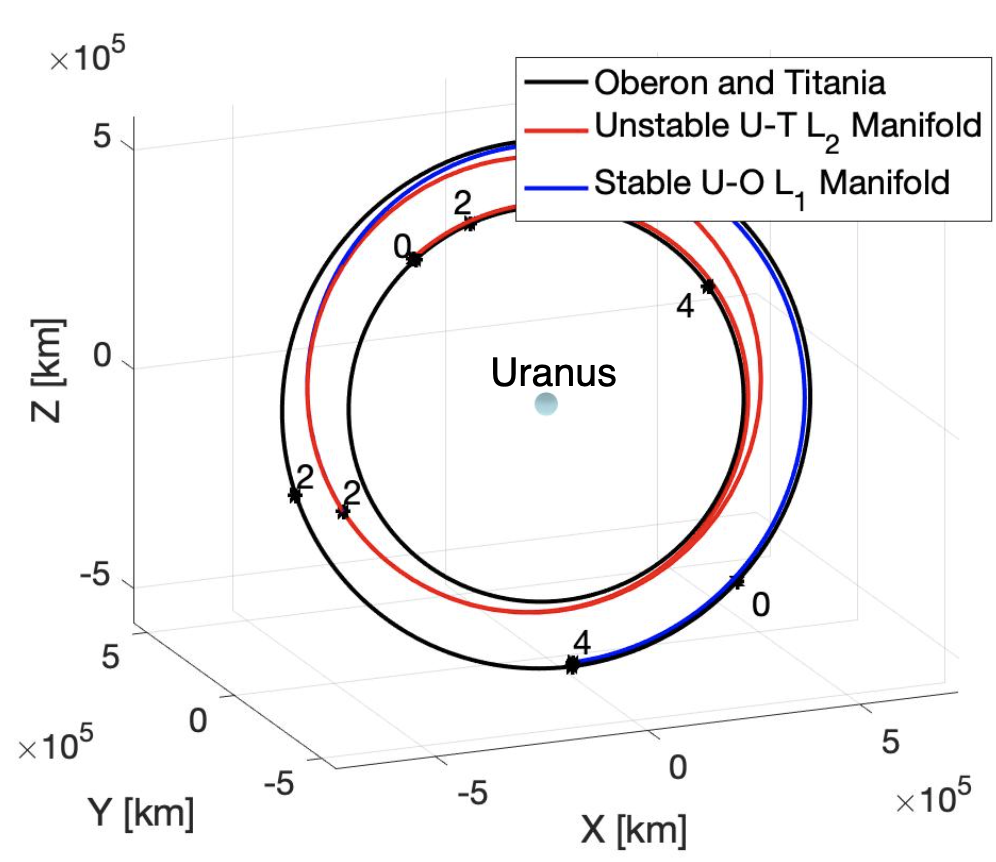}}}%
\end{minipage}\hfill{}%
\begin{minipage}[b][1\totalheight][t]{0.33\columnwidth}%
\subfigure[Uranus-Oberon rotating frame.]{\label{fig:spatialTransferCloser}{\includegraphics[width=8cm]{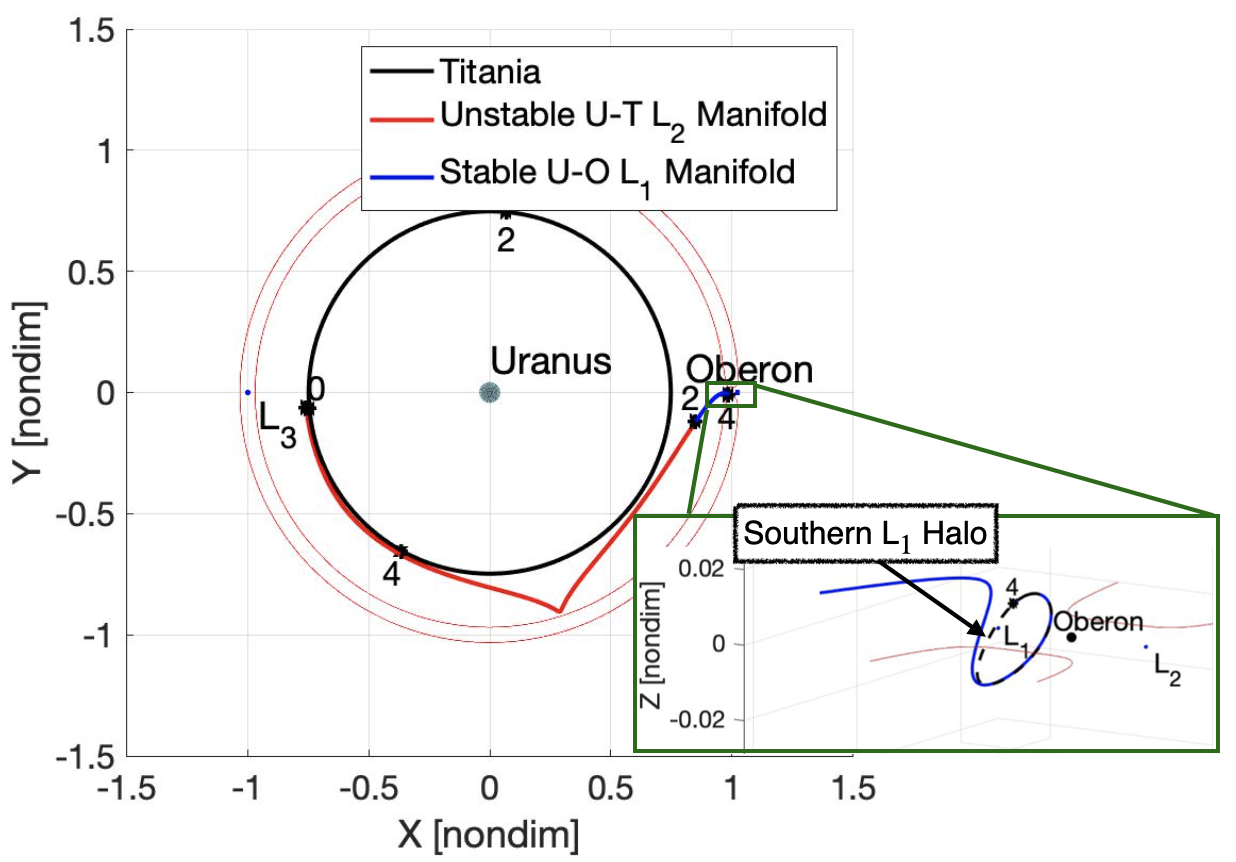}}}%
\end{minipage}\hfill{}
\caption{\label{fig:SpatialTransferResultUranus}Transfer from an L$_2$ northern halo orbit of the
Uranus-Titania system, to an L$_1$ southern halo orbit of the Uranus-Oberon
system (coupled spatial CR3BP): $\Delta v_{tot}=45.7$ m/s and $t_{tot}=28.44$ days.}
\end{figure}

Once the problem has been formulated, it is possible to directly observe potential intersections between both manifold trajectories given an
initial departure angle from Titania with respect to its ascending
node, $\theta_{0_{Tit}}$. If the given selected departure configuration leads to geometrical properties for the departure and arrival conics that meet the condition in Eq. \eqref{eq:3dConstraint}, a successful transfer occurs given an ideal
phase of Oberon at the arrival time ($\theta_{4_{Obe}}$). Nevertheless, observe in Fig. \ref{fig:RAANIncEvolution}
that, since the stable manifold trajectory crosses the 
 Oberon's SoI in 3-dimensions, $\Omega_{a}$ and $i_{a}$ for the arrival conic vary depending upon the selected arrival epoch, $\theta_{4_{Obe}}$;
of course, these angles are observably different from the values for the Oberon orbit, $\Omega_{O}$
and $i_{O}$. 
Clearly, when the rephasing technique is applied using Eq. \eqref{eq:rephasing}, the angle
$\sigma$ is defined in the plane of the conic and not in the plane of the arrival moon. Thus, when the new phase is applied, the departure and arrival conics
no longer intersect (see Fig. \ref{fig:reorientationGraph3d} for a schematic). 
Note that if we project the angle $\sigma$ onto the plane of the arrival moon, the position discontinuity between the departure and arrival conic is only of the order of $10^3$ km for this specific problem. But recall that if Eq. \eqref{eq:3dConstraint} is fulfilled,
an ideal rephasing for such a transfer to occur does exist given
a value of $\theta_{0_{Tit}}$. Therefore, $\sigma$ is projected onto the
plane of the arrival moon, offering a sufficiently close initial guess
to construct the intersection using  the differential corrections algorithm (Appendix \ref{appendix:spatialCorrections}). Since $\theta_{0_{Tit}}$ is fixed, the conic state departing the SoI is also fixed, but the propagation time at which it intersects the arrival conic plane, $t_{d}$, is now considered free. For the arrival trajectory, an initial guess for the angle $\theta_{4_{Obe}}$ is constructed using the projected value of $\sigma$ in Eq. \eqref{eq:rephasing}, which leads to values of $\Omega_{a}$ and $i_{a}$. The propagation time at which the arrival conic intersects the departure conic, $t_{a}$, is also assumed to be free. Given a sufficiently close initial guess and after applying the algorithm in Appendix \ref{appendix:spatialCorrections}, a converged solution is delivered in only a few iterations. Note that, in the case of transfers between planar periodic orbits, this step is not required, as explained in Fig. \ref{fig:MMATmethod}. Consequenly, for every $\theta_{0_{Tit}}$, a feasibility analysis similar to the one performed in the Ganymede to Europa application is produced (see Appendix \ref{appendix:FeasibilityAnalysisUranus}). Promising scenarios are, thus,
available for direct transfers, such as the minimum-$\Delta v_{tot}$ configuration represented in Fig. \ref{fig:mindVHaloOrbit}. Note that it requires about 7 days for the s/c to leave the Titania SoI, and 10.8 days measured from the Oberon SoI to arrive into the L$_1$ southern halo orbit in the Oberon vicinity. This solution is transferred to the coupled spatial CR3BP by using the initial guess obtained from the 2BP-CR3BP patched spatial model, as accomplished in the previous application. The converged solution in the coupled spatial CR3BP is produced
 in Fig. \ref{fig:SpatialTransferResultUranus}. Given this example, it is possible to observe that using the MMAT method, transfers between halo orbits near both Titania and Oberon are identified that occur in less than a month and using a total single $\Delta v_{tot}$ of less than 50 m/s. Given the reduced total $\Delta v_{tot}$ and $t_{tot}$ that are produced, new insights and possibilities emerge regarding the exploration of these two moons in the Uranian system. The MMAT method yields transfer solutions between spatial periodic orbits for two different planet-moon systems orbiting a common planet, given that Eq. \eqref{eq:3dConstraint} is fulfilled. Consequently, useful initial guesses are generated that are successfully transferred to a numerical model such as the coupled spatial CR3BP. 

\subsection{\label{subsec:epochDependence}Epoch dependence along revolutions of the departure and arrival conics}
\begin{figure}
\hfill{}\centering%
\begin{minipage}[b][1\totalheight][t]{0.45\columnwidth}%
\subfigure[Ecliptic J2000.0 Uranus-centered inertial frame.]{{\includegraphics[width=6cm]{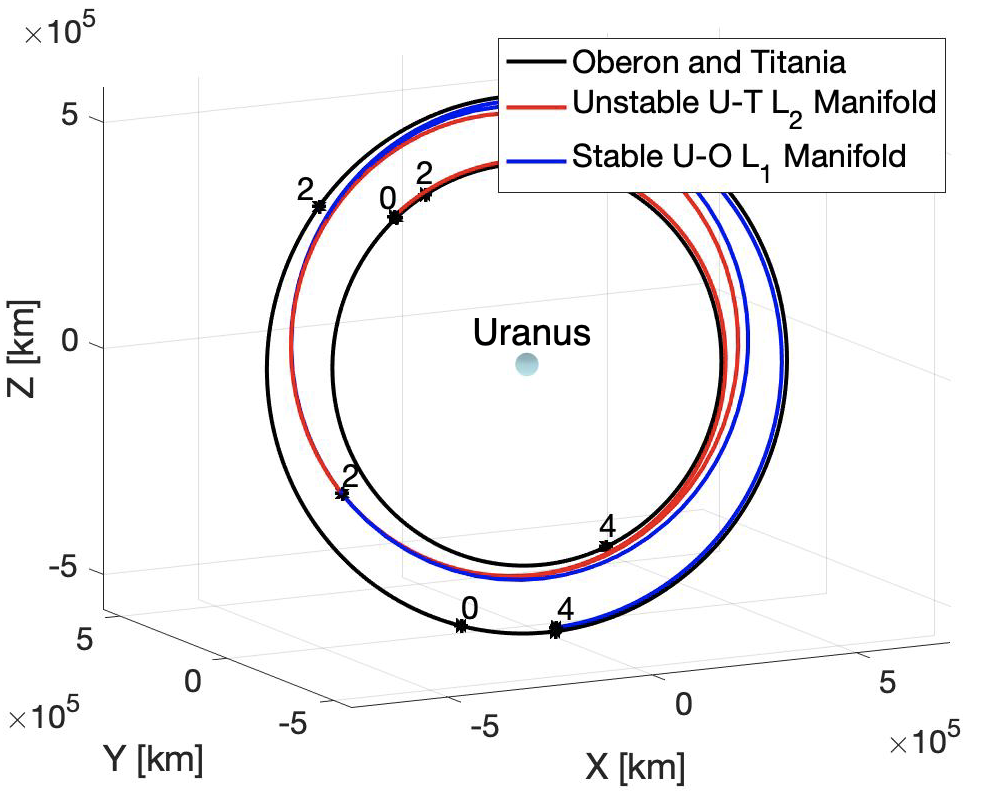}}}%
\end{minipage}\hfill{}%
\begin{minipage}[b][1\totalheight][t]{0.33\columnwidth}%
\subfigure[Uranus-Oberon rotating frame.]{{\includegraphics[width=7.5cm]{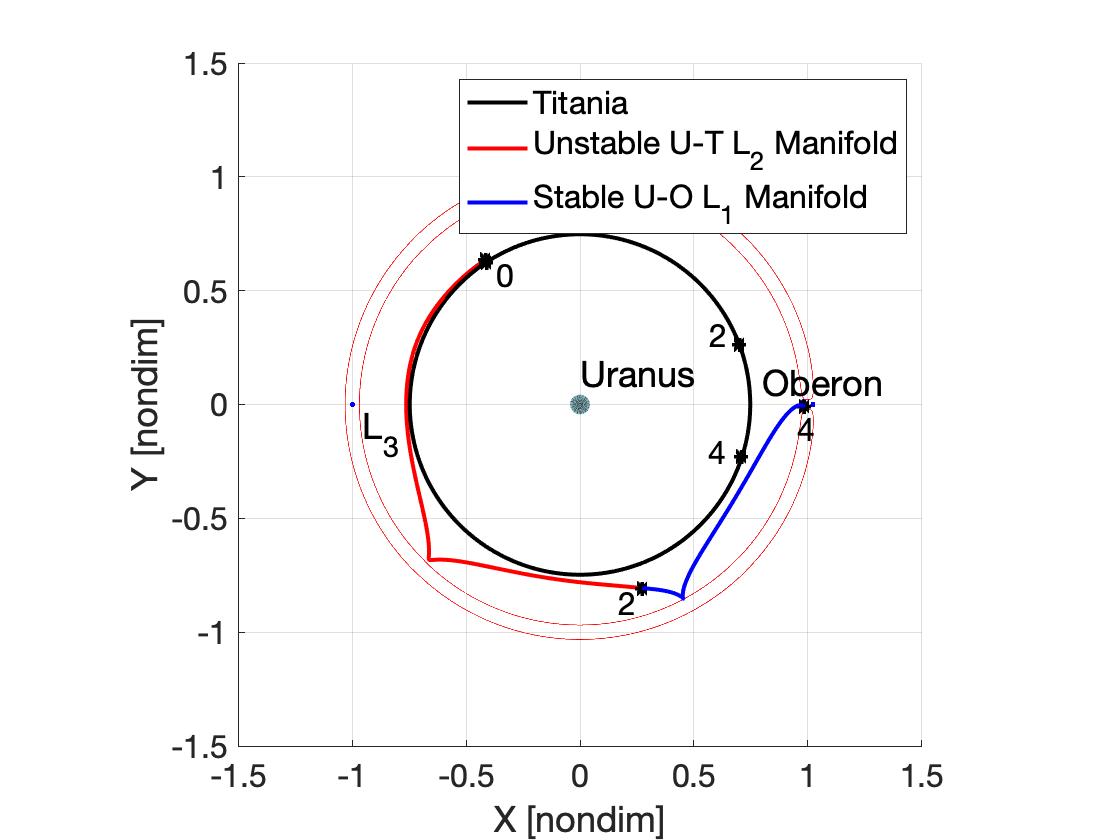}}}%
\end{minipage}\hfill{}
\caption{\label{fig:SpatialTransferResultExtraRev}Transfer from an L$_2$ northern halo orbit of the
Uranus-Titania system, to an L$_1$ southern halo orbit of the Uranus-Oberon
system, adding an extra revolution of the arrival conic (coupled spatial CR3BP): $\Delta v_{tot}=50$ m/s and $t_{tot}=39.6$ days.}
\end{figure}

Using the current methodology, it is possible to add extra revolutions to the departure and arrival conics to expand the search for feasible transfers between the moons over a wider range of epochs. For example, given the transfer in Sect. \ref{subsec:spatialApplication}, an extra revolution added to the arrival conic prior to entering Oberon's SoI delivers a new relative orientation between Oberon and Titania at departure, as represented in Fig. \ref{fig:SpatialTransferResultExtraRev}. This fact is achieved by maintaining nearly the same $\Delta v_{tot}$. Although the resulting time-of-flight increases by 10 days, incorporating such an option allows the exploration of new relative phases between the departure and arrival moons at similar $\Delta v_{tot}$ levels, thus, increasing the windows of opportunity for this direct transfer to occur. Furthermore, since the departure and arrival conics offer the same orbital properties and, thus, the same energy, adding extra revolutions along the arrival or departure conics does not significantly increase or decrease $\Delta v_{tot}$. As a result, the MMAT method adjusts to provide different epochs for a transfer between the vicinity of two moons given a departure and arrival arc, preserving $\Delta v_{tot}$ at the expense of adding time to the direct transfer. 

\subsection{\label{subsec:SoIDependence}Moon-to-moon transfer dependence on the definition of the sphere of influence}
\begin{figure}
\hfill{}\centering%
\begin{minipage}[b][1\totalheight][t]{0.4\columnwidth}%
\subfigure[$a_d$ dependence as a function of $a_{SoI}$.]{{\includegraphics[width=6cm]{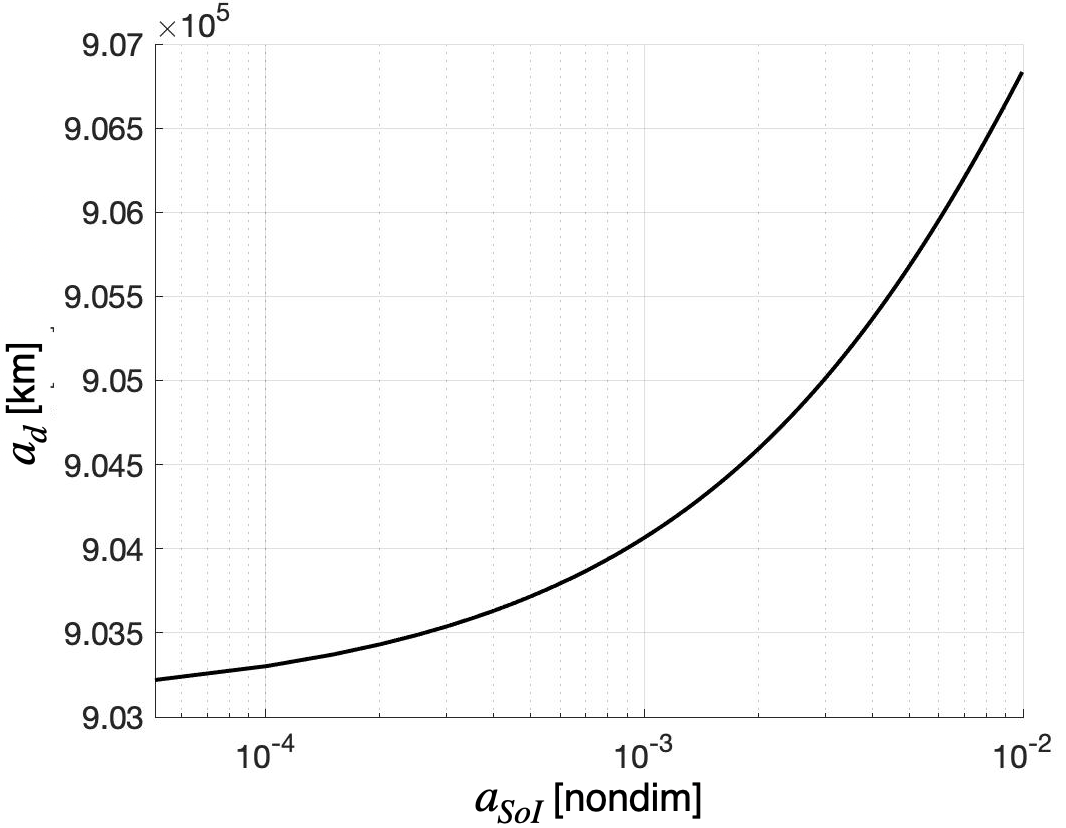}}}%
\end{minipage}\hfill{}%
\begin{minipage}[b][1\totalheight][t]{0.33\columnwidth}%
\subfigure[$e_d$ dependence as a function of $a_{SoI}$.]{{\includegraphics[width=6cm]{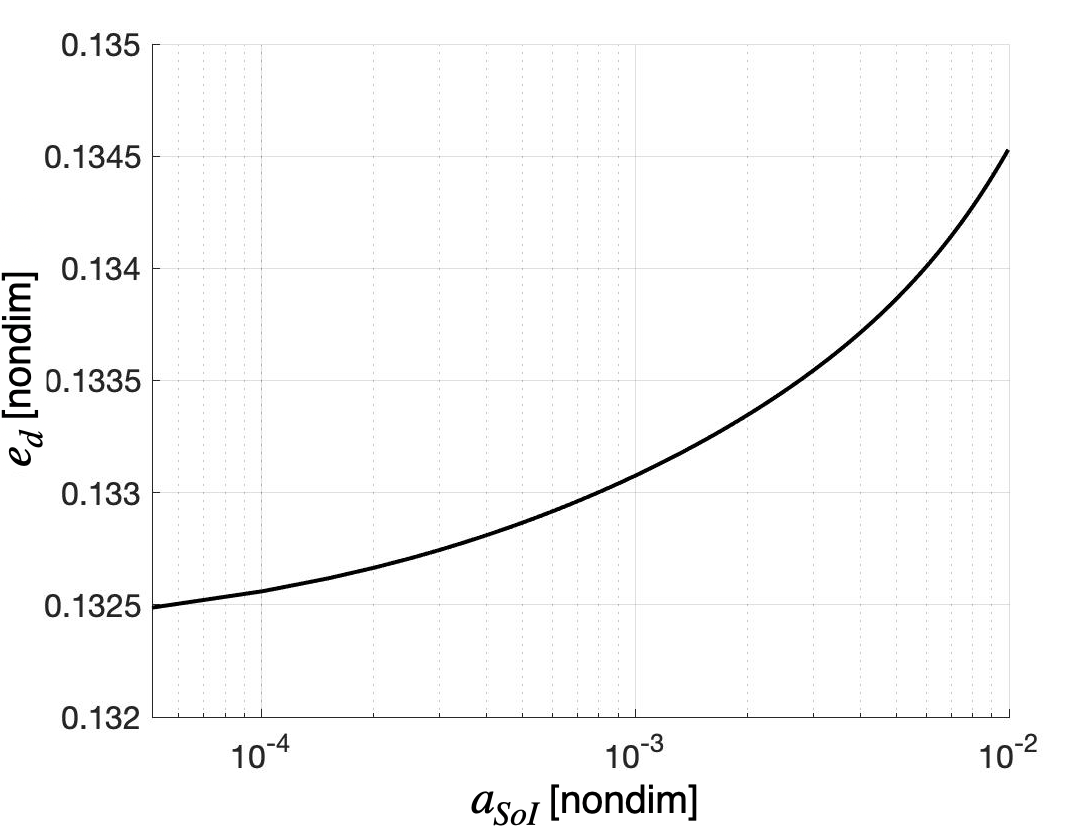}}}%
\end{minipage}\hfill{}
\hfill{}\centering%

\hfill{}\centering%
\begin{minipage}[b][1\totalheight][t]{0.4\columnwidth}%
\subfigure[$\omega_d$ dependence as a function of $a_{SoI}$.]{{\includegraphics[width=6cm]{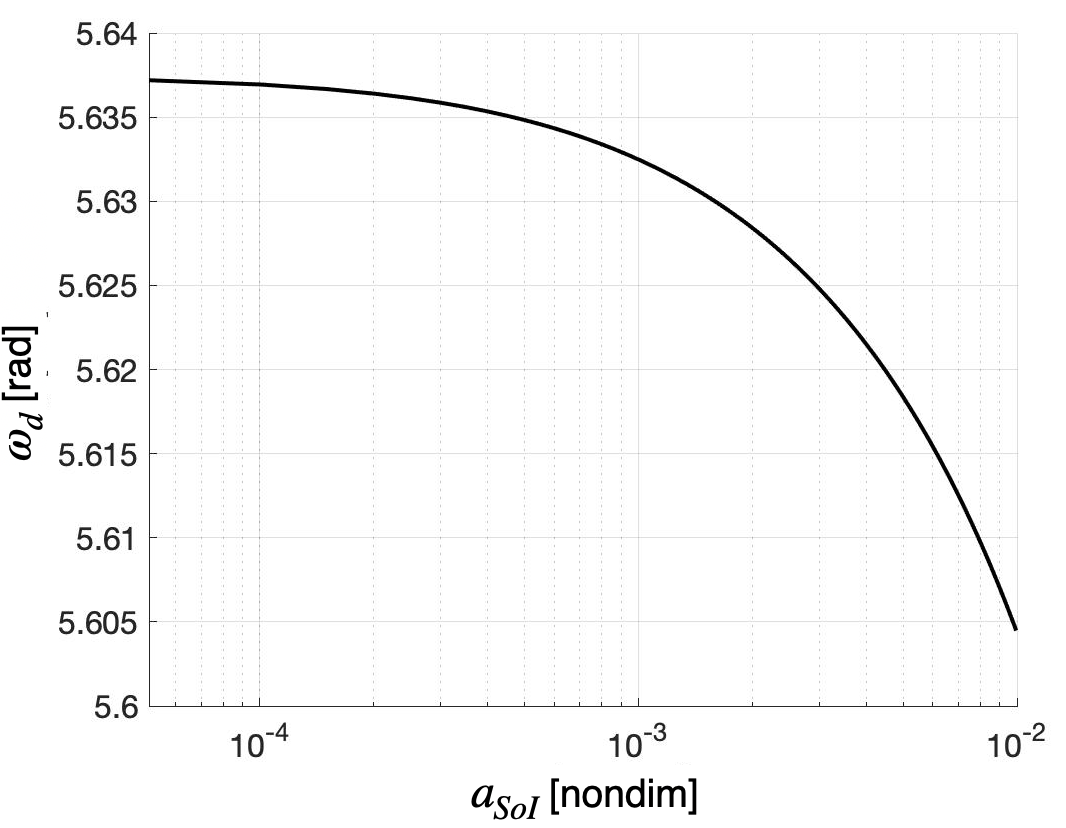}}}%
\end{minipage}\hfill{}%
\begin{minipage}[b][1\totalheight][t]{0.33\columnwidth}%
\subfigure[$\theta_{d_{SoI}}$ dependence as a function of $a_{SoI}$.]{{\includegraphics[width=6cm]{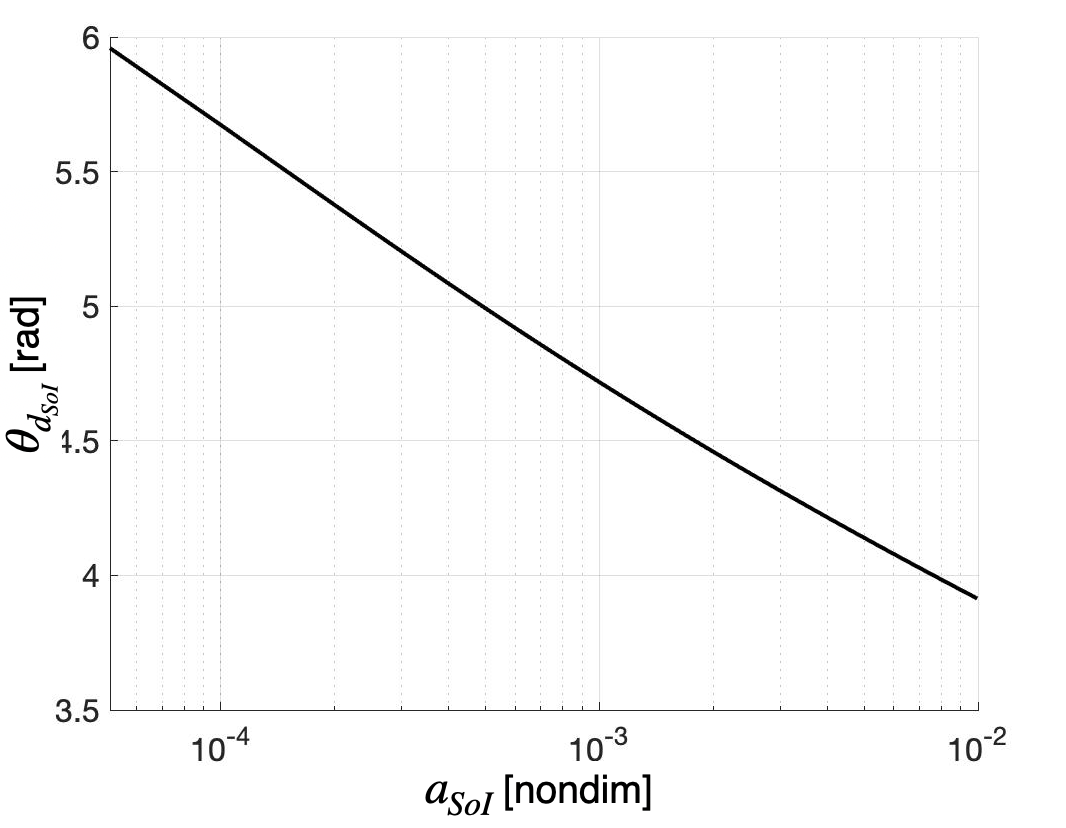}}}%
\end{minipage}\hfill{}
\hfill{}\centering%
\centering{}\caption{\label{fig:variationOrbitalElementsSoI}Variation of the instantaneous departure conic geometrical properties as a function of $a_{SoI}$ for the J-G L1 unstable manifold trajectory used in Sect. \ref{subsec:Ganymede-to-Europa}.}
\end{figure}

\begin{figure}
\hfill{}\centering%
\begin{minipage}[b][1\totalheight][t]{0.47\columnwidth}%
\subfigure[\label{deltaVDependence}$\Delta v_{tot}$ dependence as a function of Ganymede's $a_{SoI}$.]{{\includegraphics[width=6.1cm]{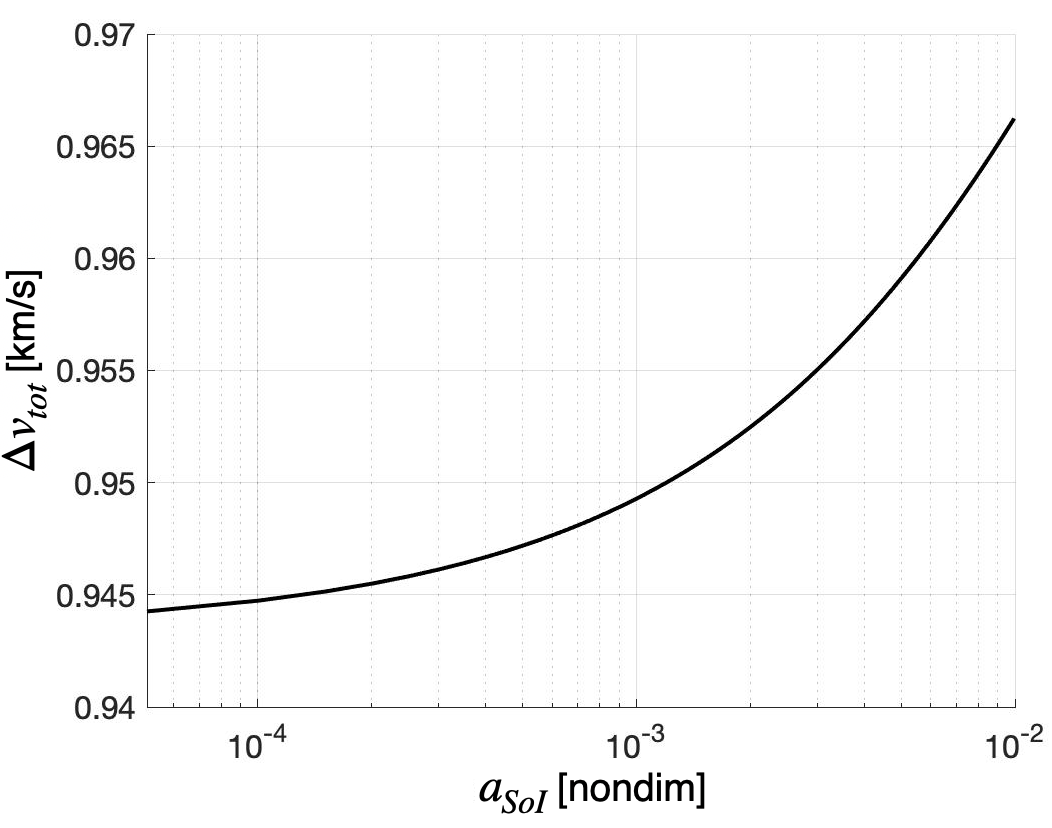}}}%
\end{minipage}\hfill{}%
\begin{minipage}[b][1\totalheight][t]{0.33\columnwidth}%
\subfigure[\label{ttotDependence}$t_{tot}$ dependence as a function of Ganymede's $a_{SoI}$.]{{\includegraphics[width=6.1cm]{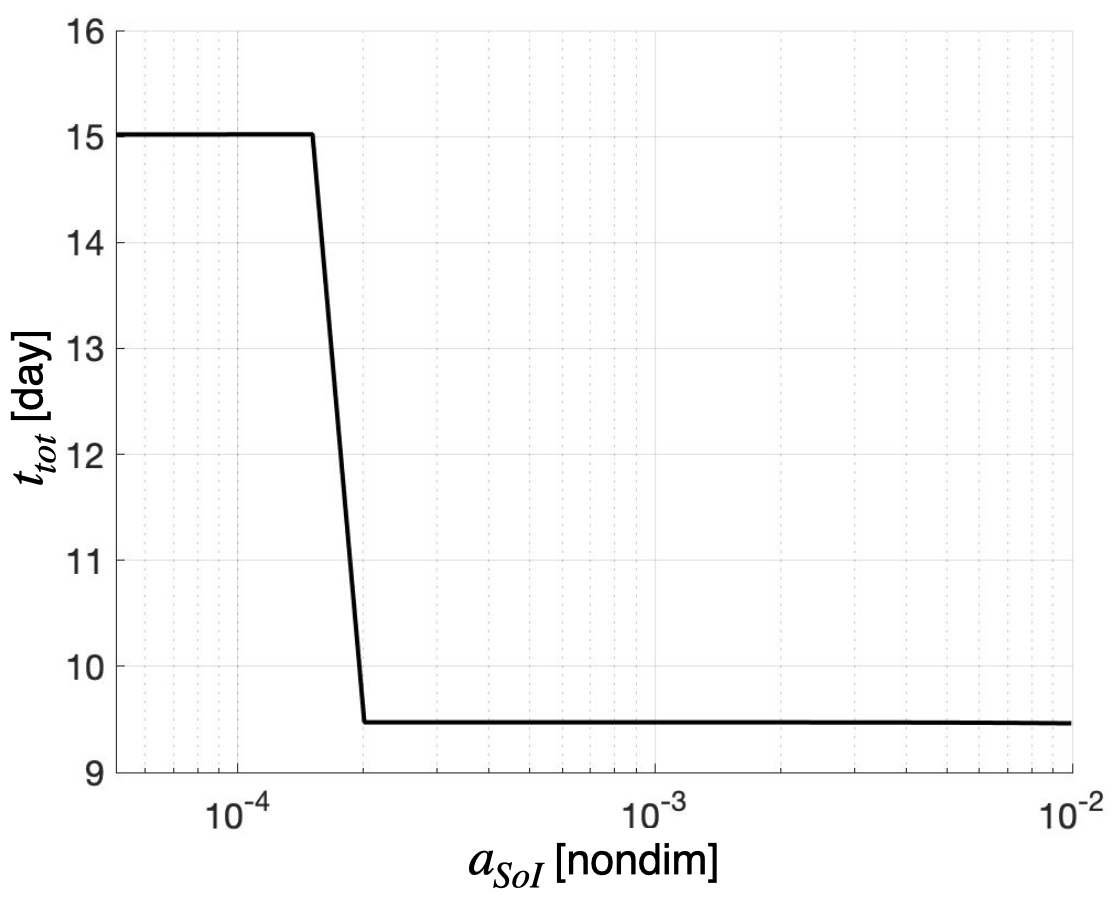}}}%
\end{minipage}\hfill{}
\hfill{}\centering%
\centering{}\caption{\label{fig:variationtransferSoI}$\Delta v_{tot}$ and $t_{tot}$ variation as a function of Ganymede's $a_{SoI}$ for the case transfer in  Sect. \ref{subsec:Ganymede-to-Europa}: L$_1$ Lyapunov in the J-G system towards an L$_2$ Lyapunov in the J-E system, assuming $a_{SoI}$ defining Europa's SoI is $a_{SoI}=5\cdot 10^{-4}$.}
\end{figure}

Arcs in the CR3BP are evaluated instantaneously as conics at the defined SoI for the departure or arrival moon. Transfers that are constructed using the MMAT method depend on the states as the CR3BP arcs cross the SoI, given that the departure and arrival conics originate at these instantaneous locations. But, as observed in Sect. \ref{subsec:2BP-CR3BP-patched-model}, the radius of the SoI is designed based on a pre-determined criterion, i.e., the ratio between the gravitational accelerations of the moon and the planet, $a_{SoI}$. For example, consider the unstable manifold trajectory departing the L$_1$ Lyapunov orbit in the Ganymede vicinity in Sect. \ref{subsec:Ganymede-to-Europa}. The variation in the instantaneous orbital elements at the SoI for the departure conic are plotted as a function of $a_{SoI}$ in Fig. \ref{fig:variationOrbitalElementsSoI}. Given that Lyapunov orbits evolve to manifold trajectories that are planar in the Jupiter-Ganymede plane, $\Omega_d$ and $i_d$ are constant. Recall that the smaller the $a_{SoI}$ ratio, the larger the distance $R_{SoI}$. Observe in Fig.  \ref{fig:variationOrbitalElementsSoI} that as $a_{SoI}$ decreases, $a_d$, $i_d$ and $\omega_d$ tend to an asymptote. This fact occurs because the acceleration of Ganymede is sufficiently small at the corresponding distance such that the motion is accurately described as a conic. Nonetheless, the true anomaly at which the conic intersects the departure SoI, $\theta_{d_{SoI}}$, does not follow the same pattern. Assume now that the sample transfer between Lyapunov orbits near Ganymede and Europa in Sect. \ref{subsec:Ganymede-to-Europa} is further explored. The same unstable and stable manifold trajectories, as well as the departure epoch $\theta_{0_{Gan}}$ that results in the minimum-$\Delta v_{tot}$ transfer in Fig. \ref{fig:minDvSpatialConic}, are now examined assuming that the Europa SoI is constant. However, $a_{SoI}$ for Ganymede is varied from $5\cdot10^{-5}$ to $1\cdot10^{-2}$. Observe in Fig. \ref{deltaVDependence} that $\Delta v_{tot}$ tends to an asymptotic value when $a_{SoI}$ decreases in a manner similar to $a_d$, $i_d$ and $\omega_d$. Therefore, the cost for the transfer strictly depends upon $a_d$, $i_d$ and $\omega_d$, since the energy required to shift from the departure to the arrival conic also tends to an asymptotic value. Nevertheless, given that $\theta_{d_{SoI}}$ does not tend to an asymptote, a value for the SoI radius that is too large may miss a possible crossing with the arrival conic and require an extra revolution, leading to an increase in the $t_{tot}$ as observed in Fig. \ref{ttotDependence}. Nonetheless, despite the increased transfer time, $t_{tot}$, $\Delta v_{tot}$ for this maneuver does not change significantly. 

\begin{figure}
\hfill{}\centering%
\begin{minipage}[b][1\totalheight][t]{0.47\columnwidth}%
\subfigure[Transfer for Titania's $a_{SoI}=5\cdot 10^{-4}$. \newline$\Delta v_{tot}=45.7$ m/s and $t_{tot}=28.44$ days.]{\includegraphics[width=6.4cm]{Fig32a_inertialFrameUranus}}%
\end{minipage}\hfill{}%
\begin{minipage}[b][1\totalheight][t]{0.33\columnwidth}%
\subfigure[Transfer for Titania's $a_{SoI}=5\cdot 10^{-3}$. \newline$\Delta v_{tot}=66.7$ m/s and $t_{tot}=	17.72$ days.]{\includegraphics[width=6.4cm]{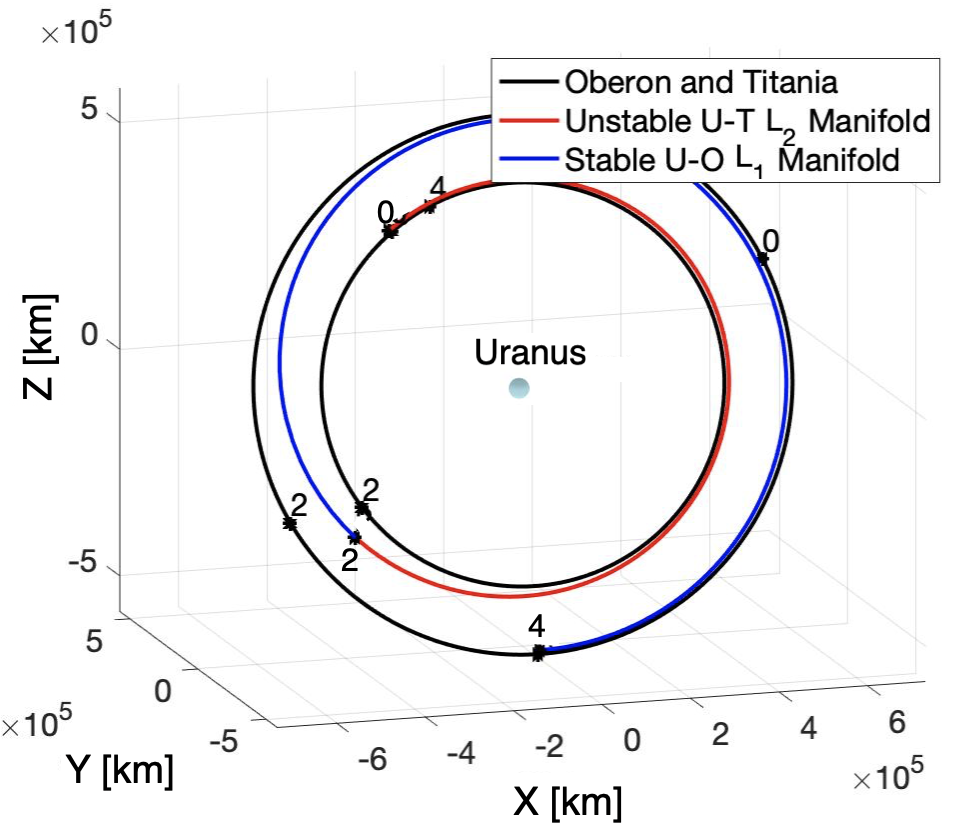}}%
\end{minipage}\hfill{}
\caption{\label{fig:soiDependenceHaloOrbitsUranus}Transfer from an L$_2$ northern halo orbit of the Uranus-Titania system to an L$_1$ southern halo orbit of the Uranus-Oberon
system. Representations in the Ecliptic J2000.0 Uranus-centered inertial frame. Model: Coupled spatial CR3BP.}
\end{figure}

The ratio value $a_{SoI}=5\cdot10^{-4}$ is selected as the threshold in the Jovian system since it delivers a sufficiently low moon gravitational acceleration for the motion to be simplified as a planet-centered conic. Nevertheless, the same ratio is also employed for the transfers between halo orbits in the vicinity of Titania and Oberon in the Uranian system. As a consequence, one feasible transfer that reduces $t_{tot}$ is, in fact, overlooked. If the $a_{SoI}$ for Titania is increased, then a shorter time-of-flight solution with similar $\Delta v_{tot}$ is obtained, as reflected in Fig. \ref{fig:soiDependenceHaloOrbitsUranus}. Consequently, from further investigation, it is apparent that the appropriate selection of the $a_{SoI}$ value depends on the system under study. If such a ratio is increased, the moon perturbation increases and the approximation
may not be valid, which leads to $\Delta v_{tot}$'s and times-of-flight that do not easily transition to the coupled spatial CR3BP. If the ratio is decreased,
links between the departure and arrival trajectories
might be missed and $t_{tot}$ increases. Also, it is demonstrated in Fig. \ref{fig:soiDependenceHaloOrbitsUranus} that it is possible to locate transfers between halo orbits near both Titania and Oberon that require about 2 weeks with a $\Delta v_{tot}$ as little as 66.7 m/s. Both transfers are also compared in Table \ref{Table:resultsMMATUranus}.

 \begin{table}[htbp!]
    \caption{Summary of the resulting transfers between halo orbits of the Titania and Oberon vicinities depending on Titania's $a_{SoI}$ (coupled spatial CR3BP).}
   \label{Table:resultsMMATUranus}
        \centering 
        \begin{tabular}{clclc}
\hline\noalign{\smallskip}
&  $\Delta v_{tot}$  [m/s]&$t_{tot}$ [days] \\
\noalign{\smallskip}\hline\noalign{\smallskip}
	$a_{SoI}=5\cdot10^{-4}$ for Titania and Oberon & 45.7& 28.44 \\
\noalign{\smallskip}\hline
	Titania $a_{SoI}=5\cdot10^{-3}$; Oberon $a_{SoI}=5\cdot10^{-4}$& 66.7& 17.72 \\
\noalign{\smallskip}\hline
\end{tabular}
\end{table}

\section{\label{sec:ephemerisValidation}Transition to a higher-fidelity ephemeris model}
Once a promising moon-to-moon transfer is produced in the coupled spatial CR3BP problem using the MMAT method, such transfers are transitioned to a higher-fidelity ephemeris model. The central body and perturbing bodies that are incorporated depend on the planetary system. Note that for representation of the transfers in the higher-fidelity ephemeris model, two new instants are defined: $t_{0\ \text{PO}}$ corresponds to the initial time along the departure periodic orbit, and $t_f$ represents the final time for the arrival of the s/c into the destination orbit. Full position and velocity state continuity are ensured throughout the transfer. However, three $\Delta v$'s are allowed across the entire transfer at: (1) instant 0 to depart the departure periodic orbit; (2) instant 2 to transfer from the departure arc towards the arrival arc; (3) instant 4 to ensure that the s/c arrives on the specified periodic orbit. 

\begin{figure}[h!]
\hfill{}\centering%
\begin{minipage}[b][1\totalheight][t]{0.55\columnwidth}%
\subfigure[Jupiter-Ganymede rotating frame.]{\label{fig:ephemerisGanymede}{\includegraphics[width=7cm]{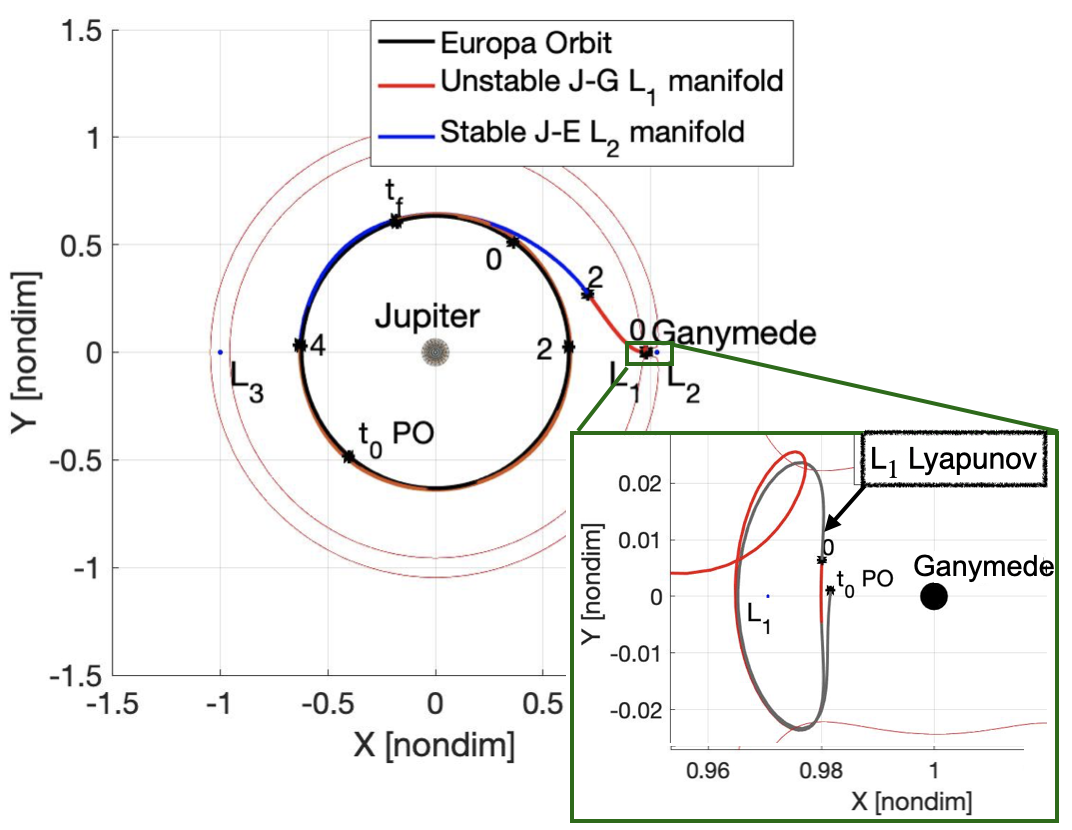}}}%
\end{minipage}\hfill{}%
\begin{minipage}[b][1\totalheight][t]{0.4\columnwidth}%
\subfigure[Jupiter-Europa rotating frame.]{\label{fig:ephemerisEuropa}{\includegraphics[width=7cm]{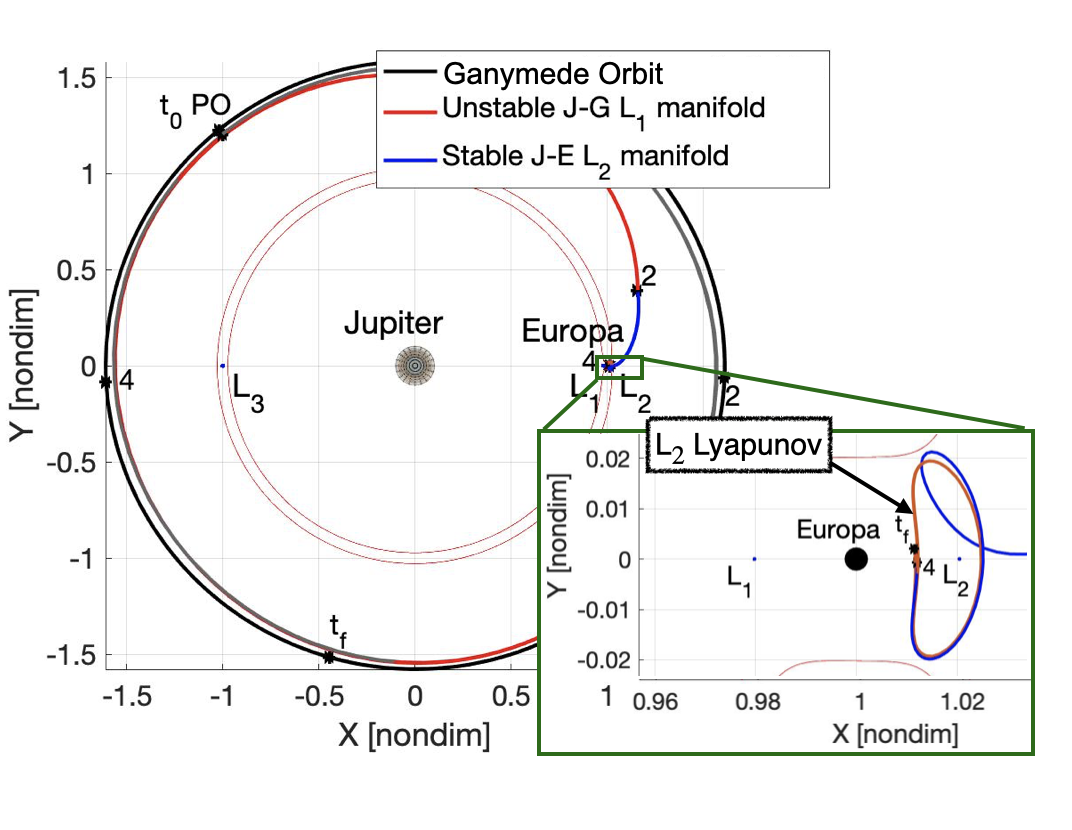}}}%
\end{minipage}\hfill{}
\caption{\label{fig:EphemerisJupiter}Transfer from an L$_1$ Lyapunov orbit of the
Jupiter-Ganymede system to an L$_2$ Lyapunov orbit of the Jupiter-Europa
system: $\Delta v_{tot}=0.966$ km/s, $t_{tot}=9.47$ days, higher-fidelity ephemeris model at epoch  JD2459323.365 (April 18th, 2021).}
\end{figure}

\begin{figure}[h!]
\hfill{}\centering%
\begin{minipage}[b][1\totalheight][t]{0.5\columnwidth}%
\subfigure[Coupled spatial CR3BP: $\Delta v_{tot}=1.08$ km/s and $t_{tot}=12.13$ days.]{\label{fig:coupledJupiterLongerTOF}{\includegraphics[width=6.4cm]{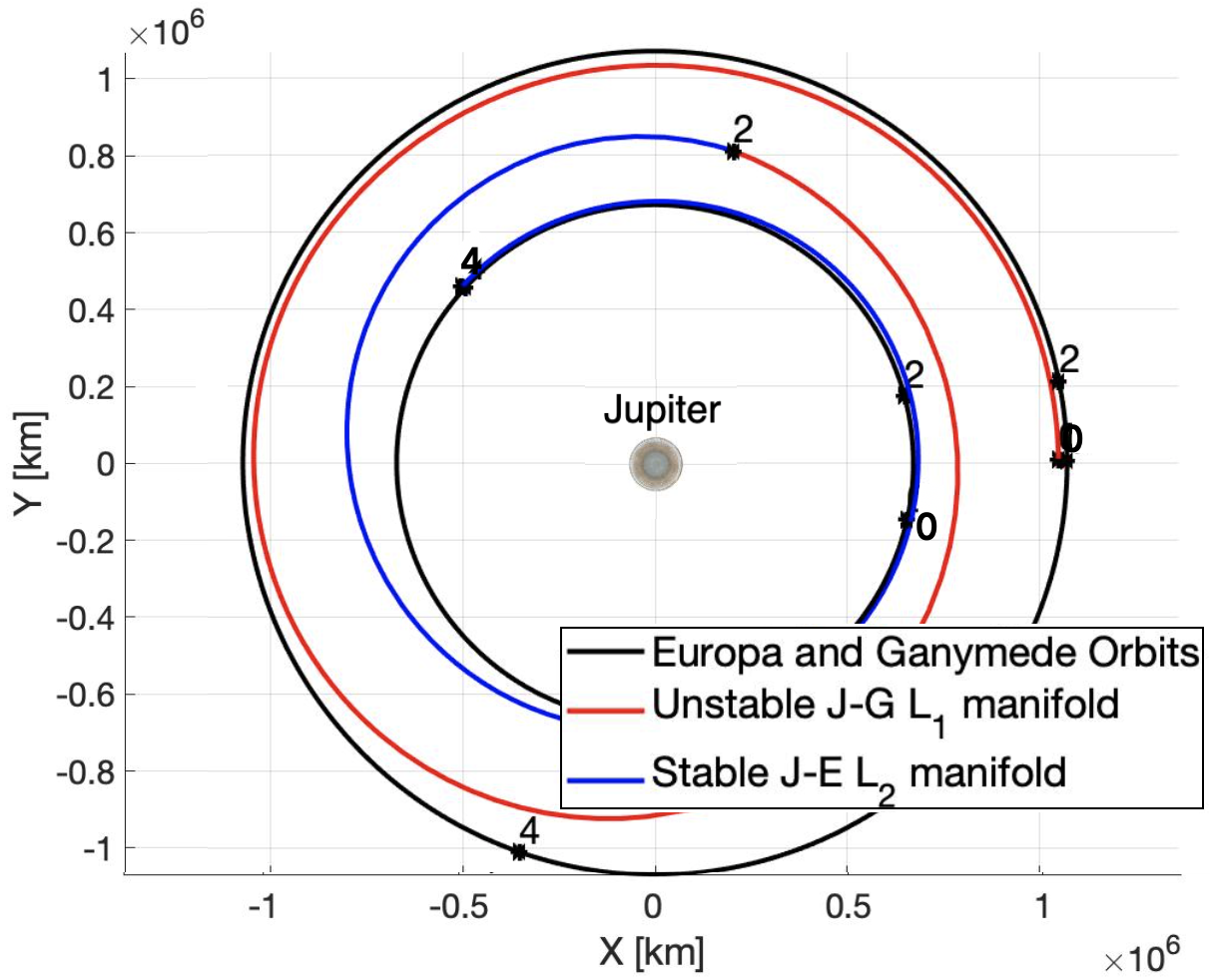}}}%
\end{minipage}\hfill{}%
\begin{minipage}[b][1\totalheight][t]{0.4\columnwidth}%
\subfigure[Higher-fidelity ephemeris model at epoch JD2459138.765  (April 16th, 2021): $\Delta v_{tot}=1.02$ km/s and $t_{tot}=12.13$ days.]{\label{fig:EphemerisJupiterLongerTOF}{\includegraphics[width=6.7cm]{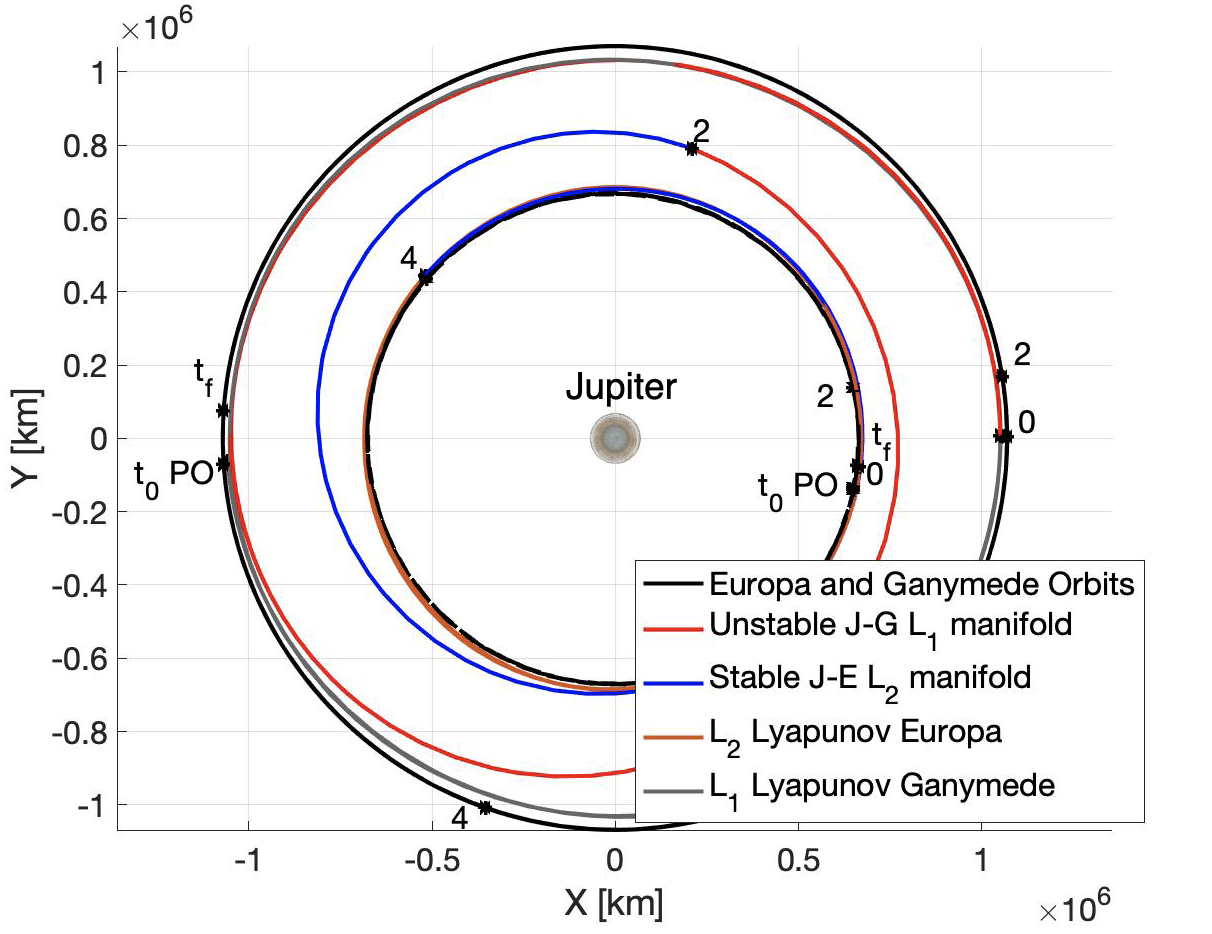}}}%
\end{minipage}\hfill{}
\caption{\label{fig:JupiterLongerTOF}Transfer from an L$_1$ Lyapunov orbit of the
Jupiter-Ganymede system to an L$_2$ Lyapunov orbit of the Jupiter-Europa
system (Ecliptic J2000.0 Jupiter-centered inertial frame).}
\end{figure}

\begin{figure}[h!]
\hfill{}\centering\includegraphics[width=13cm]{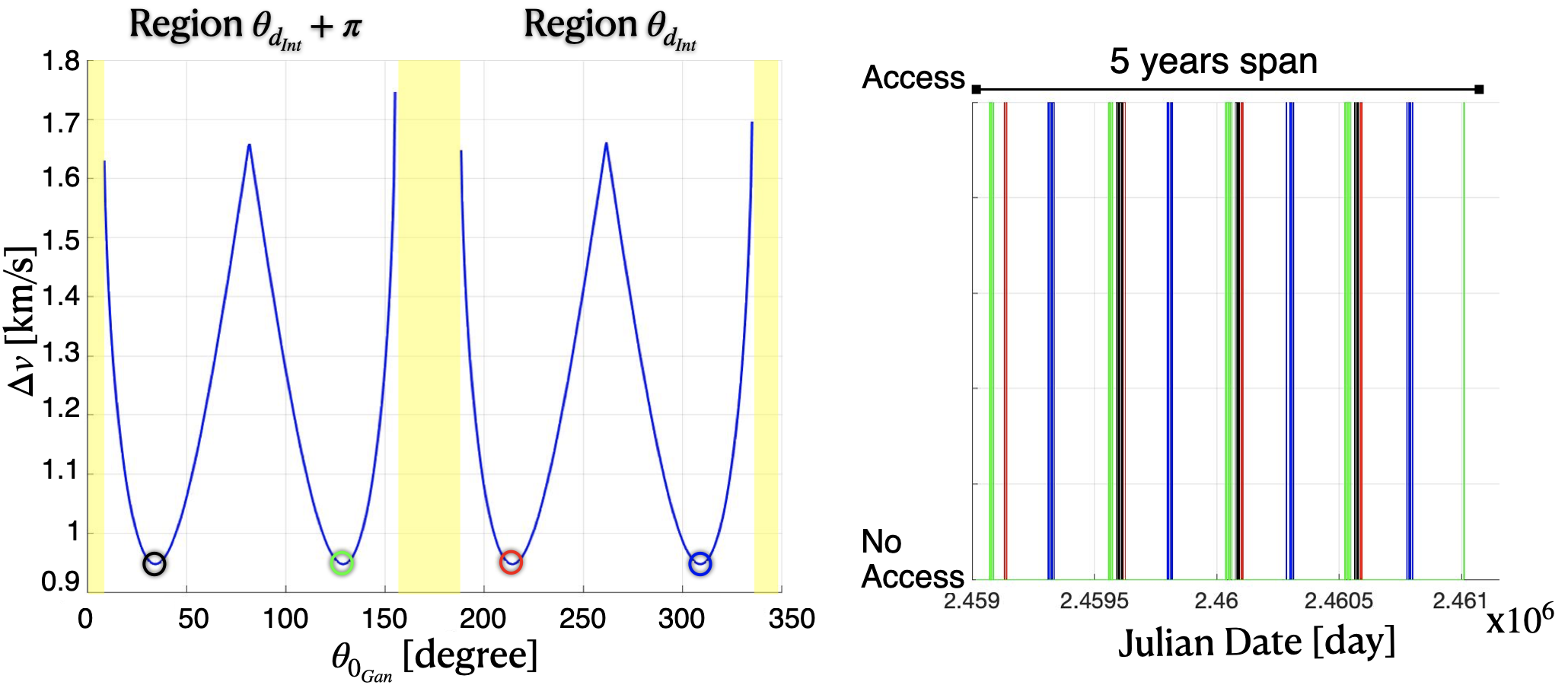}\hfill{}
\caption{\label{fig:epochRecurrence}Event recurrence of the minimum-$\Delta v_{tot}$ configurations over the span of 5 years. }
\end{figure}

The aim of this example is to transition the coupled spatial CR3BP transfer for the Ganymede to Europa transfer (Fig. \ref{fig:SpatialCoupledCR3BP-1}) to a higher-fidelity ephemeris model. Given the analytical minimum-$\Delta v$ transfer configuration from Fig. \ref{fig:thetaGanVariation}, it is possible to produce the relative phase for Europa in its respective plane given the departure epoch from Ganymede, $\theta_{0_{Gan}}$, as apparent in Fig. \ref{fig:europaVariation}, at instant 0. Note that this particular system is defined with Jupiter as the central body, and the Sun, Io, Europa, Ganymede, Callisto and Saturn as the perturbing bodies. Given the coupled spatial CR3BP transfer as an initial guess, the resulting transfer is plotted in Fig. \ref{fig:EphemerisJupiter} in the higher-fidelity ephemeris model. The total time-of-flight is preserved between the two models, and the $\Delta v_{tot}$ does not significantly increase. Notably, the geometric characteristics are generally preserved (Table \ref{tab:ephemerisTable}). Additionally, it is observed that, if a departure epoch that leads to the maximum time-of-flight is selected from Fig. \ref{fig:timeOfFLightVariation}, the solution in the coupled spatial CR3BP (Fig. \ref{fig:coupledJupiterLongerTOF}) also transitions to the higher-fidelity ephemeris result (Fig. \ref{fig:EphemerisJupiterLongerTOF}). Thus, the feasibility analysis using the MMAT method in Fig. \ref{fig:thetaGanVariation} straightforwardly transitions to the higher-fidelity ephemeris solution. Given Fig. \ref{fig:europaVariation}, it is possible to estimate the occurrence of certain configurations that render direct transfers in this sample scenario. A representation of minimum-$\Delta v_{tot}$ configurations that recur over the span of 5 years appears in Fig. \ref{fig:epochRecurrence}.

\begin{figure}[h!]
\hfill{}\centering%
\begin{minipage}[b][1\totalheight][t]{0.55\columnwidth}%
\subfigure[Uranus-Titania rotating frame.]{\label{fig:ephemerisTitania}{\includegraphics[width=7cm]{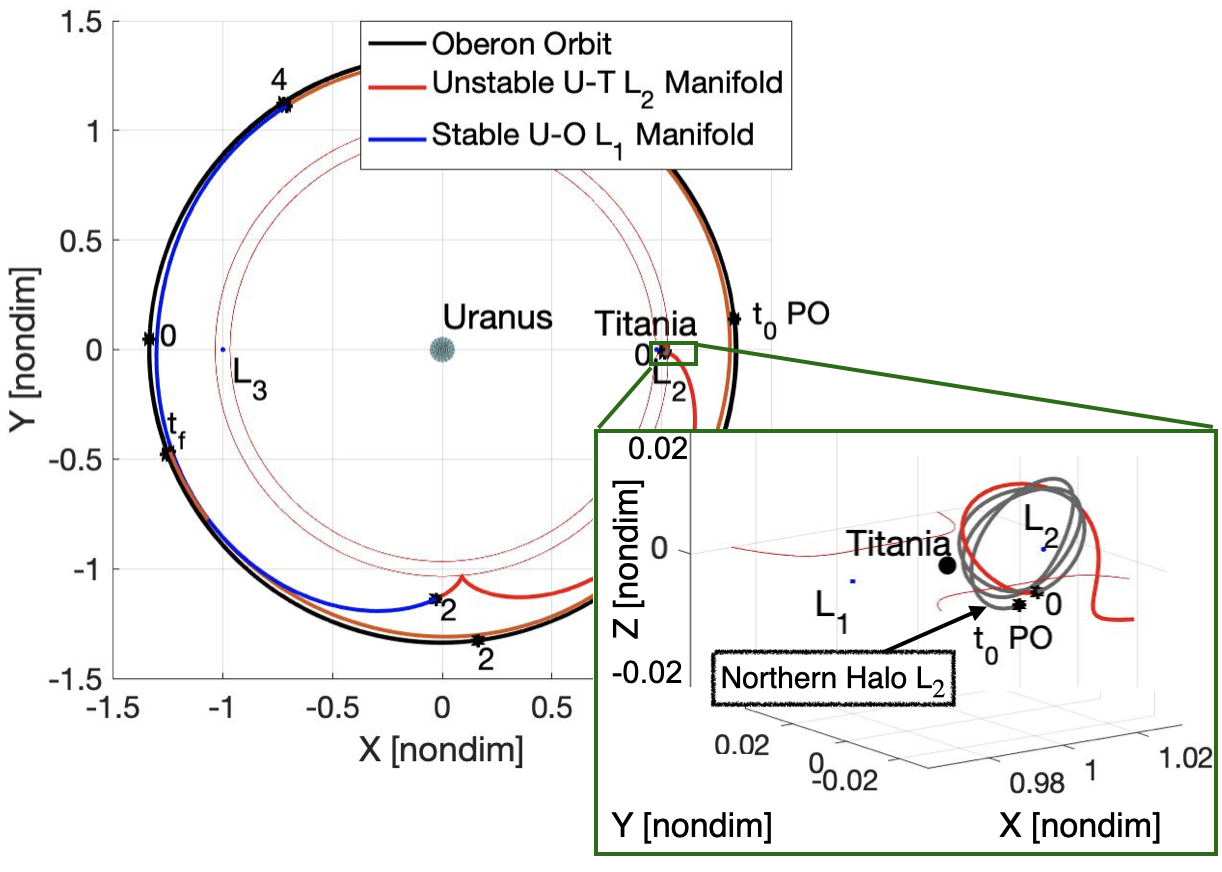}}}%
\end{minipage}\hfill{}%
\begin{minipage}[b][1\totalheight][t]{0.33\columnwidth}%
\subfigure[Uranus-Oberon rotating frame.]{\label{fig:ephemerisOberon}{\includegraphics[width=7cm]{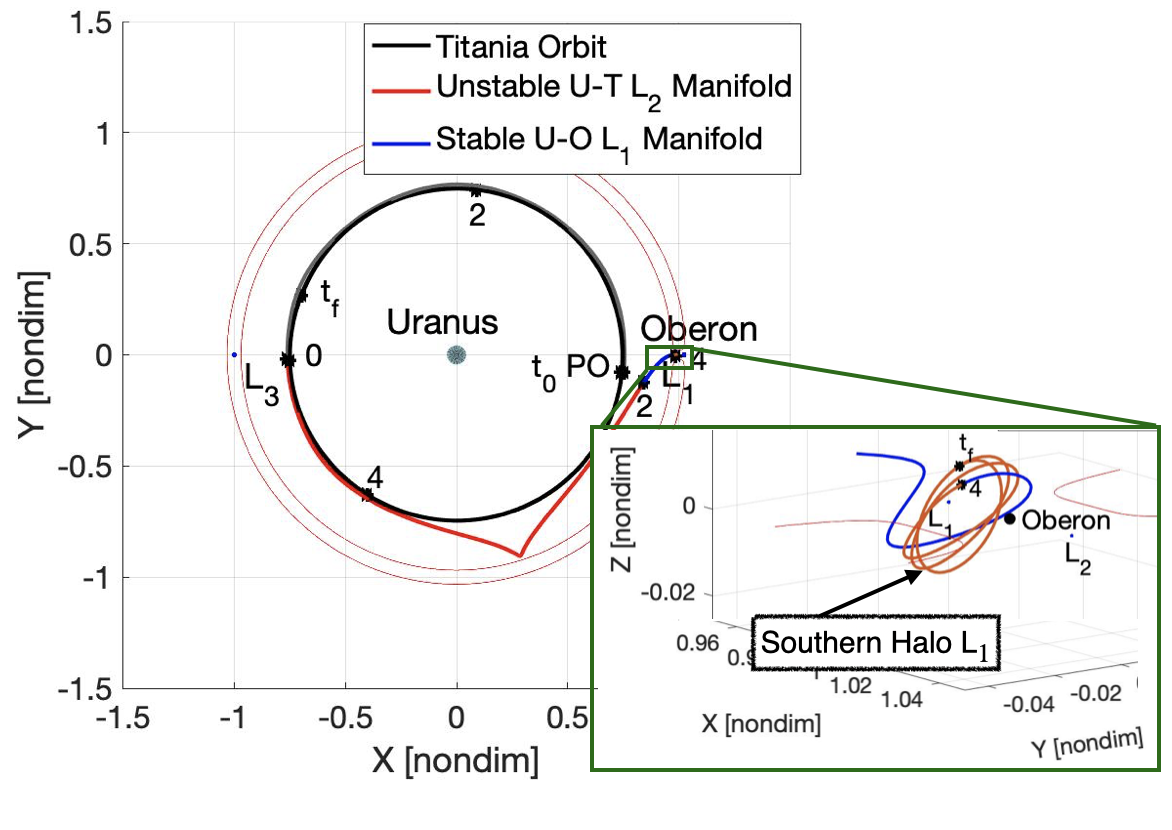}}}%
\end{minipage}\hfill{}
\caption{\label{fig:EphemerisUranus}Transfer from an L$_2$ northern halo orbit of the Uranus-Titania system to an L$_1$ southern halo orbit of the Uranus-Oberon
system: $\Delta v_{tot}=42.1$ m/s and $t_{tot}=28.44$ days and, higher-fidelity ephemeris model at epoch JD2459766.565 (July 6th, 2022).}
\end{figure}

A similar analysis is completed for the transfer between halo orbits in the Uranian system (Fig. \ref{fig:SpatialTransferResultUranus}). The higher-fidelity ephemeris model for this example is defined with Uranus as the central body, and the Sun, Ariel, Umbriel, Titania and Oberon as the perturbing bodies. Figure \ref{fig:EphemerisUranus} illustrates that, given the coupled spatial CR3BP transfer as an initial guess, a transfer between spatial periodic orbits is successfully transitioned to a higher-fidelity ephemeris model. It is therefore demonstrated that there exist transfers between halo orbits near both Titania and Oberon that can occur in less than a month and using a single $\Delta v_{tot}$ of less than 50 m/s (Table \ref{tab:ephemerisTable}).
\begin{table}[h!]
  \begin{center}
    \caption{Comparison between coupled spatial CR3BP and higher-fidelity ephemeris model for selected transfers.}
    \label{tab:ephemerisTable}
    \begin{tabular}{c|cc|cc}
      & \multicolumn{2}{c|}{Coupled spatial CR3BP model} &  \multicolumn{2}{c}{Ephemeris model}\\
      \hline
       Ganymede to Europa transfer &$\Delta v_{tot}$  [km/s]&$t_{tot}$ [days]&  $\Delta v_{tot}$  [km/s]&$t_{tot}$ [days]\\
      \hline
    Minimum-$\Delta v_{tot}$ transfer& 0.9422& 9.473&0.966&9.47\\ 
      Maximum-$t_{tot}$ transfer& 1.08&12.13&1.02&12.13\\ 
      \hline
             Titania to Oberon transfer &[m/s]&[days]&[m/s]&[days]\\
       \hline
    Minimum-$\Delta v_{tot}$ transfer& 45.7& 28.44&42.1&28.44\\ 
    \end{tabular}
  \end{center}
\end{table}

\section{\label{sec:Conclusions}Discussion and Concluding Remarks}
Trajectory design for transfers between different moons moving in the vicinity of
a common planet is a balance between diverse constraints, priorities
and requirements to enable trajectory design for successful missions. In this investigation, the MMAT method is introduced, a strategy and procedure to construct transfers between two different moons moving around a common planet within the context of the CR3BP. The analysis supports transfers between libration point orbits near different moons. Structures that arise in the CR3BP formulation are smoothly incorporated for moon tour design by blending conics from the 2BP with arcs from the CR3BP. In particular, transfers between planar and spatial libration point orbits for two different planet-moon systems are accomplished. The significant advantage of this approach is that the moon orbits are not assumed to be coplanar, thus producing direct solutions with the moons in their true orbital planes. Incorporating the 2BP-CR3BP patched model, two analytical constraints are identified for both coplanar and non-coplanar moon orbits. These analytical conditions determine the geometrical relationships between the conics and the orbital planes such that the epoch of the arrival moon is shifted for an intersection to occur. If such geometrical relationships between the conics are not met, then a direct transfer between the two moons using the selected departure and arrival arcs is not possible, and an intermediate arc to bridge the geometrical relationship is required. However, a theorem is introduced such that the fulfillment of the analytical relationship implies that there exists a relative phase between the moons that allows the intersection in space. Consequently, the MMAT method yields a phasing of the moons that is consistent with the actual moon orbits, given that the moons are located in their true orbital planes at a certain epoch. It is apparent that the analytical conditions are a requirement that must be satisfied in the search for connecting arcs between two distinct moons. By means of assessing a CR3BP state as an instantaneous conic at a sufficiently distant location from the moon, this simplification efficiently narrows the options for CR3BP arcs that allow a connection with another moon vicinity. Using the MMAT method prior to the introduction of a Poincar\'e section aids in the down-selection of relative positions between the departure and arrival moons as well as possible spatial locations where the departure and arrival arcs intersect. Therefore, the search for relative phases and locations in the coupled spatial CR3BP is reduced, saving considerable computation time. Furthermore, given the purely mathematical properties of the constraints, the strategy is applicable for any moon-to-moon transfers using CR3BP arcs, regardless of the journey being outward or inward. 

In conclusion, the MMAT method is introduced to produce a direct transfer between
two distinct moons in their true orbital planes in the coupled spatial CR3BP. A representation incorporating the various elements of the MMAT methodology is presented in Fig. \ref{fig:MMATmethod}. For any given angle of departure from one moon,
if Eq. \eqref{eq:3dConstraint} is fulfilled, a unique
natural transfer between CR3BP arcs is produced with a single $\Delta v_{tot}$, given that the unique natural configuration between moons 
in their respective planes is acquired for that given scenario, according to Theorem  \ref{theorem:spatial}. Therefore, the phase combinations between the moons that yield the most cost-efficient results emerge. Given that the root of the MMAT method is fundamentally analytical, the strategy applies for any moon-to-moon transfer around a common planet. The CR3BP arcs in this investigation are manifold trajectories, but it could also be applied to other types of arcs, such as transit orbits. Since the resulting spatial transfer is ultimately designed within the context of the coupled spatial CR3BP, an effective and smooth transition to a higher-fidelity ephemeris model is validated, reflecting potential transfers in actual systems where many perturbations are present. Also, from this analysis, new solutions with small $\Delta v_{tot}$s and total times-of-flight are offered for both transfers between Lyapunov orbits near Ganymede and Europa, as well as transfers between halo orbits in the vicinity of Titania and Oberon, using only a single maneuver. For the sake of comparison, all the solutions for the transfers in both systems are summarized in Table \ref{tab:summaryTable}. With these results and compared to the reviewed literature, the MMAT method locates cost-effective impulsive transfers between key moons of the solar system. Finally, note that the constraint in Eq. \eqref{eq:3dConstraint} is significant for designing trajectories not only in the coupled spatial
CR3BP, but also in the fundamental CR3BP when the motion is governed by the larger primary. 
\begin{table}[h!]
  \begin{center}
    \caption{Summary of the sample applications for transfers between Lyapunov orbits of Ganymede and Europa and between halo orbits of Titania and Oberon.}
    \label{tab:summaryTable}
    \begin{tabular}{c|cc|cc}
      & \multicolumn{2}{c|}{Coupled spatial CR3BP model} &  \multicolumn{2}{c}{Ephemeris model}\\
      \hline
      Transfer between Lyapunov &$\Delta v_{tot}$ &$t_{tot}$&  $\Delta v_{tot}$ &$t_{tot}$\\
      orbits of Ganymede and Europa&[km/s]&[days]&[km/s]&[days]\\
      \hline
    Minimum-$\Delta v_{tot}$ transfer& 0.9422& 9.473&0.966&9.47\\ 
      Maximum-$t_{tot}$ transfer& 1.08&12.13&1.02&12.13\\ 
      Minimum-$t_{tot}$ transfer & 0.9428& 9.472&-&- \\

      \hline
        Transfer between halo &&&&\\
         orbits of Titania and Oberon &[m/s]&[days]&[m/s]&[days]\\
       \hline
    Minimum-$\Delta v_{tot}$ transfer& 45.7& 28.44&42.1&28.44\\ 
    Minimum-$\Delta v_{tot}$ transfer&&&& \\ 
    with longer $t_{tot}$& 50& 39.6&-&-\\
    Minimum-$\Delta v_{tot}$ transfer&&&&\\ 
   reducing Titania's SoI & 66.7& 17.72&-&-\\
    \end{tabular}
  \end{center}
\end{table}

\begin{acknowledgements}
Assistance from colleagues in the Multi-Body Dynamics
Research group at Purdue University is appreciated as is the
support from the Purdue University School of Aeronautics and Astronautics
and College of Engineering including access to the Rune and Barbara Eliasen Visualization Laboratory. The authors also thank the anonymous reviewers for their thoughtful comments and suggestions. The paper is much improved as a result of their input.
\end{acknowledgements}

%
%

\bibliographystyle{spbasic}      

\bibliography{references}

\clearpage
\appendix
\section{\label{appendixRotations}Transformations between rotating and inertial frames}
\subsection{\label{appendixRotToIner}Transformation from the rotating to the inertial frame}
The transformation of a state from the rotating frame to the inertial frame is straightforward. Assuming that the
moons move on circular orbits (zero eccentricity) defined in terms of an epoch in the Ecliptic
J2000.0 frame, the position ($\bar{r}$)
and velocity ($\dot{\bar{r}}$) of the s/c are produced in the Ecliptic J2000.0 planet-centered inertial frame given $\Omega_{moon}$, $i_{moon}$, $a_{moon}$ and $\theta_{moon}$. Note that the moon location is evaluated
using $\theta_{moon}=st+\theta_{0}$, where $s$
is the angular velocity along the moon orbit given the period of the moon,
$t$ is the actual time in seconds, and $\theta_{0}$ is the angle of the moon with
respect to the ascending node line at the time of departure ($t=0$).
Given the moon position ($\bar{r}_{moon}$)
and velocity ($\dot{\bar{r}}_{moon}$) at a time $t$  in the Ecliptic J2000.0 frame, the construction of the planet-centered conic employs $\bar{h}_{moon}=\frac{\bar{r}_{moon}\times \dot{\bar{r}}_{moon}}{|\bar{r}_{moon}\times \dot{\bar{r}}_{moon}|}$, i.e.,
the angular momentum of the moon at the given time along the trajectory.
The rotation matrix  is represented by ${\bf R}=[\hat{x}^T\ \hat{y}^T\ \hat{z}^T]$,
such that $\hat{x}=\frac{\bar{r}_{moon}}{\left|\bar{r}_{moon}\right|}$, $ \hat{z}=\frac{\bar{h}_{moon}}{\left|\bar{h}_{moon}\right|}$ and $\hat{y}=\hat{z}\times\hat{x}$ define the instantaneous axes of the planet-moon rotating
frame. Note that boldface denotes a matrix. Given that the s/c rotating position and velocity states ($\bar{r}_{rot}$
and $\dot{\bar{r}}_{rot}$, respectively) are computed relative to the barycenter,
$\mu$ is added to the $x$-state of the position vector to shift the origin to the planet, but, the
velocity components remain the same after a translation. Using the basic
kinematic equation, the velocity in the inertial frame, expressed in
the rotating basis, is computed as $\dot{\bar{r}}_{in}=\dot{\bar{r}}_{rot}+\bar{s}_{rot}\times \bar{r}_{rot}$, where $\bar{s}_{rot}=[0\ 0\ s]^T$.
As a result, the position and velocity in the inertial basis are defined
by $\bar{r}={\bf R}\bar{r}_{rot}$ and $\dot{\bar{r}}={\bf R}\dot{\bar{r}}_{in}$.

\subsection{\label{appendixInerToRot}Transformation from the inertial to the rotating frame}
The transformation of a state from the Ecliptic J2000.0 planet-centered
inertial frame to the rotating frame follows a reverse rotation from the one in Appendix \ref{appendixRotToIner}. 
The rotation matrix is again defined by ${\bf R}$. Given the orbital angular
velocity of the moon, $\bar{s}_{in}={\bf R}\bar{s}_{rot}^T$,
the state of the s/c as expressed in the rotating frame is evaluated as $\bar{r}_{rot}={\bf R}^T\bar{r}$
and $\dot{\bar{r}}_{rot}={\bf R}^T(\dot{\bar{r}}-\bar{s}_{in}\times \bar{r})$.
Finally, to locate the state relative to the barycenter, $\mu$ of the arrival planet-moon CR3BP system is added to
the $x$ position component. Recall that arcs rotated from the Ecliptic J2000.0 planet-centered
inertial frame to the arrival rotating frame are scaled with the characteristic quantities of the arrival planet-moon CR3BP system. 

\section{\label{appendix:coupledCorrections}Differential corrections in the coupled spatial CR3BP}
\begin{figure}
\hfill{}\centering\includegraphics[width=12cm]{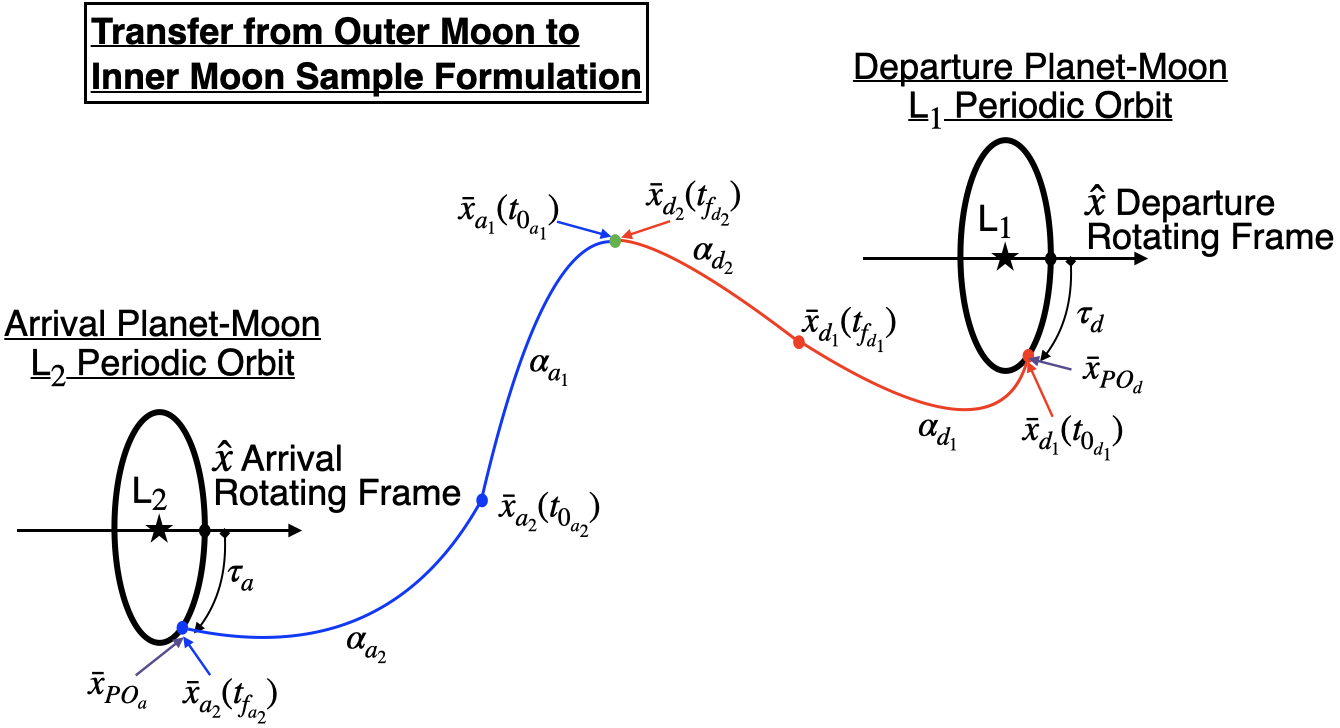}\hfill{}
\caption{\label{fig:taoAlphaScheme}Scheme that represents a sample formulation of the corrections problem to determine a  transfer between moons in the coupled spatial CR3BP using the $\tau$-$\alpha$ method introduced in \citet{Haapala2015}.}
\end{figure}

To produce transfers between moons in the coupled spatial CR3BP, a multiple shooter serves as the basis for a differential corrections algorithm. In particular, differential corrections strategies based on multi-variable Newton methods are applied in this investigation to solve boundary value problems. The multiple-shooter scheme is inspired by the $\tau$-$\alpha$ method introduced in \citet{Haapala2015}. This algorithm produces connections between unstable and stable manifold trajectories, allowing the departure or arrival location on the manifold trajectory in the periodic orbit to be free. A sample schematic of the multiple-shooter is illustrated in Fig. \ref{fig:taoAlphaScheme}. Note that, in this example, both departure and arrival manifold trajectories are subdivided into 2 segments, but the number of segments could be any positive integer number. The times $\tau_{d}$ and $\tau_{a}$ correspond to a location in position and velocity at which the manifold trajectory departs from or arrives at the periodic orbit with respect to the point where the periodic orbit originates. The $\alpha$'s correspond to the time span along each segment that discretizes the unstable and stable manifolds as reflected in Fig.  \ref{fig:taoAlphaScheme}.  The vectors $\bar{x} _{PO_{d}}$ and $\bar{x} _{PO_{a}}$ are the states at which the manifold trajectory departs or arrives at the periodic orbit, respectively. The states along the segments on the unstable manifold trajectory are denoted as $\bar{x} _{d_{k}}$, where $k=1, \dots, N_{d}$ and $ N_{d}$ is the number of segments that comprise such a trajectory. Similarly, the states along the segments on the arrival stable manifold trajectory are denoted as $\bar{x} _{a_{k}}$, where $k=1, \dots, N_{a}$ and $N_{a}$ is the number of segments along the arrival trajectory. The variables $t_{0_{d_k}}$ and $t_{f_{d_{k}}}$ are the initial and final times corresponding to a departure segment, respectively, and $t_{0_{a_{k}}}$ and $ t_{f_{a_{k}}}$ are the initial and final times on an arrival segment. Hence, the objective is the determination of a design variable vector, $\bar{X}$,  that satisfies the constraint vector, $\bar{F}(\bar{X})=\bar{0}$, and ensures position continuity at the intersection between the departure and arrival arcs, but allows a discontinuity in velocity:
\begin{equation}
\bar{X}=\left(\begin{array}{c}
\tau_{d}\\
\tau_{a}\\
\sum_{k=1}^{N_{d}}\bar{x}_{d_{k}}(t_{f_{d_{k}}})\\
\sum_{k=1}^{N_{d}}\alpha_{d_{k}}\\
\sum_{k=1}^{N_{a}}\bar{x}_{a_{k}}(t_{0_{a_{k}}})\\
\sum_{k=1}^{N_{a}}\alpha_{a_{k}}
\end{array}\right),
\end{equation}
\begin{equation}
\label{eq:constraintsShooter}
\bar{F}=\left(\begin{array}{c}
x_{a_{1}}(t_{0_{a_{1}}})-x_{d_{N_{d}}}(t_{f_{d_{N_{d}}}})\\
y_{a_{1}}(t_{0_{a_{1}}})-y_{d_{N_{d}}}(t_{f_{d_{N_{d}}}})\\
z_{a_{1}}(t_{0_{a_{1}}})-z_{d_{N_{d}}}(t_{f_{d_{N_{d}}}})\\
\sum_{k=2}^{N_{a}}(\bar{x}_{a_{k}}(t_{0_{a_{k}}})-\bar{x}_{a_{k-1}}(t_{f_{a_{k-1}}}))\\
\sum_{k=2}^{N_{d}}(\bar{x}_{d_{k}}(t_{0_{d_{k}}})-\bar{x}_{d_{k-1}}(t_{f_{d_{k-1}}}))\\
\bar{x}_{d_{1}}(t_{0_{d_{1}}})-\bar{x}_{PO_{d}}\\
\bar{x}_{a_{N_{a}}}(t_{f_{a_{N_{a}}}})-\bar{x}_{PO_{a}}
\end{array}\right)=\bar{0}.
\end{equation}
Note that $x$, $y$ and $z$ in $\bar{F}$ correspond to the position states in the arrival rotating frame in which position continuity is ensured between the last segment of the unstable manifold trajectory and the first segment of the stable manifold trajectory.

Numerical strategies are required to determine trajectories that satisfy the constraints in Eq. \eqref{eq:constraintsShooter}. To solve for $\bar{X}$, a Taylor series (truncated to first order) is expanded on $\bar{F}(\bar{X})=\bar{0}$ about the free-variable initial guess, $\bar{X}_0$. Given the relative lengths of the design and constraint vectors, the problem is solved by determining the minimum norm solution that satisfies $\bar{F}(\bar{X})=\bar{0}$ within a given tolerance. Then, ${\bf DF}=\left[ 
\frac{\partial F_i}{\partial X_j}\right]$ is the Jacobian matrix and, consequently, it is necessary to solve the problem using an iterative corrections scheme. Since, as previously mentioned, departure trajectories are rotated onto the arrival moon plane, the former become time dependent and the Jacobian matrix is, thus, solved numerically by means of  a central difference approximation. The general goal behind central difference is to numerically evaluate differential equations by approximating the slope of the solution at discretized points in time. As a result, each term inside the Jacobian matrix is computed numerically as
\begin{equation}
\frac{\partial F_i}{\partial X_j}=\frac{F_i(X_j+h)-F_i(X_j-h)}{2h},
\end{equation}
where $h$  represents a small perturbation value. To conclude, using the presented methodology, transfers in the coupled spatial CR3BP are successfully corrected given an adequate initial guess obtained from the 2BP-CR3BP patched model.

\section{\label{appendix:spatialCorrections}Differential corrections for transfers between spatial periodic orbits in the 2BP-CR3BP patched model}
The objective is the determination of a design variable vector, $\bar{X}$, defined by the arrival moon epoch at arrival, $\theta_{4_{m}}$, the departure conic time-of-flight, $t_{d}$, and the arrival conic time-of-flight, $t_{a}$, that satisfies the constraint vector, $\bar{F}(\bar{X})=\bar{0}$, to ensure position continuity at the intersection between a departure and arrival conic (or departure and arrival plane):
\begin{equation}
\bar{F}(\bar{X})=\left\{ \begin{array}{c}
x_{d}-x_{a}\\
y_{d}-y_{a}\\
z_{d}-z_{a}
\end{array}\right\} =\bar{0},\ \bar{X}=\left\{ \begin{array}{c}
\theta_{4_{m}}\\
t_{d}\\
t_{a}
\end{array}\right\} .
\end{equation}
To solve for $\bar{X}$, a Taylor series (truncated to first order) is expanded on $\bar{F}(\bar{X})=\bar{0}$ about the free-variable initial guess, $\bar{X}_0$: $\bar{F}(\bar{X}_0)+{\bf DF}(\bar{X}_0)(\bar{X}-\bar{X}_0)=\bar{0}$.  The vector $\bar{X}_0$ is defined with the values $\theta_{4_{m}}$, $t_{d}$ and $t_{a}$ obtained from the projection of $\sigma$ onto the arrival moon plane (initial guess). The Jacobian matrix ${\bf DF}$ of this particular problem corresponds to:
\begin{equation}
{\bf DF}=\left[ \begin{array}{c c c}
\frac{\partial(x_{d}-x_{a})}{\partial\theta_{4_{m}}}& \frac{\partial(x_{d}-x_{a})}{\partial t_d}&\frac{\partial(x_{d}-x_{a})}{\partial t_a}\\
\frac{\partial(y_{d}-y_{a})}{\partial\theta_{4_{m}}}& \frac{\partial(y_{d}-y_{a})}{\partial t_d}&\frac{\partial(y_{d}-y_{a})}{\partial t_a}\\
\frac{\partial(z_{d}-z_{a})}{\partial\theta_{4_{m}}}& \frac{\partial(z_{d}-z_{a})}{\partial t_d}&\frac{\partial(z_{d}-z_{a})}{\partial t_a}
\end{array}\right].
\end{equation}
Given that the arrival moon trajectories are rotated towards the Ecliptic J2000.0 planet-centered inertial frame, the {\bf DF} matrix is computed numerically by means of  a central difference approximation, which is introduced in Appendix \ref{appendix:coupledCorrections}. Note that this differential corrections scheme is only required to design direct moon-to-moon transfers between spatial periodic orbits in the 2BP-CR3BP patched model. 

\section{\label{appendixRotationsEphemeris}Transformations between rotating and inertial frames in a higher-fidelity ephemeris model}
For computing the rotations between the inertial and rotating frames in Appendix \ref{appendixRotations}, both moon orbits are assumed circular. Nevertheless, in a higher-fidelity ephemeris model, the orbits are no longer circular. Given the moon position ($\bar{r}_{m}$)
and velocity ($\dot{\bar{r}}_{m}$) at a certain epoch $t$ in the Ecliptic J2000.0 frame, an instantaneous rotating frame is constructed in an inertial planet-centered frame using the formulation introduced in Appendix \ref{appendixRotations}. Nevertheless, the characteristic time and length 'pulsate' and this fact must be incorporated when expressing states in dimensionless form, since the position and velocity of the planet and the moon with respect to the barycenter vary over time. Note that the instantaneous characteristic length and time are denoted $\tilde{l}_*$ and $\tilde{t}_*$, respectively.

\subsection{Transformation from the inertial to the rotating frame}
Given the position ($\bar{r}_{p-s/c}$) and velocity ($\dot{\bar{r}}_{p-s/c}$) of the s/c with respect to the central body in dimensional units in the inertial frame, the base point is shifted to the barycenter: $\bar{r}_{B-s/c}=\bar{r}_{p-s/c}+\bar{r}_{B-p}$ and $\dot{\bar{r}}_{B-s/c}=\dot{\bar{r}}_{p-s/c}+\dot{\bar{r}}_{B-p}$. Here, the subscript '$B$' denotes the barycenter. Given that the planet is the central body,  $\bar{r}_{B-p}=\mu \bar{r}_{p-m}$, where $\mu$ is the mass ratio for the CR3BP system and $\bar{r}_{p-m}$ is the location of the moon relative to the planet. Also, $\dot{\bar{r}}_{B-p}=\mu \dot{\bar{r}}_{p-m}$, where $\dot{\bar{r}}_{p-m}$ is the velocity of the moon with respect to the planet. The position state of the s/c in the rotating frame is expressed as $\bar{r}_{rot}={\bf R}^T \bar{r}_{B-s/c}$. Using the basic
kinematic equation, the velocity in the rotating frame, expressed in
the rotating basis, is computed as $\dot{\bar{r}}_{rot}={\bf R}^T(\dot{\bar{r}}_{B-s/c}-\frac{1}{\tilde{l}_*^2}\bar{h}_{m}\times \bar{r}_{B-s/c}-\frac{V_r}{\tilde{l}_*}\bar{r}_{B-s/c})$. The component $V_r$ is the dimensional, instantaneous radial velocity of the moon with respect to the planet: $V_r=\frac{\bar{r}_{p-m}\cdot \dot{\bar{r}}_{p-m}}{|\bar{r}_{p-m}|}$. Finally, the dimensional states are scaled by the instantaneous characteristic quantities to remove pulsation from the rotating coordinates.

\subsection{Transformation from the rotating to the inertial frame}
Given the position ($\bar{r}_{rot}$) and velocity ($\dot{\bar{r}}_{rot}$) of the s/c with respect to the central body in dimensional units in the rotating frame, and the angular momentum vector in the rotating frame $\bar{h}_{rot}={\bf R}^T\bar{h}_{m}$, it is possible to compute the velocity in the inertial frame, expressed in the instantaneous rotating basis:  $\dot{\bar{r}}_{in}=\dot{\bar{r}}_{rot}+\frac{1}{\tilde{l}_*^2}\bar{h}_{rot}\times \bar{r}_{rot}+\frac{V_r}{\tilde{l}_*}\bar{r}_{rot}$. Therefore, the position and velocity in the inertial basis are defined
by $\bar{r}_{p-s/c}={\bf R}\bar{r}_{{rot}}-\bar{r}_{B-p}$ and $\dot{\bar{r}}_{p-s/c}={\bf R}\dot{\bar{r}}_{in}-\dot{\bar{r}}_{B-p}$. Finally, the dimensional states are scaled by the characteristic quantity reflecting the rotating frame in the CR3BP system (not the instantaneous quantity).

\section{\label{appendix:FeasibilityAnalysisUranus}Feasibility analysis for successful transfers for the application between Titania and Oberon}
\begin{figure}[htbp!]
\hfill{}\centering%
\begin{minipage}[b][1\totalheight][t]{0.4\columnwidth}%
\subfigure[Evaluation of the constraint Eq. \eqref{eq:3dConstraint} at $\theta_{d_{Int}}$ and $\theta_{d_{Int}}+\pi$ .]{{\includegraphics[width=5.5cm]{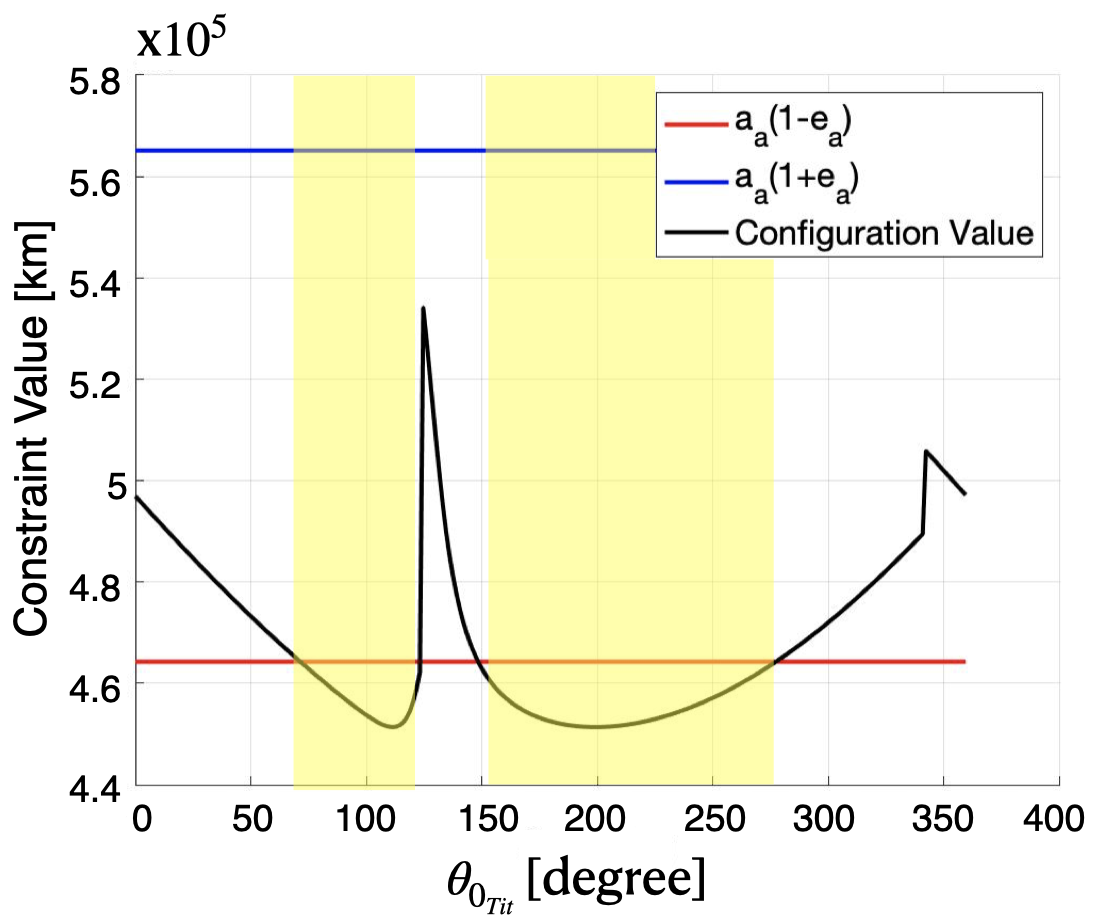}}}%
\end{minipage}\hfill{}%
\begin{minipage}[b][1\totalheight][t]{0.33\columnwidth}%
\subfigure[Phase of Titania and Oberon with respect to their ascending node direction at $t_0$.]{{\includegraphics[width=5.5cm]{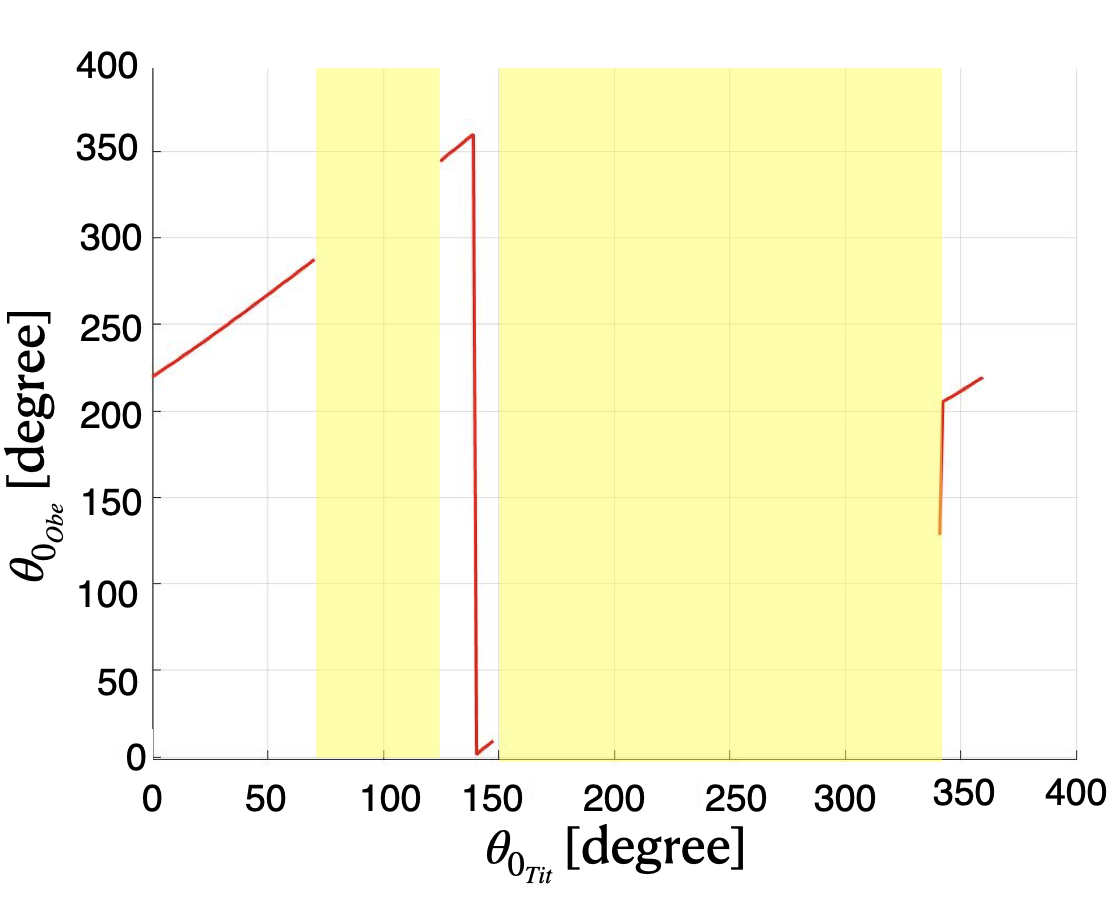}}}%
\end{minipage}\hfill{}
\hfill{}\centering%

\hfill{}\centering%
\begin{minipage}[b][1\totalheight][t]{0.4\columnwidth}%
\subfigure[Transfer $\Delta v_{tot}$ given a $\theta_{0_{Tit}}$.]{{\includegraphics[width=5.5cm]{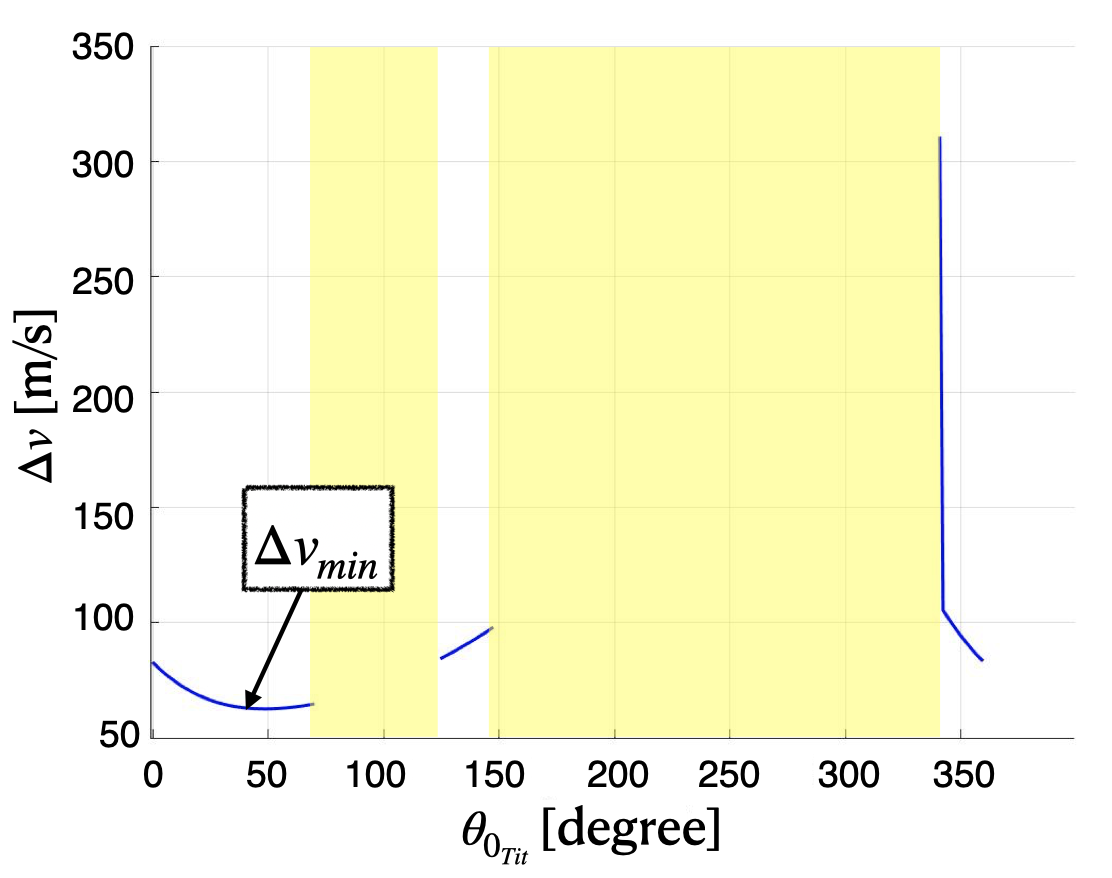}}}%
\end{minipage}\hfill{}%
\begin{minipage}[b][1\totalheight][t]{0.33\columnwidth}%
\subfigure[Transfer $t_{tot}$ given a $\theta_{0_{Tit}}$]{{\includegraphics[width=5.5cm]{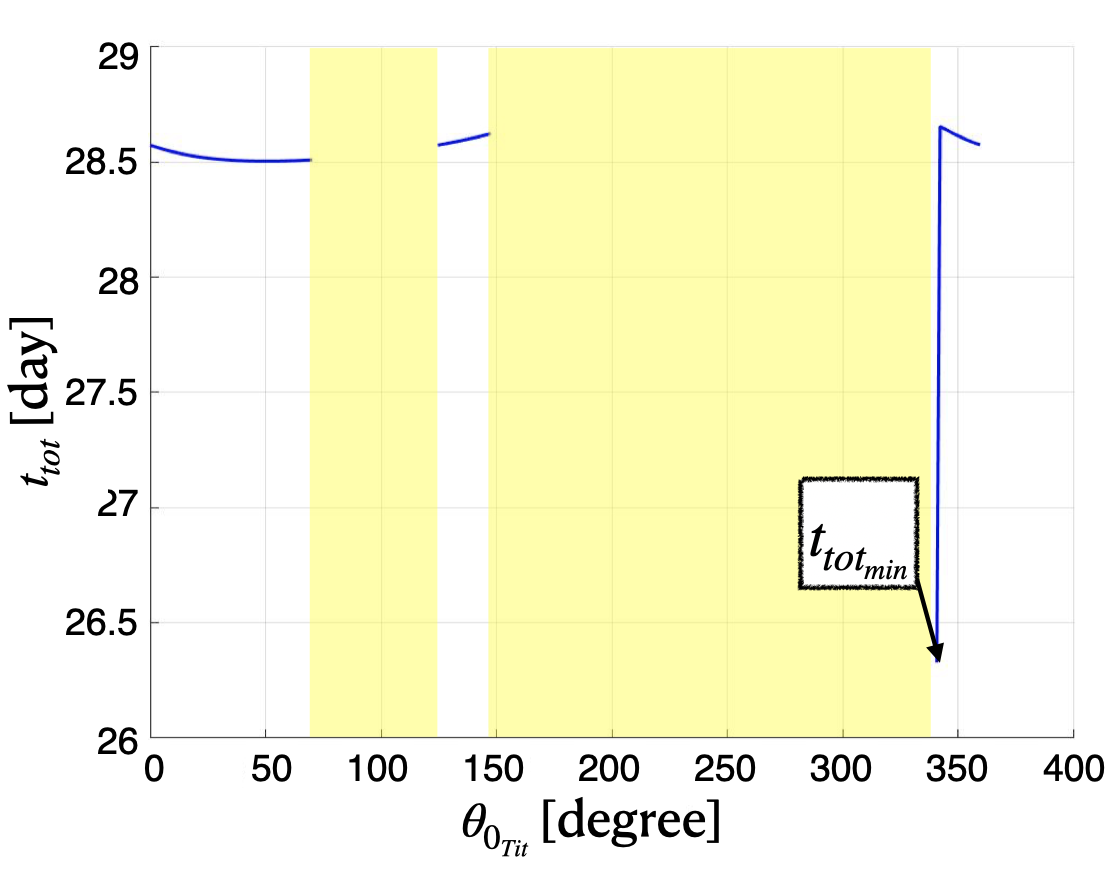}}}%
\end{minipage}\hfill{}
\hfill{}\centering%
\centering{}\caption{\label{fig:thetaTitVariationSpatial}Available successful configurations for the intersection
at $\theta_{d_{Int}}$. Note that yellow regions correspond to a phase of departure from Titania in its orbit where a direct transfer to the arrival arc cannot occur.}
\end{figure}

\begin{figure}[htbp!]
\hfill{}\centering\includegraphics[width=5.5cm]{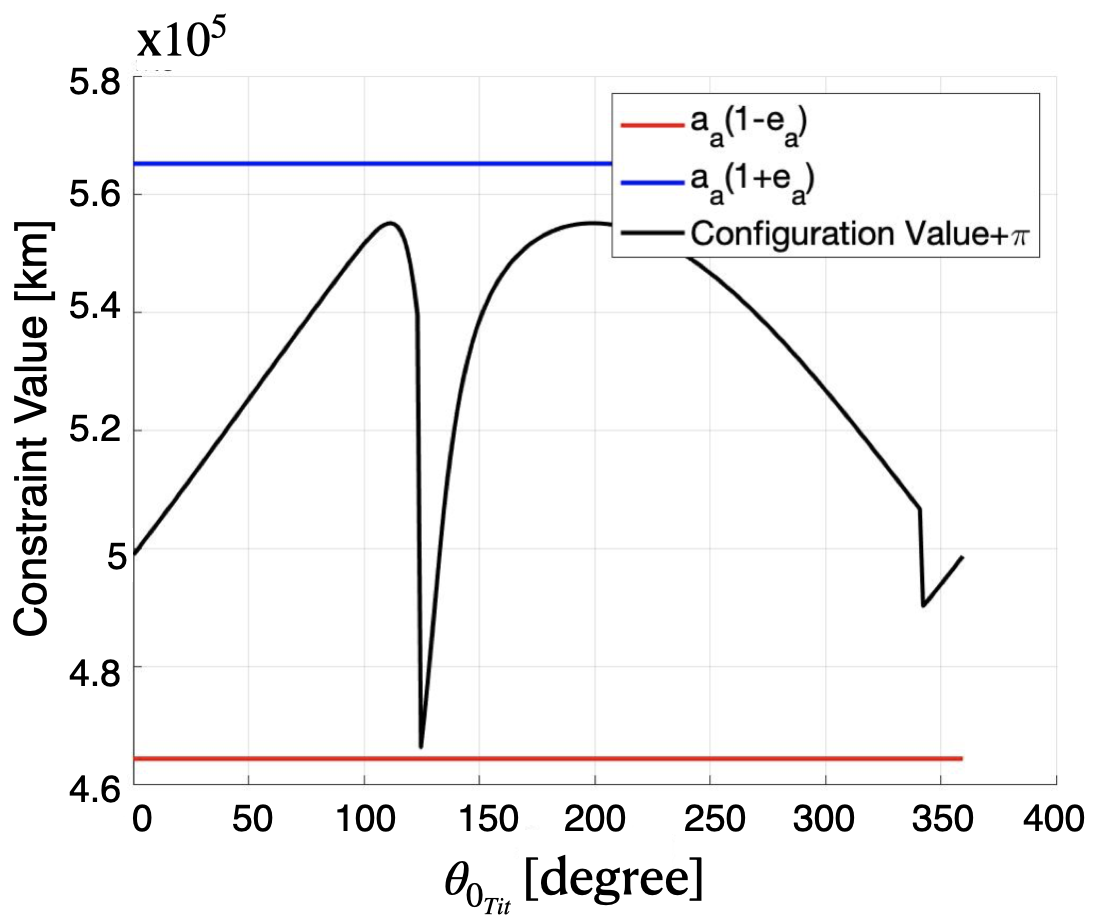}\hfill{}
\caption{\label{fig:thetaTitVariationSpatialPi} Evaluation of the constraint Eq. \eqref{eq:3dConstraint} at $\theta_{d_{Int}}+\pi$.}
\end{figure}

The feasibility analysis for the transfers between halo orbits near Titania and Oberon is represented in Fig. \ref{fig:thetaTitVariationSpatial}. Note that the analysis is presented for the $\theta_{d_{Int}}$ configuration because it is the one that supplies the minimum-$\Delta v_{tot}$ configuration. The regions that are light yellow correspond to initial departure epochs where a feasible transfer is not encountered. However, it is also apparent in Fig. \ref{fig:thetaTitVariationSpatialPi} that the case of $\theta_{d_{Int}}+\pi$ is also interesting given that it supplies successful configurations for all possible $\theta_{0_{Tit}}$. For this case, the configuration value is between the upper and lower limits of Eq. \eqref{eq:3dConstraint} (blue and red lines, respectively). 

\end{document}